\pdfoutput=1
\documentclass[preprint,3p,times]{elsarticle}

\usepackage{amssymb}
\usepackage{amsthm}
\usepackage{amsmath}
\usepackage{epsfig}
\usepackage{subfigure}
\usepackage[usenames]{color}
\usepackage{algorithm}
\usepackage{algorithmic}
\usepackage{nicefrac}
\epsfverbosetrue
\graphicspath{{fig/}{../fig/}}

\newcommand{\mb}[1]{\ensuremath{\mathbf{#1}}}

\journal{Computer Methods in Applied Mechanics and Engineering}

\begin{document}
 \setlength{\parindent}{0.0ex}
 \setcounter{secnumdepth}{4}
 \setcounter{tocdepth}{4}
\begin{frontmatter}



\title{A Finite Element Approach for the Line-to-Line Contact Interaction of Thin Beams with Arbitrary Orientation}


\author[lnm]{Christoph Meier}
\author[lnm]{Alexander Popp\corref{cor1}}
\ead{popp@lnm.mw.tum.de}
\author[lnm]{Wolfgang A. Wall}

\address[lnm]{Institute for Computational Mechanics, Technische Universit\"at M\"unchen, Boltzmannstrasse 15, D--85748 Garching b. M\"unchen, Germany}

\cortext[cor1]{Corresponding author}

\begin{abstract}
The objective of this work is the development of a novel finite element formulation describing the contact behavior of slender beams in 
complex 3D contact configurations involving arbitrary beam-to-beam orientations. It is shown by means of a mathematically concise 
investigation of well-known beam contact models based on point-wise contact forces that these formulations fail to describe 
a considerable range of contact configurations, which are, however, likely to occur in complex unstructured systems of thin fibers.
In contrary, the formulation proposed here models mechanical contact interaction of slender continua by means of 
distributed line forces, a procedure that is shown to be applicable for any geometrical contact configuration. 
The proposed formulation is based on a Gauss-point-to-segment type 
contact discretization and a penalty regularization of the contact constraint. Additionally, theoretical considerations concerning alternative 
mortar type contact discretizations and constraint enforcement by means of Lagrange multipliers are made. However, based on detailed 
theoretical and numerical investigations of these different variants, the penalty-based Gauss-point-to-segment formulation is suggested 
as the most promising and suitable approach for beam-to-beam contact.
This formulation is supplemented by a consistently linearized integration interval segmentation 
that avoids numerical integration across strong discontinuities. In combination with a smoothed contact force law and the employed $C^1$-continuous 
beam element formulation, this procedure drastically reduces the numerical integration error, an essential prerequisite for optimal spatial convergence rates. 
The resulting line-to-line contact algorithm is supplemented by contact contributions of the beam endpoints, which represent boundary minima 
of the minimal distance problem underlying the contact formulation. 
Finally, a series of numerical test cases is analyzed in order to investigate the accuracy and consistency 
of the proposed formulation regarding integration error, spatial convergence behavior and resulting contact force distributions. For one of these 
test cases, an analytical solution based on the Kirchhoff theory of thin rods is derived, which can serve as valuable benchmark for the 
proposed model but also for future beam-to-beam contact formulations. In addition to these examples, two real-world 
applications are presented in order to verify the robustness of the proposed formulation when applied to practically relevant problems.
\end{abstract}


\end{frontmatter}

% \newpage
%  
% \tableofcontents

%-------------------------------------------------------------------------------
%
\section{Introduction}
\label{sec:intro}
%
%-------------------------------------------------------------------------------
There exist many fields of application, where mechanical system behavior is crucially determined by slender fiber- or rod-like components. In technical applications, such fibers occur for example in 
industrial webbings, high-tensile ropes and cables, fiber-reinforced composite materials or synthetic polymer materials. Furthermore, also the fibers in biological systems such as muscles 
and biological tissue or the filaments in biopolymer networks \cite{cyron2012} can be identified as slender mechanical components of this type. In most cases, these fibers can be 
modeled with sufficient accuracy by applying a 1D beam theory. In the last three decades, many different types of beam element formulations have been proposed in order to discretize such beam models by means of 
the finite element method. In his recent contribution \cite{romero2008}, Romero points out the excellent performance of one specific category of beam elements denoted as geometrically 
exact beam formulations. While most of the geometrically exact beam formulations available in the literature are of Simo-Reissner type 
(see e.g. \cite{crisfield1999, eugster2013, jelenic1999, romero2004, romero2002, simo1985, simo1986, sonneville2014, zupan2003}), thus incorporating the modes of axial tension, shear, 
torsion and bending, in our recent contributions \cite{meier2014} and \cite{meier2015}, a shear-free formulation based on the Kirchhoff theory of thin rods, thus being tailored for 
the modeling of slender fibers, has been proposed. The latter formulation will also be applied within this contribution since it provides a $C^1$-continuous beam centerline 
representation, a desirable property enabling smooth contact kinematics in the context of beam-to-beam contact interaction.\\

Most of the applications mentioned above are characterized by mechanical contact interactions that significantly influence the overall system performance and by geometrically quite complex 
contact configurations, which allow for arbitrary fiber-to-fiber orientations. Despite the large number of publications concerning beam element formulations in general and despite the obvious need 
for robust and accurate beam contact formulations in many fields of application, there exists only a comparatively limited amount of literature focusing on beam-to-beam contact interaction. 
On the contrary, intensive research work has been done in the solid contact modeling of 3D continua within the last two decades. Important aspects in this field of research are for example 
the investigation of different constraint enforcement strategies (Lagrange multiplier method, penalty method, augmented Lagrange method etc.), 
types of contact discretization (node-to-segment/collocation-point-to-segment, Gauss-point-to-segment, mortar-like formulations), 
efficient contact search and active set strategies, procedures for Lagrange multiplier condensation and accurate integration schemes. 
Exemplarily, the reader is referred to the monographs \cite{laursen2002, wriggers2006} and to the review articles \cite{wohlmuth2011, popp2014}.\\

In contrast to contact formulations for 3D continua, which are typically based on a 2D contact traction field acting on the contact surfaces, the arguably most popular beam contact 
formulation \cite{wriggers1997} known in the literature models mechanical beam-to-beam contact interaction by means of a discrete contact force acting at the closest point between the 
two space curves representing the contacting 
beams (with circular cross-sections). This model, in the following denoted as point-to-point contact formulation, results in an elegant and efficient numerical formulation, which subsequently 
has been extended to frictional problems considering friction forces \cite{zavarise2000} and friction torques \cite{konjukhov2010}, rectangular beam cross-sections 
\cite{litewka2002,litewka2002b}, smoothed centerline geometries \cite{litewka2007}, constraint enforcement via Lagrange multipliers \cite{litewka2005} and adhesion effects \cite{kulachenko2012}. Quite recently, 
it has been applied to self-contact problems \cite{neto2015}. However, one of the limitations of these point-to-point contact formulations can be attributed to the question 
whether contact between beams enclosing small contact angles, i.e. 
nearly parallel or entangled beams, should rather be modeled by means of a distributed line force instead of a discrete point force from a mechanical point of view. This question has been 
addressed by the recent publications \cite{litewka2013} and \cite{litewka2015}, which propose additional contact points located in the neighborhood of the closest point in order to somewhat distribute 
the contact force in such configurations. Nevertheless, this formulation still relies on the existence of a locally unique closest point projection between the two contacting beams.\\

It is precisely this requirement that represents the second and essential limitation of point-to-point type beam contact formulations. In very general scenarios, such as in the applications mentioned 
in the beginning, where arbitrary beam-to-beam orientations can occur, a unique closest point projection cannot be guaranteed for all potential contact regions. Consequently, some mechanically relevant
contact points might be missed leading to large nonphysical penetrations or even to an entirely undetected crossing of the considered beams. There exist only a few alternative beam contact formulations 
available in the literature today that can overcome this limitation. One of these alternatives is the contact formulation developed by Durville 
\cite{durville2004}, \cite{durville2007}, \cite{durville2010}, \cite{durville2012}, which is based on a collocation-point-to-segment type formulation and the definition of proximity zones 
on an intermediate geometry. A second alternative proposed by Chamekh et al. \cite{chamekh2009}, \cite{chamekh2014} is based on a 
Gauss-point-to-segment type formulation and primarily investigates self-contact problems of beams. What these two formulations have in common is that the contact forces are 
\textit{distributed} along the two beams. Consequently, these types of formulations will be denoted as line-to-line contact formulations in the following.\\

The mentioned limitations of the point-to-point contact formulations were our motivation to perform mathematically concise and rigorous investigations concerning the existence 
of the corresponding closest-point-projection. In contrary to Konjukhov et al. \cite{konjukhov2008},\cite{konjukhov2010} who have already treated this question by means of geometrical criteria, 
we derive a very general analytical criterion that is valid for arbitrary contact configurations and that is based on proper and easy-to-determine control quantities. 
Based on this analytical criterion, we can conclude that the standard point-to-point contact formulation is not applicable in a considerable range of practically relevant 
contact configurations. This result encouraged us to develop new improved line-to-line contact formulations, which are applicable beyond the standard point contact model, 
and which are inspired by some well-known and successful techniques known from contact mechanics for 3D solids. Specifically, we propose a novel beam contact formulation based on a 
Gauss-point-to-segment type contact discretization and a penalty regularization of the contact constraint. Additionally, we make theoretical considerations concerning alternative 
constraint enforcement strategies by means of Lagrange multipliers and alternative contact discretizations based on mortar methods. However, detailed theoretical and numerical 
investigations of these different approaches suggest the penalty-based Gauss-point-to-segment formulation as the variant that is most suitable for beam-to-beam contact and as the method of 
choice for the applications considered within this work.\\ 

In contrast to existing line-to-line beam contact formulations, our approach is extended by a consistently linearized integration interval segmentation that avoids numerical integration 
across strong discontinuities. It is verified by means of suitable numerical examples that precisely this component in combination with a smoothed contact force law and the applied $C^1$-continuous 
beam element formulation leads to a drastic reduction of the numerical integration error. This, in turn, improves spatial convergence rates and in many cases only enables optimal convergence behavior 
under uniform mesh refinement. Furthermore, the resulting line-to-line 
contact algorithm is supplemented by contact contributions of the beam endpoints, whereas all existing contact formulations, no matter if point-to-point or line-to-line, typically search 
for minimal distance solutions only within the beams interior, but not for possible boundary minima. On the basis of a suitable numerical example, it is shown that in many applications 
these endpoint contact scenarios can appear with considerable frequency. Although, the influence of these endpoint forces on the overall solution quality might be of secondary interest, 
it is shown that neglecting these contributions will drastically reduce the robustness of the nonlinear solution scheme in many cases and may even prohibit convergence at all.
Finally, a numerical test case suitable for 
line-to-line contact scenarios has been designed and a corresponding analytical solution based on the 
Kirchhoff theory of thin rods has been derived. This test case and the associated analytical solution can serve as valuable benchmark for the proposed formulation 
but also for future beam-to-beam contact approaches.\\

The remainder of this paper is organized as follows. In Section~\ref{sec:beamformulation}, we briefly repeat the main constituents of the applied beam element formulation initially proposed in 
\cite{meier2014, meier2015} and 
extend the formulation to elastodynamics. In Section~\ref{sec:point}, the theory of standard point-to-point contact formulations is presented, followed by 
an analytical investigation of the existence and uniqueness of the required closest point projection and the derivation of a simple but mathematically concise criterion for solvability. 
In Section~\ref{sec:line}, the proposed line-to-line contact formulation is introduced and intensively compared to alternative methods known from the field of solid contact mechanics. 
The overall contact algorithm is completed by the contact contributions arising from the beam endpoints in Section~\ref{sec:endpoint} before a detailed numerical verification is performed 
in Section~\ref{sec:numerical_examples}. While the first four examples presented in Sections~\ref{sec:examples_example1}~-~\ref{sec:examples_example4} aim at investigating the accuracy and consistency 
of the new formulation regarding integration error, spatial convergence behavior and contact force evolutions, the final two examples in 
Sections~\ref{sec:examples_example5}~and~\ref{sec:examples_example6} represent possible real-world applications in order to verify the robustness of the proposed formulation when applied to 
practically relevant problems.

%-------------------------------------------------------------------------------
%
\section{Applied beam formulation}
\label{sec:beamformulation}
%
%-------------------------------------------------------------------------------

In our recent contributions \cite{meier2014, meier2015}, a geometrically exact beam element formulation according to the geometrically nonlinear Kirchhoff theory of thin rods 
incorporating the modes of axial tension, torsion and non-isotropic bending has been proposed. The underlying beam theory and the resulting finite element formulation are tailored to 
deal with problems involving highly slender fibers and the numerical challenges (e.g. membrane locking) resulting from such high beam slenderness ratios. In addition to the general 
element formulation, a reduced element formulation neglecting the mode of torsion has been proposed in \cite{meier2015} and has been shown to deliver identical results as the general beam 
element formulation when restricting the considered structures to initially straight beams with circular cross-sections and excluding axial/torsional moments from 
the set of external loads. Since these restrictions are easily fulfilled for the numerical examples considered in this contribution, we will exclusively resort to this simple and efficient 
torsion-free variant. However, the transfer 
of the following derivations from the torsion-free to the general element formulation is trivial, since the beam centerline representations of both are identical. The constituents 
of the static, torsion-free formulation presented in \cite{meier2015} will be summarized and extended to elastodynamics in the following.

%-------------------------------------------------------------------------------
%
\subsection{Continuum formulation}
\label{sec:beam_continuum}
%
%-------------------------------------------------------------------------------

The current configuration of the torsion-free beam is completely described by the beam centerline represented via a parametrized space curve $(s,t) \rightarrow \mb{r}(s,t) \in \Re^3$. 
Here, $s \in [0,l] \subset \Re$ 
and $l \in \Re$ represent an arc-length parametrization of the curve and the beam length in the initial configuration, respectively, and $(.)^{\prime}\!=\!\frac{\partial}{\partial s}(.)\!=\!(.)_{,s}$ 
denotes the derivative with respect to this arc-length coordinate. Furthermore, $t \in [0,t_{end}] \subset \Re$ represents the time and $\dot{(.)}\!=\!\frac{\partial}{\partial t}(.)\!=\!(.)_{,t}$ denotes 
the corresponding time derivative. If we neglect rotational inertia contributions, which is common practice and mechanically sensible when considering highly slender beams, the extension of the 
torsion-free formulation according to \cite{meier2015} to dynamic problems is straightforward. In this case, the kinetic and hyper-elastic stored energies are:
\begin{align}
\label{storedenergyfunction}
\Pi_{kin}:= \! \int \limits_{s=0}^l \frac{1}{2} \rho A v^2 ds, \,\, \,\,
\Pi_{int}:= \! \int \limits_{s=0}^l \left[\frac{1}{2}EA \epsilon^2 + \frac{1}{2}EI \kappa^2\right] ds \,\, \,\, \text{with} \, \, \,\,
v=||\dot{\boldsymbol{r}}||, \, 
\epsilon=||\mb{r}^{\prime}||-1, \, \, 
\kappa=||\boldsymbol{\kappa}||, \, \, 
\boldsymbol{\kappa}=\frac{\mb{r}^{\prime} \times \mb{r}^{\prime  \prime}}{||\mb{r}^{\prime}||^2}.
\end{align}
Here, $\rho$ is the mass density, $A$ the cross-section area, $I$ the moment of inertia and $E$ the Young´s modulus. Furthermore, $v=v(s)$ represents the material velocity field,
while $\epsilon=\epsilon(s)$ and $\kappa=\kappa(s)$ are the fields of axial tension and bending curvature. 
The corresponding weak form of the dynamic balance equations of the considered beam reads
\begin{align}
\label{weakform}
\int \limits_0^l
  \Bigg[
  \delta \epsilon EA \epsilon 
 +\delta \boldsymbol{\kappa} EI \boldsymbol{\kappa} 
 +\delta \mb{r}^T \rho A \ddot{\boldsymbol{r}}
  \Bigg] ds
  -\int \limits_0^l
  \Bigg[ 
 \delta \mb{r}^T \mb{\tilde{f}}
 +\delta \boldsymbol{\theta^T_{\perp}} \mb{\tilde{m}_{\perp}}
  \Bigg] ds
 -\Bigg[\delta \mb{r}^T \bar{\mb{f}}+ \delta \boldsymbol{\theta^T_{\perp}} \bar{\mb{m}}_{\perp} \Bigg]_{\Gamma_{\sigma}} \hspace{-0.3cm}
 = 0. 
\end{align}
Here, $\mb{\tilde{f}}$ and $\mb{\tilde{m}_{\perp}}$ denote distributed forces and moments, whereas $\bar{\mb{f}}$ and $\bar{\mb{m}}_{\perp}$ denote discrete point forces and 
moments on the Neumann boundary $\Gamma_{\sigma}$ of the beam. Furthermore, we have applied the following additional abbreviations:
\begin{align}
\label{abbriviations}
  \delta \epsilon = \frac{\delta \mb{r}^{\prime T} \mb{r}^{\prime}}{||\mb{r}^{\prime}||}, \quad
  \delta \boldsymbol{\kappa} = \frac{
                               ||\mb{r}^{\prime}||^2 \left(\delta \mb{r}^{\prime} \times \mb{r}^{\prime  \prime} + \mb{r}^{\prime} \times \delta \mb{r}^{\prime  \prime}\right)
                               -2 \left( \delta \mb{r}^{\prime T}\mb{r}^{\prime} \right) \left( \mb{r}^{\prime} \times \mb{r}^{\prime  \prime} \right)
                               }{||\mb{r}^{\prime}||^4} \quad \text{and} \quad
  \delta \boldsymbol{\theta_{\perp}}=\frac{\mb{r}^{\prime} \times \delta \mb{r}^{\prime}}{||\mb{r}^{\prime}||^2}.
\end{align}
As indicated by the subscript $(.)_{\perp}$, the torsion-free beam theory is only applicable if the external moment vectors contain no components parallel to the centerline 
tangent vector, i.e. $\mb{r}^{\prime T}(s)\mb{\tilde{m}_{\perp}}(s) \equiv 0 \, \forall \, s \in [0,l]$ and $\mb{r}^{\prime T}\bar{\mb{m}}_{\perp} \equiv 0 \, \text{on} \,\Gamma_{\sigma}$.

%-------------------------------------------------------------------------------
%
\subsection{Spatial discretization}
\label{sec:beam_spatial}
%
%-------------------------------------------------------------------------------

After having defined the weak form of the dynamic equilibrium equations, corresponding boundary and initial conditions and proper spaces of trial and test functions, 
i.e. $\mb{r} \in  \mathcal{S} \subset \Re^3$ satisfying the essential boundary conditions on the Dirichlet boundary $\Gamma_{u}$ and $\delta \mb{r} \in \mathcal{V} \subset \Re^3$ 
with $\delta \mb{r} = \mb{0} \, \text{on} \, \varGamma_{u}$, the space- and time-continuous problem setting is completed. Spatial discretization is performed by replacing the 
test and trial spaces by finite-dimensional subsets, i.e. $\mb{r} \approx \mb{r}_h \in  \mathcal{S}_h \subset \mathcal{S}$ and $\delta \mb{r} \approx \delta \mb{r}_h \in 
\mathcal{V}_h \subset \mathcal{V}$. Here and in the following, the index $h$ denotes the spatially discretized version of a quantity. However, in the following, this index 
will often be omitted when there is no danger of confusion. Concretely, we follow a Bubnov-Galerkin approach leading to the following discretized beam centerline:
\begin{align}
\label{interpolation}
\mb{r}_h(\xi) = \sum_{i=1}^{2} N^i_{d}(\xi) \mb{\hat{d}}^i + \frac{l_{ele}}{2} \sum_{i=1}^{2} N^i_{t}(\xi) \mb{\hat{t}}^i =: \mb{N}(\xi) \mb{d} \quad \text{and} \quad 
\delta \mb{r}_{h}(\xi) = \sum_{i=1}^{2} N^i_{d}(\xi) \delta \mb{\hat{d}}^i + \frac{l_{ele}}{2} \sum_{i=1}^{2} N^i_{t}(\xi) \delta \mb{\hat{t}}^i  =: \mb{N}(\xi) \delta \mb{d}\, ,
\end{align}
where $\mb{\hat{d}}^i, \mb{\hat{t}}^i \! \in \! \Re^3$ are positions and tangent vectors at the two element nodes ($i=1,2$), $\delta \mb{\hat{d}}^i, \delta \mb{\hat{t}}^i \! \in \! \Re^3$ represent 
their variations, $l_{ele}$ is the initial 
length of the initially straight beam element and $\xi \in [-1;1]$ is an element parameter coordinate. If considering initially straight beams, the latter can explicitly be related to the 
arc-length coordinate according to $s(\xi)=s_0 + (\xi+1)l_{ele}/2$ and $(.)_{,s}=(.)_{,\xi}\cdot J_{ele}^{-1}(\xi)$. Here, $s_0$ represents the
arc-length coordinate of the first node and $J_{ele}(\xi)=l_{ele}/2$ the element Jacobian. Similar to the abbreviation $(.)^{\prime}\!=\!(.)_{,s}$ for the arc-length derivative, 
we will use the notation $(.)^{\shortmid}\!=\!(.)_{,\xi}$ for the derivative with respect to the element parameter coordinate. The third order Hermite shape functions 
$N^i_{d}(\xi)$ and $N^i_{t}(\xi)$ (see \cite{meier2014} for further information concerning their properties) are defined as
\begin{align}
\label{shapefunctions}
N^1_{d}(\xi) = \frac{1}{4}(2+\xi)(1-\xi)^2, \,\,\, N^2_{d}(\xi) = \frac{1}{4}(2-\xi)(1+\xi)^2, \,\,\,
N^1_{t}(\xi) = \frac{1}{4}(1+\xi)(1-\xi)^2, \,\,\, N^2_{t}(\xi) = -\frac{1}{4}(1-\xi)(1+\xi)^2.
\end{align}
They provide a $C^1$-continuous beam centerline representation, thus enabling smooth contact kinematics. This property will be very beneficial for the derivation of the contact 
formulation in the following sections. The abbreviations $\mb{d}$, $\delta \mb{d}$ and $\mb{N}(\xi)$ appearing in \eqref{interpolation} represent proper element-wise vector- and matrix-valued 
assemblies of the nodal variables and shape functions. In order to avoid membrane locking in the range of very high slenderness ratios as considered in this contribution, we additionally apply 
the so-called MCS method introduced in \cite{meier2015}, where the original axial strain field $\epsilon$ and its variation $\delta \epsilon$ occurring in \eqref{weakform} are replaced by the following 
re-interpolations
\begin{align}
\label{MCS}
\bar{\epsilon}(\xi)=\sum_{k=1}^{3} L^k(\xi)\epsilon(\xi^k)
\quad \text{and} \quad \delta \bar{\epsilon}(\xi)=\sum_{k=1}^{3} L^k(\xi) \delta \epsilon(\xi^k) \quad \text{with} \quad \xi^1=-1, \xi^2=0, \xi^3=1,
\end{align}
which are based on second-order Lagrange polynomials $L^k(\xi)$. The resulting element residual contributions $\mb{r}_{int}, \mb{r}_{kin}$ and $\mb{r}_{ext}$
of the internal, inertia and external forces and their linearizations are summarized in \ref{anhang:reslin_beamelement}. An assembly of these quantities and the corresponding element-wise 
contact contributions $\mb{r}_{con}$ presented in the next sections leads to the following global system of equations representing the spatially discretized version of 
\eqref{weakform}, viz.
\begin{align}
\label{global_system}
\mb{R}_{tot}=\mb{M} \ddot{\mb{D}}+\mb{R}_{int}(\mb{D})+\mb{R}_{con}(\mb{D})-\mb{R}_{ext}(\mb{D})=\mb{0},
\end{align}
where $\mb{D}$ is the assembled global vector of primary variables containing the nodal degrees of freedom 
$\mb{\hat{d}}^k,\mb{\hat{t}}^k$ of all $n_{node}$ nodes with $k = 1,...,n_{node}$. It is worth to mention that the presented torsion-free beam formulation is a geometrically exact representation of a real 
``cross-section-reduced'' structural model based on a $1$D continuum theory, but it neither requires the application of any rotational primary degrees of freedom nor the enforcement of additional director constraints. Furthermore, 
the global inertia forces are composed of a constant symmetric mass matrix $\mb{M}$ and the global acceleration vector $\ddot{\mb{D}}$, while general geometrically exact beam formulations 
usually lead to nonlinear inertia force contributions. Finally, as long as no external moments are acting, i.e. $\mb{\tilde{m}_{\perp}}(s) \equiv \mb{0} \, \forall \, s \in [0,l]$ and 
$\bar{\mb{m}}_{\perp} \equiv \mb{0} \, \text{on} \,\Gamma_{\sigma}$, the global stiffness matrix $\mb{K}_{int}=d \mb{R}_{int}/ d \mb{D}$ is symmetric (see \ref{anhang:reslin_beamelement}). 
Due to the absence of rotational degrees of freedom, any time discretization scheme suitable for second-order ODEs can be applied to~\eqref{global_system}.

%-------------------------------------------------------------------------------
%
\section{Point-to-point contact formulation and limitations}
\label{sec:point}
%
%-------------------------------------------------------------------------------

\begin{figure}[ht!]
 \centering
  \includegraphics[width=0.5\textwidth]{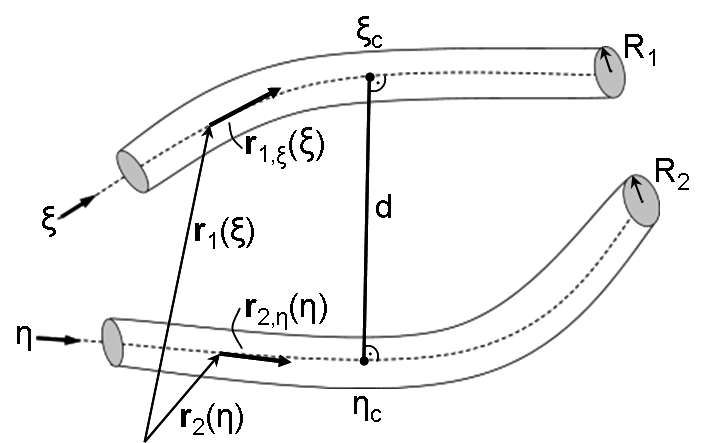}
  \caption{Kinematic quantities defining the point-to-point contact problem of two beams}
  \label{fig:point_problemdescription1}
\end{figure}

Within this section, we briefly repeat the main constituents of a standard point-to-point beam contact formulation as introduced in \cite{wriggers1997}. Thereto, we consider two arbitrarily 
curved beams with cross-section radii $R_1$ and $R_2$, respectively. The beam centerlines are represented by two parametrized curves~$\mb{r}_{1}(\xi)$ and $\mb{r}_{2}(\eta)$ with curve 
parameters~$\xi$ and $\eta$. Furthermore, $\mb{r}_{1,\xi}(\xi)= \mb{r}_1^{\shortmid}(\xi)$ and $\mb{r}_{2,\eta}(\eta)=\mb{r}_2^{\shortmid}(\eta)$ denote the tangents to these curves at positions 
$\xi$ and $\eta$, respectively. In what follows, we assume that the considered space curves are at least $C^1-$continuous, thus providing a unique tangent 
vector at every position $\xi$ and $\eta$. The kinematic quantities introduced above are illustrated in Figure~\ref{fig:point_problemdescription1}.

%-------------------------------------------------------------------------------
%
\subsection{Contact formulation and contribution to weak form}
\label{sec:point_weakform}
%
%-------------------------------------------------------------------------------

The point-to-point beam contact formulation enforces the contact constraint by prohibiting penetration of the two beams at the closest point positions $\xi_c$ and $\eta_c$. Here and in the following, 
the subscript $c$ indicates, that a quantity is evaluated at the closest point coordinate $\xi_c$ or $\eta_c$, respectively. These closest point 
coordinates are determined as solution of the bilateral (''bl``) minimal distance problem, also denoted as bilateral closest point projection, with
\begin{align}
\label{point_mindist}
  d_{bl}:=\min_{\xi, \eta} d(\xi, \eta) = d(\xi_c, \eta_c) \quad \text{with} \quad d(\xi, \eta)=||\mb{r}_{1}(\xi)-\mb{r}_{2}(\eta)||.
\end{align}
This leads to two orthogonality conditions that have to be solved for the unknown closest point coordinates $\xi_c$ and $\eta_c$:
\begin{align}
\label{point_orthocond}
\begin{split}
  p_{1}(\xi,\eta)&=\mb{r}^T_{1,\xi}(\xi)\left( \mb{r}_{1}(\xi)-\mb{r}_{2}(\eta) \right) \quad \rightarrow \, p_{1}(\xi_c,\eta_c) \dot{=} 0, \\
  p_{2}(\xi,\eta)&=\mb{r}^T_{2,\eta}(\eta)\left( \mb{r}_{1}(\xi)-\mb{r}_{2}(\eta) \right) \quad \rightarrow \, p_{2}(\xi_c,\eta_c) \dot{=} 0.
\end{split}
\end{align}
The contact condition of non-penetration at the closest point is formulated by means of the inequality constraint
\begin{align}
  g \geq 0 \quad \text{with} \quad g:=d_{bl}-R_1-R_2,
\end{align}
where $g$ is the gap function. This constraint can be included into our variational problem setting via a penalty potential
\begin{align}
\label{point_penaltypotential}
     \quad \Pi_{c\varepsilon}=\frac{1}{2} \varepsilon \langle g\rangle^2 \quad \text{and} \quad 
           \langle x\rangle=\left\{\begin{array}{ll}
                                 x, & x \leq 0 \\
                                 0, & x > 0
                    \end{array}\right.
\end{align}
or alternatively via a contact contribution in terms of a corresponding Lagrange multiplier potential
\begin{align}
\label{point_lmpotential}
     \Pi_{c\lambda}= \lambda g \quad \text{and} \quad \lambda \geq 0, \,\,\, g \geq 0, \,\,\, \lambda g = 0.
\end{align}
Throughout this work, we solely apply constraint enforcement via penalty regularization according to \eqref{point_penaltypotential} (see also our remarks in 
Section~\ref{sec:line_penaltyvslagrange}). 
Variation of \eqref{point_penaltypotential} leads to the contribution of one contact point to the weak form:
\begin{align}
\label{point_weakform}
  \delta \Pi_{c\varepsilon} =
      \varepsilon \langle g\rangle \delta g =
      \varepsilon \langle g\rangle \left(\delta \mb{r}_{1c} - \delta \mb{r}_{2c} \right)^T \mb{n}. 
\end{align}
In \eqref{point_weakform}, we can identify the contact force vector $\mb{f}_{c\varepsilon}$ as well as the normal vector $\mb{n}$. The two are defined as:
\begin{align}
\label{point_contactforce}
  \mb{f}_{c\varepsilon}= \underbrace{-\varepsilon \langle g\rangle}_{=:f_{c\varepsilon}} \mb{n}, \quad  \mb{n}:=\frac{\mb{r}_{1}(\xi_c)-\mb{r}_{2}(\eta_c)}{||\mb{r}_{1}(\xi_c)-\mb{r}_{2}(\eta_c)||}.
\end{align}
According to \eqref{point_contactforce}, the point-to-point beam contact formulation models the contact force $\mb{f}_{c\varepsilon}$ that is transferred between the two beams 
as a discrete point force acting at the respective closest points of the beam centerlines.\\

\hspace{0.5 cm}
\begin{minipage}{15.0 cm}
\textbf{Remark:} Since the contact point parameter coordinates $\xi_c$ and $\eta_c$ are deformation-dependent, the total variation or linearization of a quantity $X(\xi,\eta)$ can 
be split up into the following three contributions:
\begin{align*}
  \delta \left(X(\xi,\eta)\right) = X_{,\xi} \delta \xi + X_{,\eta} \delta \eta + \delta X \quad \text{and} 
  \quad \Delta \left( X(\xi,\eta) \right) = X_{,\xi} \Delta \xi + X_{,\eta} \Delta \eta + \Delta X.
\end{align*}
Here, the first two contributions denote the change in $X(\xi,\eta)$ due to a change in the parameter coordinates $\xi$ and $\eta$, whereas the contributions $\delta X$/$\Delta X$ represent 
the variation/linearization of $X(\xi,\eta)$ at fixed parameter coordinates. As already mentioned 
in \cite{wriggers1997}, the total variation of the gap simplifies according to
\begin{align*}
  \delta g = \mb{n}^T \! \left( \delta \left( \mb{r}_{1c}\right) - \delta \left(\mb{r}_{2c} \right) \right)= \mb{n}^T \! \left( \delta \mb{r}_{1c} + \mb{r}_{1c}^{\shortmid} \delta \xi - 
  \delta \mb{r}_{2c} - \mb{r}_{2c}^{\shortmid} \delta \eta \right)
           = \mb{n}^T \! \left( \delta \mb{r}_{1c} - \delta \mb{r}_{2c} \right) \,\, \text{since} \,\, \mb{n}^T \! \mb{r}_{1c}^{\shortmid}=\mb{n}^T \! \mb{r}_{2c}^{\shortmid}=0,
\end{align*}
which is a consequence of the orthogonality conditions~\eqref{point_orthocond} satisfied at the closest points $\xi_c$ and $\eta_c$.\\
\end{minipage}

For later use, we also define the so-called contact angle as the angle between the tangent vectors at the contact point:
\begin{align}
\label{point_contactangle}
    \alpha = \arccos{ \left(z \right) }
  \quad \text{with} \quad z=\frac{ ||\mb{r}_1^{\shortmid T}(\xi_c) \mb{r}_2^{\shortmid}(\eta_{c})|| }{ ||\mb{r}_1^{\shortmid}(\xi_c)|| \cdot ||\mb{r}_2^{\shortmid}(\eta_{c})|| },
  \quad \alpha \in [0;90^{\circ}].
\end{align}
In a next step, spatial discretization has to be performed. Since, for simplicity, we only consider the contact contribution of one contact point, the indices $1$ and $2$ are directly transferred 
to the two finite elements where the point contact takes place.
Inserting the spatial discretization \eqref{interpolation} into the orthogonality conditions \eqref{point_orthocond} allows to solve the latter for the unknown closest point parameter 
coordinates $\xi_c$ and $\eta_c$. Since, in general, the system of equations provided by \eqref{point_orthocond} is nonlinear in $\xi$ and $\eta$, a local Newton-Raphson scheme is applied for 
its solution. The corresponding linearizations of \eqref{point_orthocond} can for example be found in \cite{wriggers1997}. Inserting equations 
\eqref{interpolation} into equation \eqref{point_weakform} leads to the following contact residual contributions $\mb{r}_{con,1}$ and $\mb{r}_{con,2}$ of the two considered elements:
\begin{align}
\label{point_discreteweakform}
  \delta \Pi_{c\varepsilon}=\delta \mb{d}_1^T \underbrace{\varepsilon \langle g\rangle \mb{N}_{1}^T(\xi_c) \mb{n}}_{=:\mb{r}_{con,1}} - 
  \delta \mb{d}_2^T \underbrace{\varepsilon \langle g\rangle \mb{N}_{2}^T(\eta_c) \mb{n}}_{=:\mb{r}_{con,2}}.
\end{align}

%-------------------------------------------------------------------------------
%
\subsection{Limitations of point-to-point contact formulation}
\label{sec:point_limitationspoint}
%
%-------------------------------------------------------------------------------
The point-to-point contact formulation provides an elegant and efficient contact model as long as sufficiently large contact angles are considered. However, its limitation 
lies in the requirement of a unique closest point solution according to \eqref{point_orthocond}, which cannot be guaranteed for arbitrary geometrical configurations. In \cite{konjukhov2010},
the authors have already treated the question of uniqueness and existence of the closest point projection by means of geometrical criteria based on so-called projection domains. Within this section, 
we want to analyze this question from a different perspective: This procedure will allow us to define easy-to-evaluate control quantities and to derive
proper upper and lower bounds of these control quantities within which a unique closest point solution can be guaranteed in a mathematically rigorous manner.
In the following, it will be derived that the contact angle $\alpha$ defined in \eqref{point_contactangle}, the closest point distance $d_{bl}$ as well as the 
geometrical (or mathematical) curvature $\bar{\kappa}$ of the beam centerline according to
\begin{align}
\label{limitationspoint_kappa}
  \bar{\kappa}:= \frac{\kappa}{||\mb{r}^{\prime}||}=\frac{||\mb{r}^{\prime} \times \mb{r}^{\prime \prime}||}{||\mb{r}^{\prime}||^3} = ||\mb{r}_{,\tilde{s}\tilde{s}}|| 
\quad \text{since} \quad || \mb{r}_{,\tilde{s}}||=1,
\end{align}
are such suitable control quantities. We have introduced the parameter coordinate $\tilde{s} \in [0;\tilde{l}]$ representing 
the arc-length of the current, deformed beam centerline and $\tilde{l}$ denoting the corresponding current length. For the following analytical derivations, which are based on the space-continuous problem 
setting, we use the current arc-length parameters $\tilde{s}_1$ and $\tilde{s}_2$ instead of the initial arc-length parameters $s_1$ and $s_2$ (required for the space-continuous problem setting 
of the beam element formulation) or the normalized element parameters $\xi$ and $\eta$ (required for the spatially discretized problem setting). This choice simplifies many steps due to 
the essential property $|| \mb{r}_{1,\tilde{s_1}}||=||\mb{r}_{2,\tilde{s_2}}||=1$. Moreover, we define the maximal cross-section to curvature radius ratio 
$\mu_{max}$ according to
\begin{align}
\label{limitationspoint_kappa2}
  \mu_{max}=\frac{R}{min \, (\bar{r})} \ll 1 \quad \text{with} \quad \bar{r}=\frac{1}{\bar{\kappa}},
\end{align}
i.e. as the quotient of the cross-section radius $R$ and the minimal radius of curvature $\bar{r}$ occurring in the deformed geometry. The application of beam theories in general, particularly the application 
of the Kirchhoff beam theory, is only justified for problems exhibiting small values of this ratio, i.e. $\mu_{max} \ll 1$. This property will be useful later on in this section.
In order to simplify the following derivations, we anticipate the definition of the unilateral (``ul'') distance function field $d_{ul}(\tilde{s}_1)$ presented in Section~\ref{sec:line}, which 
assigns a closest partner point $\tilde{s}_{2c}$ of the second beam (in this context also denoted as master beam) for every given point $\tilde{s}_1$ on the first beam 
(in this context also denoted as slave beam) by means of the following unilateral closest point projection (see Figure~\ref{fig:line_problemdescription_contiuous} for an illustration):
\begin{align}
\label{limitationspoint_mindista}
  d_{ul}(\tilde{s}_{1})=\min_{\tilde{s}_2} d(\tilde{s}_1,\tilde{s}_2) = d(\tilde{s}_1,\tilde{s}_{2c})
  \quad \text{with} \quad
  d(\tilde{s}_1,\tilde{s}_2)=||\mb{r}_1(\tilde{s}_1)-\mb{r}_2(\tilde{s}_2)||.
\end{align}
Next, one has to realize that the bilateral closest point projection \eqref{point_mindist} represents a special case of 
the unilateral closest point projection~\eqref{limitationspoint_mindista}. Concretely, the closest point coordinates \eqref{point_mindist} are found 
through minimization of the minimal distance function $d_{ul}(\tilde{s}_1)$ according to \eqref{limitationspoint_mindista} with respect to the slave beam parameter $\tilde{s}_1$, viz.:
\begin{align}
\label{limitationspoint_mindist}
  d_{bl} = \min_{\tilde{s}_1} d_{ul}(\tilde{s}_1) = d_{ul}(\tilde{s}_{1c}).
\end{align}
Now, in a first step, we want to examine the requirements for the existence of a unique solution of the unilateral closest point projection. As soon as we can guarantee a unique distance function 
$d_{ul}(\tilde{s}_1)$, the investigation of the existence and uniqueness of the bilateral closest point projection simplifies from the analysis of a function with 2D support occurring in \eqref{point_mindist} 
to the analysis of a function with 1D support according to \eqref{limitationspoint_mindist}. For a given point with coordinate vector $\mb{r}_1(\tilde{s}_1)$, the unilateral closest point projection 
according to \eqref{limitationspoint_mindista} searches for the corresponding closest point coordinate $\tilde{s}_{2c}$ on the space curve $\mb{r}_2(\tilde{s}_{2})$. In case of $C^1$-continuous curves, which is guaranteed by 
the applied Hermite shape functions and which leads to a uniquely defined tangent vector field, a necessary condition for the existence of the minimal distance solution~\eqref{limitationspoint_mindista} 
is satisfied in case the requirement of a vanishing first derivative is fulfilled, i.e.
\begin{align}
\label{limitationspoint_requirementsecondderiv0}
  d_{,\tilde{s}_{2}}(\tilde{s}_{1},\tilde{s}_{2c})
  =-\frac{\mb{r}^T_{2,\tilde{s}_{2}}(\tilde{s}_{2c})\left( \mb{r}_{1}(\tilde{s}_{1})-\mb{r}_{2}(\tilde{s}_{2c}) \right)}{||\mb{r}_{1}(\tilde{s}_{1})-\mb{r}_{2}(\tilde{s}_{2c})||} \dot{=} 0 
  \quad \rightarrow \quad \mb{r}^T_{2,\tilde{s}_{2}}(\tilde{s}_{2c})\left( \mb{r}_{1}(\tilde{s}_{1})-\mb{r}_{2}(\tilde{s}_{2c}) \right) \dot{=} 0,
\end{align}
which, in turn, is guaranteed by the second equation of \eqref{point_orthocond}. A sufficient condition for the existence of a locally unique 
closest point solution is~\eqref{limitationspoint_requirementsecondderiv0} together with the requirement of a positive second derivative of the distance function:
\begin{align}
\label{limitationspoint_requirementsecondderiv1}
  d_{,\tilde{s}_2 \tilde{s}_2}(\tilde{s}_1,\tilde{s}_{2c})
     = -\frac{\mb{r}^T_{2,\tilde{s}_{2} \tilde{s}_{2c}}(\tilde{s}_{2c}) \left( \mb{r}_{1}(\tilde{s}_1)-\mb{r}_{2}(\tilde{s}_{2c}) \right) 
     -\mb{r}^T_{2,\tilde{s}_{2}} (\tilde{s}_{2c})\mb{r}_{2,\tilde{s}_{2}}(\tilde{s}_{2c})}{||\mb{r}_{1}(\tilde{s}_1)-\mb{r}_{2}(\tilde{s}_{2c})||}
     - \underbrace{\mb{r}^T_{2,\tilde{s}_{2}}(\tilde{s}_{2c})\left( \mb{r}_{1}(\tilde{s}_{1})-\mb{r}_{2}(\tilde{s}_{2c}) \right)}_{= 0} \cdot (...) \dot{>} 0.
\end{align}
Together with the auxiliary relation $\mb{r}_{2,\tilde{s}_{2}}^T(\tilde{s}_{2c})\mb{r}_{2,\tilde{s}_{2}}(\tilde{s}_{2c})=1$, relation~\eqref{limitationspoint_requirementsecondderiv1} 
leads to the following requirement:
\begin{align}
\label{limitationspoint_requirementsecondderiv2}
      \rightarrow \quad \underbrace{\mb{r}^T_{2,\tilde{s}_2 \tilde{s}_2}(\tilde{s}_{2c})}_{\bar{\kappa}_{2}(\tilde{s}_{2c}) \bar{\mb{n}}_{2}(\tilde{s}_{2c})} 
      \underbrace{\left( \mb{r}_{1}(\tilde{s}_{1})-\mb{r}_{2}(\tilde{s}_{2c}) \right)}_{d_{ul}(\tilde{s}_{1})\mb{n}(\tilde{s}_{1})} - 1 \dot{<} 0.
\end{align}
Making use of the definition of the geometrical curvature according to \eqref{limitationspoint_kappa} and the additional definitions
\begin{align}
\label{limitationspoint_definitions2}
  \bar{\mb{n}}_{2}(\tilde{s}_{2c}):= \frac{\mb{r}_{2,\tilde{s}_{2} \tilde{s}_{2}}(\tilde{s}_{2c})}{||\mb{r}_{2,\tilde{s}_{2} \tilde{s}_{2}}(\tilde{s}_{2c})||},
  \quad \mb{n}(\tilde{s}_{1}):=\frac{\mb{r}_{1}(\tilde{s}_{1})-\mb{r}_{2}(\tilde{s}_{2c})}{||\mb{r}_{1}(\tilde{s}_{1})-\mb{r}_{2}(\tilde{s}_{2c})||}
  \quad \beta_2(\tilde{s}_{1}):= \arccos \left( \mb{n}^T \!(\tilde{s}_{1}) \bar{\mb{n}}_{2}(\tilde{s}_{2c}) \right)
\end{align}
of the Frenet-Serret unit normal vector $\bar{\mb{n}}_{2}(\tilde{s}_{2c})$ aligned to the curve representing the master beam and the angle $\beta_2(\tilde{s}_{1})$ between this vector and 
the normal vector $\mb{n}(\tilde{s}_{1})$ (which is defined similarly to \eqref{point_contactforce}), \eqref{limitationspoint_requirementsecondderiv2} can be reformulated as:
\begin{align}
\label{limitationspoint_requirementsecondderiv3}
  \bar{\kappa}_{2}(\tilde{s}_{2c})  d_{ul}(\tilde{s}_{1}) \cos(\beta_2(\tilde{s}_{1}))\dot{<}1.
\end{align}
In case the two beams are close enough so that the sought-after closest point $\tilde{s}_{2c}$ is relevant in terms of active contact forces 
($g(\tilde{s}_{1}) = 0 \, \rightarrow d_{ul}(\tilde{s}_{1}) = 2R$) and under 
consideration of the worst case $\cos(\beta_2)(\tilde{s}_{1})=1$, we obtain the following final requirement for a unique solution of the unilateral closest point projection 
according to~\eqref{limitationspoint_mindista}:
\begin{align}
\label{limitationspoint_requirementsecondderiv4}
  2 \frac{R}{\bar{r}_{2}(\tilde{s}_{2c})} \leq 2 \mu_{max} \dot{<} 1 \quad \square
\end{align}
As a consequence of the maximal cross-section to curvature radius ratio $\mu_{max} \ll 1$, a uniquely defined unilateral distance function $d_{ul}(\tilde{s}_{1})$ can be guaranteed as 
long as the beams are sufficiently close. A corresponding criterion for arbitrary distances defined via $d_{ul}(\tilde{s}_{1}) =: k \cdot R$ can be derived by replacing the factor $2$ by $k$ 
in~\eqref{limitationspoint_requirementsecondderiv4}. 
In a second step, we want to investigate the requirements for a unique bilateral closest point solution according to \eqref{limitationspoint_mindist}, based on a uniquely 
defined distance function $d_{ul}(\tilde{s}_{1})$ (which is provided as consequence of \eqref{limitationspoint_requirementsecondderiv4}). Again, the first derivative
\begin{align}
\label{limitationspoint_requirementsecondderiv5}
    \frac{d \, d_{ul}(\tilde{s}_{1})}{d \tilde{s}_{1}} \!=\!
    \frac{d \, d(\tilde{s}_{1},\tilde{s}_{2c}(\tilde{s}_{1}))}{d \tilde{s}_{1}} \!=\!
    \frac{\partial d}{\partial \tilde{s}_{1}} 
    \!+\! \underbrace{\frac{\partial d}{\partial \tilde{s}_{2c}}}_{\equiv 0}\! 
    \frac{\partial \tilde{s}_{2c}}{\partial \tilde{s}_{1}}
    \!=\!\frac{\mb{r}^T_{1,\tilde{s}_{1}}(\tilde{s}_{1})\left( \mb{r}_{1}(\tilde{s}_{1})-\mb{r}_{2}(\tilde{s}_{2c}) \right)}{||\mb{r}_{1}(\tilde{s}_{1})-\mb{r}_{2}(\tilde{s}_{2c})||}
    \, \rightarrow \, \mb{r}^T_{1,\tilde{s}_{1}}(\tilde{s}_{1c}) \left( \mb{r}_{1}(\tilde{s}_{1c})-\mb{r}_{2}(\tilde{s}_{2c}) \right) \dot{=} 0
\end{align}
has to vanish. This is satisfied at the closest point $\tilde{s}_{1c}$ by the first line of \eqref{point_orthocond}. Furthermore, the additional identity 
$\partial d / \partial \tilde{s}_{2c} \equiv 0 \, \forall \, \tilde{s}_{1} \, \in \, [0;\tilde{l}_1]$ is fulfilled as consequence of the second line of \eqref{point_orthocond}. 
Again, a locally unique solution of 
the minimal distance problem \eqref{limitationspoint_mindist} additionally requires a positive second derivative. Differentiation of \eqref{limitationspoint_requirementsecondderiv5} yields:
\begin{align}
\label{limitationspoint_requirementsecondderiv6}
    \frac{d^2 \, d_{ul}(\tilde{s}_{1})}{d \tilde{s}_{1}^2} \Bigg|_{(\tilde{s}_{1c},\tilde{s}_{2c})} =
    \frac{d^2 \, d(\tilde{s}_{1},\tilde{s}_{2c}(\tilde{s}_{1}))}{d \tilde{s}_{1}^2} \Bigg|_{(\tilde{s}_{1c},\tilde{s}_{2c})} =
    \left(\frac{\partial^2 d}{\partial \tilde{s}_{1}^2}
    +\frac{\partial^2 d}{\partial \tilde{s}_{1} \, \partial \tilde{s}_{2c}}\frac{\partial \tilde{s}_{2c}}{\partial \tilde{s}_{1}} \right) \Bigg|_{(\tilde{s}_{1c},\tilde{s}_{2c})} \dot{>} 0.
\end{align}
The derivative $\partial \tilde{s}_{2c} / \partial \tilde{s}_{1}$ appearing in \eqref{limitationspoint_requirementsecondderiv6} can be derived by consistently linearizing the orthogonality 
condition~\eqref{limitationspoint_requirementsecondderiv0}:
\begin{align}
  \left[ \mb{r}^T_{2,\tilde{s}_{2} \tilde{s}_{2}} \left( \mb{r}_{1}-\mb{r}_{2} \right) - \mb{r}^T_{2,\tilde{s}_{2}} \mb{r}_{2,\tilde{s}_{2}} \right] \delta \tilde{s}_{2c}
  +\mb{r}^T_{2,\tilde{s}_{2}}\mb{r}_{1,\tilde{s}_{1}} \, \delta \tilde{s}_{1}=0 \quad \rightarrow \quad 
   \frac{\partial \tilde{s}_{2c}}{\partial \tilde{s}_{1}}=
   \frac{\mb{r}^T_{2,\tilde{s}_{2}}\mb{r}_{1,\tilde{s}_{1}}}{\mb{r}^T_{2,\tilde{s}_{2}}\mb{r}_{2,\tilde{s}_{2}}-\mb{r}^T_{2,\tilde{s}_{2} \tilde{s}_{2}}\left( \mb{r}_{1}-\mb{r}_{2} \right)}.
\end{align}
After making use of this result and calculating the derivatives of \eqref{limitationspoint_requirementsecondderiv5} with respect to $\tilde{s}_{1}/\tilde{s}_{2c}$, 
requirement \eqref{limitationspoint_requirementsecondderiv6} yields:
\begin{align}
\label{limitationspoint_requirementsecondderiv7a}
     \frac{
           \mb{r}^T_{1,\tilde{s}_{1} \tilde{s}_{1}} \left( \mb{r}_{1}-\mb{r}_{2} \right) + \mb{r}^T_{1,\tilde{s}_{1}} \mb{r}_{1,\tilde{s}_{1}} 
           -\mb{r}^T_{1,\tilde{s}_{1}} \mb{r}_{2,\tilde{s}_{2}} \cdot 
           \frac{\mb{r}^T_{2,\tilde{s}_{2}}\mb{r}_{1,\tilde{s}_{1}}}{\mb{r}^T_{2,\tilde{s}_{2}}\mb{r}_{2,\tilde{s}_{2}}-\mb{r}^T_{2,\tilde{s}_{2} \tilde{s}_{2}}\left( \mb{r}_{1}-\mb{r}_{2} \right)}
          }{||\mb{r}_{1}-\mb{r}_{2}||} \,\, \Bigg|_{(\tilde{s}_{1c},\tilde{s}_{2c})} \dot{>} 0.
\end{align}
Using the quantities defined in \eqref{limitationspoint_definitions2}, the contact angle $\alpha$ according to \eqref{point_contactangle} and the additional definitions
\begin{align}
\label{limitationspoint_definitions3}
  \bar{\mb{n}}_{1}(\tilde{s}_{1c}):= \frac{\mb{r}_{1,\tilde{s}_{1} \tilde{s}_{1}}(\tilde{s}_{1c})}{||\mb{r}_{1,\tilde{s}_{1} \tilde{s}_{1}}(\tilde{s}_{1c})||},
   \quad \beta_1(\tilde{s}_{1c}):= \arccos \left( \mb{n}^T(\tilde{s}_{1c}) \bar{\mb{n}}_{1}(\tilde{s}_{1c}) \right),
\end{align}
condition~\eqref{limitationspoint_requirementsecondderiv7a} can be reformulated. Due to the strictly positive denominator, we only have to consider the 
numerator:
\begin{align}
\label{limitationspoint_requirementsecondderiv7}
     1 + \bar{\kappa}_1 d_{bl} \cos(\beta_1) - \frac{\cos(\alpha)^2}{1-\bar{\kappa}_2 d_{bl} \cos(\beta_2)} \dot{>} 0 \Leftrightarrow
      \underbrace{\big(1 + \overbrace{\bar{\kappa}_1 d_{bl}}^{ \in \, [0;1[} \cos(\beta_1)\big)}_{>0} \underbrace{\big( 1-\overbrace{\bar{\kappa}_2 d_{bl}}^{ \in \, [0;1[} \cos(\beta_2) \big)}_{>0} \dot{>} \cos(\alpha)^2,
\end{align}
where we have assumed sufficiently close beams $d_{bl}=k \cdot R$ satisfying $\bar{\kappa}_1 d_{bl} < 1$ and $\bar{\kappa}_2 d_{bl} < 1$ as consequence of \eqref{limitationspoint_kappa2}. 
In case the two beams are close enough so that the sought-after closest point pair 
$(\tilde{s}_{1c}, \tilde{s}_{2c})$ is relevant in terms of active contact forces 
($g = 0 \, \rightarrow \, d_{bl} = 2R$), the inequality \eqref{limitationspoint_requirementsecondderiv7} can be reformulated by means of worst case estimates:
\begin{align}
\label{limitationspoint_requirementsecondderiv8a}
     \left(1 + 2\bar{\kappa}_1 R \cos(\beta_1)\right) \left( 1-2\bar{\kappa}_2 R \cos(\beta_2) \right)
      \geq \left(1 - 2\bar{\kappa}_1 R \right) \left( 1-2\bar{\kappa}_2 R \right)
     \geq \left(1 - 2 \mu_{max} \right)^2
     \dot{>} \cos(\alpha)^2.
\end{align}
Since we solely consider positive contact angles $\alpha \in [0;90^{\circ}]$, only the positive branch of the quadratic inequality~\eqref{limitationspoint_requirementsecondderiv8a} 
has to be considered. Consequently, we end up with the following lower bound for the contact angle:
\begin{align}
\label{limitationspoint_requirementsecondderiv8}
     \alpha \dot{>} \alpha_{min}=\arccos \left(1 - 2 \mu_{max} \right).
\end{align}
The importance of the final requirement in \eqref{limitationspoint_requirementsecondderiv8} is quite obvious: As long as we can provide an upper bound $\mu_{max}$ for the admissible ratio 
of cross-section to curvature radius, we will directly obtain from \eqref{limitationspoint_requirementsecondderiv8} a lower bound for the admissible contact angles above which the closest 
point solution is unique. Again, condition~\eqref{limitationspoint_requirementsecondderiv8} can be expanded to general, but still sufficiently small ($\bar{\kappa} d_{bl} < 1$!), 
distances $d_{bl}=k \cdot R$ by replacing the factor $2$ by $k$. 
\begin{figure}[ht]
 \centering
   \subfigure[Two parallel beams]
   {
    \includegraphics[height=0.17\textwidth]{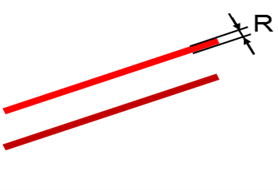}
    \label{fig:line_limitations_parallel}
   }
   \hspace{0.1\textwidth}
   \subfigure[Straight + circular beam]
   {
    \includegraphics[height=0.17\textwidth]{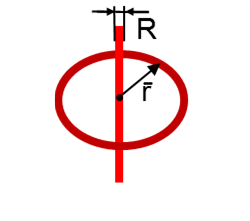}
    \label{fig:limitations_circle}
   }
   \hspace{0.1\textwidth}
   \subfigure[Straight + helical beam]
   {
    \includegraphics[height=0.20\textwidth]{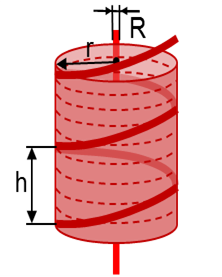}
    \label{fig:limitations_helix}
   }
  \caption{Contact interaction of two beams: Different geometrical configurations concerning contact angle and curvature}
  \label{fig:limitations}
\end{figure}
The three examples illustrated in Figure \ref{fig:limitations} shall visualize the important 
result in~\eqref{limitationspoint_requirementsecondderiv8}: If only straight rigid beams are considered ($\mu_{max}=0$, see 
Figure~\ref{fig:line_limitations_parallel}), we obtain the trivial requirement $\alpha>0$, which reflects the well-known singularity of the closest point projection for parallel beams. 
If we consider a straight beam and a circular beam, both being oriented in a centrical manner as depicted in Figure \ref{fig:limitations_circle}, we observe a constant gap $g(\tilde{s}_{1})=const.$ between both 
beams, thus leading to a non-unique bilateral closest point solution, but this time at a contact angle of $\alpha=90^{\circ}$. However, this case is not practically relevant, since contact in such a 
scenario can only occur if $\bar{r} \approx 2R$, therefore leading to a cross-section to curvature radius ratio $\mu \approx 0.5$, which is not supported by the considered beam theory, anyway.
The third situation (Figure \ref{fig:limitations_helix}) is similar to the example that will later be numerically investigated in Section~\ref{sec:examples_example2}. The contact interaction between a 
straight beam 
and a helical beam again leads to a constant gap function $g(\tilde{s}_{1})=const.$ and consequently to a non-unique bilateral closest point solution. With decreasing slope $h$, the ratio of cross-section 
to curvature radius as well as the contact angle at which this non-unique solution appears increase. This is in perfect agreement with \eqref{limitationspoint_requirementsecondderiv8}. In this context, 
the helix represents an intermediate configuration between the case of two straight parallel beams according to Figure \ref{fig:line_limitations_parallel} (slope $h \rightarrow \infty$) and the 
case of a straight and a circular beam according to Figure \ref{fig:limitations_circle} (slope $h = 0$). In Section~\ref{sec:examples_example2} it will be shown 
that for such geometries a comparatively large scope of contact angles $\alpha \in [0^{\circ};\alpha_{min}]$ can not be modeled by means of the standard point-to-point contact formulation. 
In practical simulations, the lower bound \eqref{limitationspoint_requirementsecondderiv8} has to be supplemented by a proper safety factor in order to guarantee for a unique closest point solution 
not only when contact actually occurs ($g=0$) but already for a sufficient range of 
small positive gaps $g>0$. Furthermore, too small angles $\alpha$ marginally above the lower bound \eqref{limitationspoint_requirementsecondderiv8} might lead to an ill-conditioned system of 
equations in~\eqref{point_orthocond} even if a unique analytical solution exists. Thus, the important result of this section is that the standard point-to-point contact formulation 
is not only unfeasible for examples including strictly parallel beams, but rather for a considerable range of small contact angles, since no locally unique closest point solution is 
existent in this range. According to \eqref{limitationspoint_requirementsecondderiv8}, the size of this range depends on the ratio of the maximal bending curvature amplitude 
expected for the considered mechanical problem and the cross-section radius.\\

So far, we have only used mathematical arguments to show why the point-to-point beam contact formulation cannot be applied in the range of small contact angles. However, it is also questionable 
from a physical or mechanical point of view if the model of ``point-to-point contact`` itself is suitable to describe the contact interaction of beams enclosing small angles at all. On the one hand, it is clear 
that configurations providing a strictly constant distance function, i.e. $d_{ul,\tilde{s}_{1}}(\tilde{s}_{1}) \equiv 0$, are best modeled by a line-to-line and not by a point-to-point contact formulation. 
On the other hand, if an exact constraint enforcement of beams with rigid cross-sections is assumed, a pure point-to-point contact situation would already occur for non-constant distance functions with 
very small slopes, i.e. $0 < ||d_{ul,\tilde{s}_{1}}(\tilde{s}_{1})|| \ll 1$. However, this is a pure consequence of the rigid cross-section assumption inherent to the employed beam model, while a 
$3D$ continuum approach would naturally lead to \textit{distributed} contact tractions. Consequently, also in the context of $1D$ continuum theories, such scenarios should better be modeled by a 
line-to-line rather than a point-to-point contact formulation. In the next section, a novel line-to-line contact formulation, which is capable of modeling arbitrary beam contact scenarios 
spanning the entire range of possible contact angles $\alpha \in [0^{\circ};90^{\circ}]$ and which is particularly beneficial for small contact angles 
and nearly constant distance functions $d_{ul}(\tilde{s}_{1})$, will be proposed.

%-------------------------------------------------------------------------------
%
\section{Line-to-line contact formulation}
\label{sec:line}
\begin{figure}[ht]
 \centering
  \subfigure[Space continuous problem setting]
   {
    \includegraphics[width=0.4\textwidth]{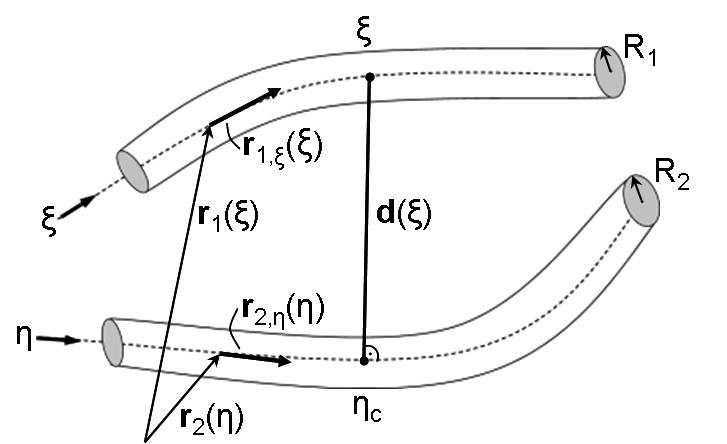}
    \label{fig:line_problemdescription_contiuous}
   }
   \hspace{0.05 \textwidth}
   \subfigure[Discretized problem setting]
   {
    \includegraphics[width=0.4\textwidth]{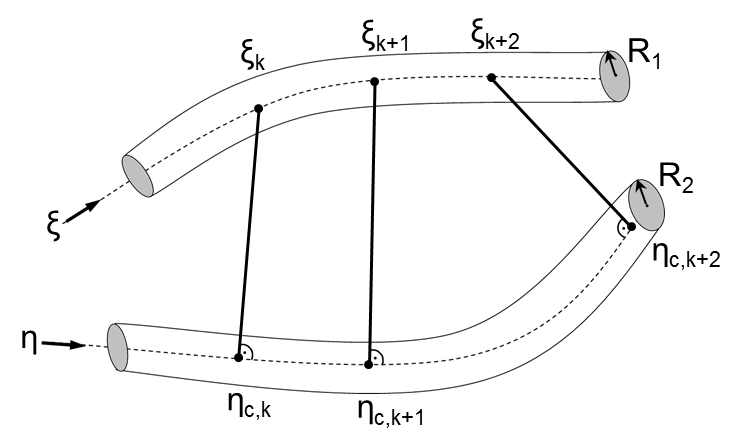}
    \label{fig:line_problemdescription_discrete}
   }
  \caption{Kinematic quantities defining the line-to-line contact problem of two close beams}
  \label{fig:line_problemdescription}
\end{figure}

%
%-------------------------------------------------------------------------------
In the following, we present a novel line-to-line contact formulation that does not formulate the contact condition in form of 
a point-constraint at the closest points anymore, 
but rather as a line constraint enforced along the entire beam length. Consequently, we do not search for one closest point pair, but rather for a closest point field $\eta_c(\xi)$ 
on the second beam (master) assigned to the parameter coordinate field $\xi$ on the first beam (slave). The relevant kinematic quantities of this approach are illustrated in 
Figure~\ref{fig:line_problemdescription_contiuous}.
The closest master point $\eta_c$ to a given slave point $\xi$ is determined as solution of the following unilateral (``ul'') minimal distance problem:
\begin{align}
\label{line_mindist}
  d_{ul}(\xi):=\min_{\eta} d(\xi,\eta)= d(\xi,\eta_c) \quad \text{with} \quad d(\xi,\eta) = ||\mb{r}_{1}(\xi)-\mb{r}_{2}(\eta)||.
\end{align}
It has already been shown in Section~\ref{sec:point_limitationspoint} (see \eqref{limitationspoint_requirementsecondderiv4}), that a unique unilateral closest point solution 
according to~\eqref{line_mindist} can be guaranteed in case the considered beams are close enough so that contact can occur (i.e. $d_{ul}(\xi) \approx 2R$).
Condition~\eqref{line_mindist} leads to one orthogonality condition that has to be solved for the unknown parameter coordinate $\eta_c$:
\begin{align}
\label{line_orthocond}
\begin{split}
  p_2(\xi,\eta)&=\mb{r}^T_{2,\eta}(\eta)\left( \mb{r}_{1}(\xi)-\mb{r}_{2}(\eta) \right) \quad \rightarrow \, p_{2}(\xi,\eta_c) \dot{=} 0
\end{split}
\end{align}
Thus, in contrary to the procedure of the last section, the normal vector is still perpendicular to the second beam but not to the first beam anymore. Furthermore, in the context of line contact 
the subscript $c$ indicates that a quantity is evaluated at the closest master point $\eta_c$ of a given slave point $\xi$. Now, the contact condition of non-penetration
\begin{align}
\label{line_constraint}
  g(\xi) \geq 0 \, \forall \, \xi \quad \text{with} \quad g(\xi):=d_{ul}(\xi)-R_1-R_2
\end{align}
is formulated by means of an inequality-constraint for the gap function field $g(\xi)$ along the entire slave beam. 

%-------------------------------------------------------------------------------
%
\subsection{Constraint enforcement and contact residual contribution}
\label{sec:line_weakform}
%
%-------------------------------------------------------------------------------
In the following, we apply a constraint enforcement strategy based on the space-continuous penalty potential:
\begin{align}
\label{line_pen_totalpotential}
     \Pi_{c\varepsilon}=\frac{1}{2} \varepsilon \int \limits_0^{l_1} \langle g(\xi) \rangle^2 ds_1.
\end{align}
In Section~\ref{sec:line_alternatives}, it will be shown that this strategy is preferable in beam-to-beam contact applications as compared to alternative methods known 
from contact modeling of $3D$ continua. The space-continuous penalty potential in~\eqref{line_pen_totalpotential} does not only serve as purely mathematical tool for constraint enforcement, 
but also has a physical interpretation: It can be regarded as a mechanical model for the flexibility of the surfaces and/or cross-sections 
of the contacting beams. Variation of the penalty potential defined in~\eqref{line_pen_totalpotential} leads to the following contact contribution to the weak form:
\begin{align}
\label{line_pen_weakform}
  \delta \Pi_{c\varepsilon} = \varepsilon \int \limits_0^{l_1} \langle g(\xi)\rangle \delta g (\xi)  ds_1
  \quad \text{and} \quad \delta g(\xi) = ( \delta \mb{r}_1(\xi) - \delta \mb{r}_2(\xi) )^T \mb{n}(\xi).
\end{align}
In the virtual work expression~\eqref{line_pen_weakform}, we can identify the contact force vector $\mb{f}_{c\varepsilon}(\xi)$ and the normal vector $\mb{n}(\xi)$:
\begin{align}
\label{line_pen_contactforce}
  \mb{f}_{c\varepsilon}(\xi)= \underbrace{- \varepsilon \langle g(\xi)\rangle }_{=:f_{c\varepsilon}(\xi)} \mb{n}(\xi), \quad  \mb{n}(\xi):=\frac{\mb{r}_{1}(\xi)-\mb{r}_{2}(\eta_c)}{||\mb{r}_{1}(\xi)-\mb{r}_{2}(\eta_c)||}.
\end{align}
According to \eqref{line_pen_contactforce}, the line-to-line beam contact formulation models the contact force $\mb{f}_{c\varepsilon}(\xi)$ that is transferred between the beams as a distributed 
line force. For comparison reasons, we can again define the contact angle field as:
\begin{align}
\label{line_contactangle}
  \alpha(\xi) = \arccos{ \left(z(\xi) \right) }
  \quad \text{with} \quad z(\xi)=\frac{ ||\mb{r}_1^{\shortmid T}(\xi) \mb{r}_2^{\shortmid}(\eta_{c})|| }{ ||\mb{r}_1^{\shortmid}(\xi)|| \cdot ||\mb{r}_2^{\shortmid}(\eta_{c})|| },
  \quad \alpha \in [0;90^{\circ}].
\end{align}
Next, spatial discretization has to be performed. For simplicity, we only consider the contact contribution stemming from one finite element on the slave beam and one finite element on 
the master beam being assigned to the former via projection~\eqref{line_orthocond}. Therefore, the indices $1$ of the slave beam and $2$ of the master beam will in the following also be used in order to 
denote the two considered finite elements lying on these beams. Inserting the spatial discretization in~\eqref{interpolation} into the orthogonality 
condition \eqref{line_orthocond} allows to solve the latter for the unknown closest point parameter coordinate $\eta_c(\xi)$ for any given slave coordinate $\xi$. The linearizations 
of \eqref{line_orthocond}  required for an iterative solution procedure can be found in~\ref{anhang:linearizationlinecontact}. Inserting the discretization \eqref{interpolation} into equation 
\eqref{line_pen_weakform} and replacing the analytical integral by a Gauss quadrature finally leads to the following contributions of element $1$ and $2$ to the discretized weak form:
\begin{align}
\label{line_pen_discreteweakform}
  \delta \mb{d}_1^T \underbrace{ \sum \limits_{k=1}^{n_{GP}} w_k J(\xi_k) \varepsilon \langle g(\xi_k) \rangle \mb{N}_{1}^T(\xi_k) 
  \mb{n}(\xi_k)}_{=:\mb{r}_{con,1}} + 
 \delta \mb{d}_2^T \underbrace{\sum \limits_{k=1}^{n_{GP}} -w_k J(\xi_k) \varepsilon \langle g(\xi_k) \rangle \mb{N}_{2}^T(\eta_{c}(\xi_k)) \mb{n}(\xi_k)}_{=:\mb{r}_{con,2}}.
\end{align}
Here, $n_{GP}$ is the number of Gauss points per slave element, $w_k$ are the corresponding Gauss weights, $\xi_k$ are the 
Gauss point coordinates in the parameter space $\xi \in [-1;1]$ and finally $\eta_{c,k}$ is the closest master point coordinate assigned to the Gauss point coordinate $\xi_k$ 
on the slave beam (see also Figure~\ref{fig:line_problemdescription_discrete}). 
The Jacobian $J(\xi_k)$ maps between the slave beam arc-length increment $ds_1$ and an increment in the parameter space used for numerical integration (see also Section \ref{sec:line_integrationsegments}). 
Furthermore, $\mb{r}_{con,1}$ and $\mb{r}_{con,2}$ are the residual contributions of the slave ($1$) and master ($2$) element.\\

\hspace{0.5 cm}
\begin{minipage}{15.0 cm}
\textbf{Remark:} In \eqref{line_pen_weakform}, we derived a similar expression for the variation of the gap as for the point-to-point contact case. This time, the variation $\delta \xi$ is zero 
since $\xi$ remains fixed, and, again, the contribution due to the variation of $\eta$ vanishes as a consequence of the orthogonality condition on the slave side:
\begin{align*}
\begin{split}
  \delta g(\xi) & = \mb{n}^T(\xi) \left( \delta (\mb{r}_{1}(\xi)) - \delta (\mb{r}_{2}(\eta_c)) \right)= \mb{n}^T(\xi) \left( \delta \mb{r}_{1}(\xi) 
  - \delta \mb{r}_{2}(\eta_c) - \mb{r}_{2,\eta}(\eta_c) \delta \eta \right) \\
           & = \mb{n}^T(\xi) \left( \delta \mb{r}_{1}(\xi) - \delta \mb{r}_{2}(\eta_c) \right) \quad \text{since} \quad \mb{n}^T(\xi) \mb{r}_{2,\eta}(\eta_c)=0.\\
\end{split}
\end{align*}
\end{minipage}
\vspace{0.5cm}

\hspace{0.5 cm}
\begin{minipage}{15.0 cm}
\textbf{Remark:} The gap function in~\eqref{line_constraint} describes the exact value of the minimal beam surface-to-surface distance at a given coordinate $\xi$, 
only if the contact normal vector is perpendicular to both beam centerlines:
\begin{align}
\label{conditionexactgap}
  \mb{r}^T_{1,\xi}(\xi) \mb{n}(\xi) = 0 \quad \text{and} \quad \mb{r}^T_{2,\eta}(\eta_c) \mb{n}(\xi) = 0.
\end{align}
While both conditions in~\eqref{conditionexactgap} are exactly satisfied at the closest point of the point-to-point contact formulation per definition, only the second condition is 
fulfilled for an arbitrary contact point $\xi$ within an active line-to-line contact segment. However, on the one hand, when considering non-constant evolutions of the centerline distance 
field along the considered beams, i.e. $d_{ul}(\xi)\neq const.$, the region of active line-to-line contact contributions characterized by $g(\xi) < 0$, decreases 
with increasing penalty parameter. In the limit $\epsilon \rightarrow \infty$, the line-to-line contact formulation converges towards the point-to-point contact formulation, where 
both conditions~\eqref{conditionexactgap} are fulfilled exactly. Thus, for a sensibly chosen penalty parameter, the gap function definition~\eqref{line_constraint} provides also a 
good approximation for the line-to-line contact formulation. On the other hand, in configurations with constant centerline distance field $d_{ul}(\xi)= const.$, 
i.e. a range where no unique bilateral closest point solution exists and the point-to-point contact formulation cannot be applied, the two orthogonality conditions~\eqref{conditionexactgap} 
are exactly fulfilled for the entire beam anyway.
\end{minipage}

%-------------------------------------------------------------------------------
%
\subsection{Integration segments}
\label{sec:line_integrationsegments}
\begin{figure}[ht]
 \centering
   \subfigure[Subsegments at all master element boundaries]
   {
    \includegraphics[height=0.082\textwidth]{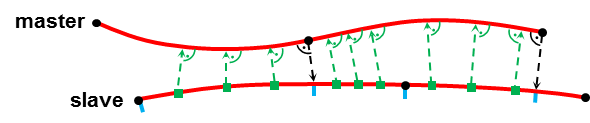}
    \label{fig:line_integrationsegments_full}
   }
   \hspace{0.005\textwidth}
   \subfigure[Subsegments only at master beam endpoints]
   {
    \includegraphics[height=0.082\textwidth]{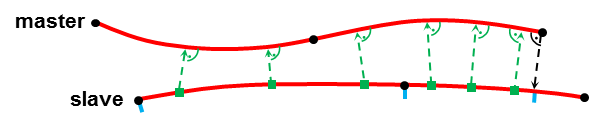}
    \label{fig:line_integrationsegments_endpoints}
   }
   \hspace{0.02\textwidth}
   \subfigure
   {
    \includegraphics[height=0.075\textwidth]{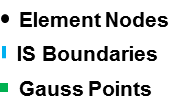}
   }
  \caption{Creation of integration sub-segments on the slave beam in order to avoid discontinuities of the integrand}
  \label{fig:line_integrationsegments}
\end{figure}

%
%-------------------------------------------------------------------------------
From a pratical point of view, it is desirable to decouple the beam discretization and the contact discretization. This can be achieved by allowing for 
$n_{II} \geq 1$ contact integration intervals per slave beam element with $n_{GR}$ integration points defining a Gauss rule of order $p=2 n_{GR}-1$ on each of these integration intervals, thus leading to
$n_{GP}=n_{II} \cdot n_{GR}$ integration points per slave element. In order to realize such a procedure, one has to introduce $n_{II}$ further parameter spaces 
$\bar{\xi}_i \in [-1;1]$ with $i=1,...,n_{II}$ on each slave element:
\begin{align}
\label{line_segmentparametrization}
\begin{split}
  \xi(\bar{\xi}_i)=\frac{1.0-\bar{\xi}_i}{2}\xi_{1,i} + \frac{1.0+\bar{\xi}_i}{2}\xi_{2,i} \quad \text{with} \quad i=1,...,n_{II}.
\end{split}
\end{align}
In the simplest case, the parameter coordinates $\xi_{1,i}$ and $\xi_{2,i}$ confining the $i^{th}$ integration interval are chosen equidistantly within the slave element. Further information 
on the general determination of $\xi_{1,i}$ and $\xi_{2,i}$ is provided later on in this section.
The total Jacobian $J(\xi(\bar{\xi}_i))= ds_1 / d \bar{\xi}_i$ follows directly from \eqref{line_segmentparametrization} and reads
\begin{align}
\label{line_totaljacobian}
  J(\xi(\bar{\xi}_i))=\frac{ds_1}{d \bar{\xi}_i}=\frac{\partial s_1}{\partial \xi} \cdot \frac{\partial \xi}{\partial \bar{\xi}_i}
  =J_{ele}(\xi(\bar{\xi}_i)) \cdot \frac{\xi_{2,i}-\xi_{1,i}}{2} \quad \text{with} \quad i=1,...,n_{II},
\end{align}
where the mapping $J_{ele}(\xi(\bar{\xi}_i))$ from the arc-length space $s_1$ to the element parameter space $\xi$ results from the applied beam element formulation. Additionally, the sum over the number 
of Gauss points appearing in \eqref{line_pen_discreteweakform} has to be split:
\begin{align}
\label{line_pen_discreteweakform_multipleIS}
\begin{split}
  \mb{r}_{con,1} \! = \! \sum \limits_{i=1}^{n_{II}} \sum \limits_{j=1}^{n_{GR}}  \underbrace{w_j J(\xi_{ij},\xi_{1,i},\xi_{2,i}) \varepsilon \langle g(\xi_{ij}) \rangle \mb{N}_{1}^T(\xi_{ij}) \mb{n}(\xi_{ij})}_{\mb{r}_{con,1}^{ij}}, \,\,
  \mb{r}_{con,2} \! = \! \sum \limits_{i=1}^{n_{II}} \sum \limits_{j=1}^{n_{GR}} \underbrace{-w_j J(\xi_{ij},\xi_{1,i},\xi_{2,i}) \varepsilon \langle g(\xi_{ij}) \rangle \mb{N}_{2}^T(\eta_{c}(\xi_{ij})) \mb{n}(\xi_{ij})}_{\mb{r}_{con,2}^{ij}}.
\end{split}
\end{align}
Here, the terms $\mb{r}_{con,1}^{ij}$ and $\mb{r}_{con,2}^{ij}$ denote the residual contributions of one individual Gauss point $j$ in the integration interval $i$ and the element parameter 
coordinates $\xi_{ij}$ are evaluated according to \eqref{line_segmentparametrization} at the Gauss point coordinates $\bar{\xi}_j$:  
\begin{align}
\label{line_integrationparamvalues}
\begin{split}
  \xi_{ij}=\frac{1.0-\bar{\xi}_j}{2}\xi_{1,i} + \frac{1.0+\bar{\xi}_j}{2}\xi_{2,i} \quad \text{for} \quad i=1,...,n_{II}, \,\,\, j=1,...,n_{GR}.
\end{split}
\end{align}
Similar to the Gauss weights $w_j$, these Gauss point coordinates $\bar{\xi}_j$ are constant, i.e. not deformation-dependent, and identical for all integration intervals in case the same Gauss rule is 
applied in each of these intervals.
The Gauss quadrature applied for integration of \eqref{line_pen_discreteweakform_multipleIS} guarantees for exact integration of polynomials up to order $p=2 n_{GR}-1$ when using an integration rule with 
$n_{GR}$ quadrature points per integration interval. However, by simply integrating across the element boundaries of two successive master elements associated with the considered integration interval 
via the closest point projection \eqref{line_orthocond}, the integrand would not have a closed-form polynomial representation anymore and the mentioned polynomial order of exact integration can not 
be guaranteed. On the one hand, the integrands occurring in \eqref{line_pen_discreteweakform_multipleIS} are not of purely polynomial nature, a fact, that precludes exact integration anyway. 
On the other hand, strong discontinuities in the integrand, such as e.g. jumps in the contact force from a finite value to zero at the master beam endpoints, might increase the integration 
error drastically. In the following, we try to find a compromise between integration accuracy and computational efficiency. Thereto, we subdivide the integration intervals introduced above into sub-segments 
whenever the projections of master beam endpoints lie within the considered integration interval. With this integration interval segmentation, we avoid integration across strong discontinuities at the master beam 
endpoints (see Figure~\ref{fig:line_integrationsegments_endpoints}).
However, we do not create integration segments at all master element boundaries (see Figure~\ref{fig:line_integrationsegments_full}), where weak discontinuities in the integrand might occur. 
A further example for locations showing weak discontinuities in the integrand are the boundaries of active contact zones, i.e. locations where the contact line force decreases from a positive value to zero. 
As we will see later, the integration across this kind of discontinuities is rather unproblematic due to the applied beam formulation being $C^1$-continuous at the element boundaries 
(see Section \ref{sec:beamformulation}) and an applied quadratic penalty law regularization (see Section~\ref{sec:line_penaltylaws}) that leads to a smoother transition between contact and 
non-contact zones along the beam length. In order to find the boundary coordinate $\xi_{B}$ of an integration sub-segment created at a given master beam endpoint $\eta_{EP}$, the latter has 
to be projected onto the slave beam according to the following rule (with $p_{2}$ according to~\eqref{line_orthocond}):
\begin{align}
\label{line_segmentprojection}
\begin{split}
  p_{2}(\xi_B,\eta_{EP}) \dot{=} 0,
\end{split}
\end{align}
where the given parameter coordinate $\eta_{EP}$ can take on the values $-1.0$ and $1.0$ and $\xi_B$ is in general found via an iterative solution of \eqref{line_segmentprojection}. 
The derivative $p_{2,\xi}$ needed for such an iterative solution procedure can be found in~\ref{anhang:linearizationlinecontact}. In the worst, yet very unlikely, case that two master 
beam endpoints have valid projections according to \eqref{line_segmentprojection} within one integration interval, this interval has to be subdivided into three sub-segments. In this case, 
for one of these three sub-segments, both boundary coordinates $\xi_{1,i}$ and $\xi_{2,i}$ are determined via \eqref{line_segmentprojection} and are consequently deformation-dependent. 
Thus, in general, the boundary coordinates $\xi_{1,i}$ and $\xi_{2,i}$ introduced in \eqref{line_segmentparametrization} can be determined by:
\begin{align}
\label{line_integrationboundarycoordinates}
\begin{split}
  \xi_{1,i} & =
    \left\{\begin{array}{ll}
	    -1+(i-1) \cdot \frac{2}{n_{II}} & \text{if no valid master beam endpoint projection exists}  \\
	    \xi_{B1}(\eta_{EP},\mb{d}_{12}) & \text{if a valid master beam endpoint projection exists}
           \end{array}\right. 
           \quad \text{for} \quad i=1,...,n_{II}, \\
  \xi_{2,i} & =
    \left\{\begin{array}{ll}
	    -1+i \cdot \frac{2}{n_{II}} & \hspace{0.05\textwidth} \text{if no valid master beam endpoint projection exists}  \\
	    \xi_{B2}(\eta_{EP},\mb{d}_{12}) & \hspace{0.05\textwidth} \text{if a valid master beam endpoint projection exists}
           \end{array}\right.           
          \quad \text{for} \quad i=1,...,n_{II},
\end{split}
\end{align}
with $\mb{d}_{12}:=\left(\mb{d}_1^T,\mb{d}_2^T\right)^T$. Thus, in the standard case, these boundary coordinates are equidistantly distributed and constant. In case a valid projection of a master 
beam endpoint onto an integration interval exists, 
$\xi_{B1}(\eta_{EP},\mb{d}_{12})$ denotes the resulting deformation-dependent lower boundary of an created sub-segment, whereas $\xi_{B2}(\eta_{EP},\mb{d}_{12})$ denotes the corresponding upper boundary. 
Equation~\eqref{line_integrationboundarycoordinates} together with equations \eqref{line_integrationparamvalues} and \eqref{line_totaljacobian} provide all information necessary in order to 
evaluate the element residual contributions according to \eqref{line_pen_discreteweakform_multipleIS}. The linearization of the contributions $\mb{r}_{con,1}^{ij}$ and $\mb{r}_{con,2}^{ij}$ 
of one individual Gauss point on element $1$ can be formulated by means of the following total differential:
\begin{align}
\label{line_linearization}
\begin{split}
\mb{k}_{con,l}^{ij} =\dfrac{d \mb{r}_{con,l}^{ij}}{d \mb{d}_{12}} & =
  \dfrac{\partial \mb{r}_{con,l}^{ij}}{\partial \mb{d}_{12}} 
 +\dfrac{\partial \mb{r}_{con,l}^{ij}}{\partial \xi_{ij}}\dfrac{d \xi_{ij}}{d \mb{d}_{12}} 
 +\dfrac{\partial \mb{r}_{con,l}^{ij}}{\partial \eta_c}\dfrac{d \eta_c}{d \mb{d}_{12}}
 +\dfrac{\partial \mb{r}_{con,l}^{ij}}{\partial \xi_{1,i}}\dfrac{d \xi_{1,i}}{d \mb{d}_{12}}
 +\dfrac{\partial \mb{r}_{con,l}^{ij}}{\partial \xi_{2,i}}\dfrac{d \xi_{2,i}}{d \mb{d}_{12}}, \quad l=1,2 \\
  \text{with} \quad \dfrac{d \xi_{ij}}{d \mb{d}_{12}} & = \dfrac{\partial \xi_{ij}}{\partial \xi_{1,i}}\dfrac{d \xi_{1,i}}{d \mb{d}_{12}}+\dfrac{\partial \xi_{ij}}{\partial \xi_{2,i}}\dfrac{d \xi_{2,i}}{d \mb{d}_{12}} \\
  \text{and} \quad \dfrac{d \eta_c}{d \mb{d}_{12}} & = \dfrac{\partial \eta_{c}}{\partial \xi_{ij}}\dfrac{d \xi_{ij}}{d \mb{d}_{12}}+\dfrac{\partial \eta_{c}}{\partial \mb{d}_{12}}.
\end{split}
\end{align}
It should be emphasized that \textit{no} summation convention applies to the repeated indices appearing in \eqref{line_linearization}. 
Again, all basic linearizations appearing in \eqref{line_linearization} are summarized in~\ref{anhang:linearizationlinecontact}. The linearization in~\eqref{line_linearization} 
represents the most general case where the upper and lower boundary of an integration interval are deformation-dependent. However, this is only the case for slave elements with valid master beam 
endpoint projections according to~\eqref{line_segmentprojection} with $\xi_B \in [-1;1]$. In practical simulations, for the vast majority of contact element pairs this is not the case, 
i.e. $d \xi_{1,i} / d \mb{d}_{12} = \mb{0}$ and $d \xi_{2,i} / d \mb{d}_{12} = \mb{0}$, 
thus leading to the following remaining linearization contributions of an individual Gauss point:
\begin{align}
\label{line_linearization_compact}
\begin{split}
\mb{k}_{con,l}^{ij} =\dfrac{d \mb{r}_{con,l}^{ij}}{d \mb{d}_{12}} & =
  \dfrac{\partial \mb{r}_{con,l}^{ij}}{\partial \mb{d}_{12}} 
 +\dfrac{\partial \mb{r}_{con,l}^{ij}}{\partial \eta_c}\dfrac{\partial \eta_c}{\partial \mb{d}_{12}}, \quad l=1,2.
\end{split}
\end{align}
The combination of a line-to-line type contact model with a consistently linearized integration interval segmentation at the beam end points as presented in this section, a quadratically regularized 
smooth penalty law and a $C^1$-continuous smooth beam centerline representation is a distinctive feature of the proposed contact formulation. The benefits of these additional means are a drastical 
reduction of the integration error which enables a consistent spatial convergence behavior for a low number of Gauss points 
(see Section~\ref{sec:numerical_examples} for verification), an increase of the 
algorithmic robustness as well as a reduction of possible contact force/energy jumps without significantly increasing the computational effort.

%-------------------------------------------------------------------------------
%
\subsection{Penalty Laws}
\label{sec:line_penaltylaws}
\begin{figure}[ht]
 \centering
  \subfigure[Standard linear penalty law]
   {
    \includegraphics[width=0.35\textwidth]{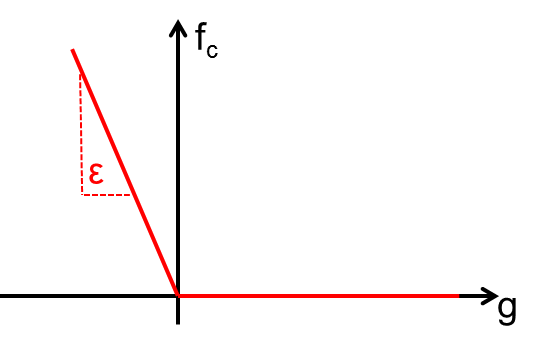}
    \label{fig:penlaw_lin}
   }
   \hspace{0.05 \textwidth}
   \subfigure[Linear penalty law with quadratic regularization]
   {
    \includegraphics[width=0.35\textwidth]{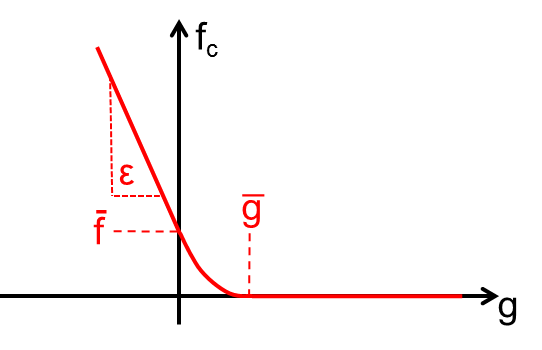}
    \label{fig:penlaw_linposquad}
   }
  \caption{Graphical visualization of standard and quadratically regularized penalty law}
  \label{fig:penlaws}
\end{figure}

%
%-------------------------------------------------------------------------------
Up to now, we have considered the following linear penalty law as introduced in \eqref{line_pen_contactforce} and illustrated in Figure \ref{fig:penlaw_lin}:
\begin{align}
\label{penlaw_lin}
  f_{c\varepsilon}(g)=\left\{\begin{array}{ll}
			-\varepsilon \cdot g, & g \leq 0 \\
			0, & g > 0 \\
	  \end{array}\right.
\end{align}
In practical simulations, one often applies regularized penalty laws that allow for a smooth contact force transition 
as illustrated in Figure \ref{fig:penlaw_linposquad}. This second variant is favorable from a numerical point of view:  First of all, it may improve the performance of tangent-based iterative solution schemes 
applied to the nonlinear system of equations stemming from the considered discretized problem, since a unique tangent exists at the transition point $\bar{g}$ between the states of ``contact'' 
and ``non-contact''. Secondly, the time integration scheme applied in dynamic simulations benefits from such a smooth contact force law. And thirdly, also numerical integration of the line-to-line 
contact forces along the beam length (see Section~\ref{sec:line_integrationsegments}) becomes more accurate if a smooth force law is used.
The quadratically regularized penalty law applied within this contribution has the following analytical representation:
\begin{align}
\label{penlaw_linposquad}
  f_{c\varepsilon}(g)=\left\{\begin{array}{lll}
			\bar{f} - \varepsilon \cdot g, & g \leq 0 \\
			\frac{\varepsilon \bar{g} - \bar{f}}{\bar{g}^2} \cdot g^2 -\varepsilon \cdot g + \bar{f}, & 0 < g \leq \bar{g} \\
			0, & g > \bar{g} \\
	  \end{array}\right.
	  \quad \text{with} \quad \bar{f}=\frac{\varepsilon \bar{g}}{2}.
\end{align} 
For simplicity, all theoretical derivations within this work are still based on a linear penalty law according to \eqref{penlaw_lin}. However, a more general form of these 
equations that is valid for arbitrary penalty laws can easily be derived 
by simply replacing all linear force-like expressions of the form $ -\varepsilon \langle g\rangle $ by the generic expression $f_{c\varepsilon}(g)$.

%-------------------------------------------------------------------------------
%
\subsection{Alternative constraint enforcement strategies}
\label{sec:line_alternatives}
%
%-------------------------------------------------------------------------------
Similar to point-to-point contact formulations, the constraint equation resulting from the line-to-line contact formulation can be considered within a variational framework by means of a 
Lagrange multiplier potential or by means of a penalty potential. In contrast to the point-to-point case, however, the constraint in~\eqref{line_constraint} is not only defined at a 
single point but rather on a parameter interval $\xi \in [\xi_a,\xi_b]$. According to Section~\ref{sec:line_weakform}, the penalty method, which introduces
no additional degrees of freedom, can be directly applied in terms of a space-continuous penalty potential, see~\eqref{line_pen_totalpotential}, that can alternatively be interpreted as a 
simple hyper-elastic stored-energy function representing the accumulated cross-section stiffness of the contacting beams. The final contact formulation resulting from such a procedure 
after spatial discretization and numerical integration is often denoted as Gauss-point-to-segment type formulation.\\

In contrary, the Lagrange multiplier method applied to the constraint in~\eqref{line_constraint} introduces an additional primary variable field $\lambda(\xi)$, which is typically discretized 
in a manner consistent to the spatial discretization of the displacement variables (discrete inf-sup stable pairing). Eventually, the nodal primary variables resulting from the discretization of 
the Lagrange multiplier field 
can be considered as additional unknowns or be eliminated by means of a penalty regularization (applied to a spatially discretized version of~\eqref{line_constraint}). 
Both variants are typically denoted as mortar-type formulations 
(see e.g. \cite{popp2009}, \cite{popp2010}). In Section~\ref{sec:line_lmdiscretization}, the main steps of applying a mortar formulation to beam contact problems, 
thus representing an alternative to the formulation of Section~\ref{sec:line_weakform}, are provided. Finally, in Sections~\ref{sec:line_comparison} and~\ref{sec:line_penaltyvslagrange}, 
a detailed comparison and evaluation of the variants "Gauss-point-to-segment" versus "mortar" and "penalty method" versus "Lagrange multiplier method", respectively, 
is performed in the context of beam-to-beam contact.

%-------------------------------------------------------------------------------
%
\subsubsection{Constraint enforcement based on consistent Lagrange multiplier discretization}
\label{sec:line_lmdiscretization}
%
%-------------------------------------------------------------------------------

As an alternative to Section~\ref{sec:line_weakform}, we now consider constraint enforcement via a Lagrange multiplier potential:
\begin{align}
\label{line_lm_totalpotential}
     \Pi_{c\lambda}=\int \limits_0^{l_1} \lambda(\xi) g(\xi) ds_1 
     \quad \text{with} \quad \lambda(\xi) \geq 0, \,\,\, g(\xi) \geq 0, \,\,\, \lambda(\xi) g(\xi) = 0.
\end{align}
Variation of the Lagrange multiplier potential leads to the following contact contribution to the weak form:
\begin{align}
\label{line_lm_weakform}
  \delta \Pi_{c\lambda} = \int \limits_0^{l_1} \left[ \lambda (\xi) \delta g (\xi) + \delta \lambda (\xi) g (\xi) \right]  ds_1
  \quad \text{and} \quad \delta g(\xi) = \left[ \delta \mb{r}_1(\xi) - \delta \mb{r}_2(\xi) \right]^T \mb{n}(\xi).
\end{align}
In \eqref{line_lm_weakform}, the contact force $\mb{f}_{c \lambda}(\xi)= -\lambda (\xi) \mb{n}(\xi)=:f_{c \lambda}(\xi) \mb{n}(\xi)$ transferred between the two beams can again be interpreted 
as a distributed line force. This time, the Lagrange multiplier field represents the magnitude of this line force. Next, spatial discretization has to be performed. Again, we consider the 
contribution of one slave beam element $1$ and one master beam element $2$.
In addition to the spatial discretization~\eqref{interpolation}, a trial space 
$\lambda \! \approx \! \lambda_h \! \in \! \mathcal{S}_{\lambda h} \! \subset \! \mathcal{S}_{\lambda} \! \subset \! \Re$
and a weighting space $\delta \lambda \! \approx \delta \lambda_h \! \in \! \mathcal{V}_{\lambda h} \! \subset \! \mathcal{V}_{\lambda} \! \subset \! \Re$ have to 
be defined for the field of Lagrange multipliers, too: 
\begin{align}
\label{line_discretization}
  \lambda(\xi) \approx \lambda_{h}(\xi) = \sum \limits_{j=1}^{n_{\lambda}} N_{\lambda,1}^j(\xi) \hat{\lambda}_1^j =:\mb{N}_{\lambda,1}(\xi) \boldsymbol{\hat{\lambda}}_1, 
  \,\,\,\,\,\,\,\,\,\,\, \delta \lambda(\xi) \approx \delta \lambda_{h}(\xi) = 
  \sum \limits_{j=1}^{n_{\lambda}} N_{\lambda,1}^j(\xi) \delta \hat{\lambda}_1^j =:\mb{N}_{\lambda,1}(\xi) \delta \boldsymbol{\hat{\lambda}}_1.
\end{align}
Here, $n_{\lambda}$ represents the number of nodes of the Lagrange multiplier discretization per slave element, the vector $\mb{N}_{\lambda,1}(\xi)$ collects the corresponding test and trial 
functions with support on slave beam $1$, and $\boldsymbol{\hat{\lambda}_1}$ as well as 
$\delta \boldsymbol{\hat{\lambda}}_1$ contain the corresponding discrete nodal Lagrange multipliers and their variations, respectively 
(see e.g. \cite{wohlmuth2001} concerning a proper choice of the spaces $\mathcal{S}_{\lambda h}$ and $\mathcal{V}_{\lambda h}$). 
Inserting~\eqref{interpolation} and~\eqref{line_discretization} into~\eqref{line_lm_weakform} 
and replacing the analytical integral by a Gauss quadrature finally leads to the following contribution of elements $1$ and $2$ to the discretized weak form:
\begin{align}
\label{line_lm_discreteweakform}
\begin{split}
           \delta \mb{d}_1^T \underbrace{\sum \limits_{k=1}^{n_{GP}}  w_k J(\xi_k) \lambda(\xi_k) \mb{N}_{1}^T(\xi_k) \mb{n}(\xi_k)}_{=\mb{r}_{con,1}}
         + \delta \mb{d}_2^T \underbrace{\sum \limits_{k=1}^{n_{GP}} -w_k J(\xi_k) \lambda(\xi_k) \mb{N}_{2}^T(\eta_{c,k}) \mb{n}(\xi_k)}_{=\mb{r}_{con,2}}
         + \delta \boldsymbol{\hat{\lambda}}_1^T \underbrace{\sum \limits_{k=1}^{n_{GP}} w_k J(\xi_k) \mb{N}_{\lambda,1}^T(\xi_k) g(\xi_k)}_{=\mb{r}_{\lambda,1,2}}.
\end{split}
\end{align}
Again, $\mb{r}_{con,1}$ and $\mb{r}_{con,2}$ represent the contact force residual contributions of slave element $1$ and master element $2$, whereas $\mb{r}_{\lambda,1,2}$ denotes the corresponding 
residual contribution stemming from constraint equation \eqref{line_constraint}. 
Based on \eqref{line_lm_discreteweakform}, different strategies of constraint enforcement are possible: Considering the nodal Lagrange multipliers $\boldsymbol{\hat{\lambda}}_1$ 
as additional unknowns would lead to an exact satisfaction of the discrete version of the constraints~\eqref{line_constraint}. Alternatively, these discrete constraint equations can be 
regularized by means of a penalty approach. Let $n_{ele,s}$ denote the total number of slave elements. Then, one typically defines so-called nodal gaps $\hat{g}^j$ according to
\begin{align}
\label{line_lm_nodalgaps}
    \hat{g}^j:= \sum \limits_{e=1}^{n_{ele,s}} \sum \limits_{k=1}^{n_{GP}} w_k J(\xi_k) N_{\lambda, 1}^j(\xi_k) g(\xi_k) \quad \text{for} \quad j=1,...,n_{\lambda}.
\end{align}
In \eqref{line_lm_nodalgaps}, a summation over all slave elements with support of the shape function $N_{\lambda, 1}^j(\xi)$ assigned to the nodal 
gap $\hat{g}_j$ is sufficient. Consequently, each nodal gap according to~\eqref{line_lm_nodalgaps} represents one line of the total residual contribution 
$\mb{R}_{\lambda}$ resulting from constraint equation \eqref{line_constraint}. Now, one can replace the nodal Lagrange multipliers by nodal penalty forces: 
\begin{align}
\label{line_lm_nodalpenforces}
    \hat{\lambda}_{\varepsilon,1}^j=\varepsilon \langle \hat{g}^j \rangle \quad \text{for} \quad j=1,...,n_{\lambda}.
\end{align}
Inserting the nodal penalty forces instead of the unknown nodal Lagrange multipliers into \eqref{line_lm_discreteweakform} finally results in:
\begin{align}
\label{line_lm_discreteweakformpenalty}
\begin{split}
           \delta \mb{d}_1^T \underbrace{\sum \limits_{k=1}^{n_{GP}}  w_k J(\xi_k) \lambda_{\varepsilon}(\xi_k) \mb{N}_{1}^T(\xi_k) \mb{n}(\xi_k)}_{=\mb{r}_{con,1}}
         + \delta \mb{d}_2^T \underbrace{\sum \limits_{k=1}^{n_{GP}} -w_k J(\xi_k) \lambda_{\varepsilon}(\xi_k) \mb{N}_{2}^T(\eta_{c,k}) \mb{n}(\xi_k)}_{=\mb{r}_{con,2}}
         \quad \text{with} \quad \lambda_{\varepsilon}(\xi)= \sum \limits_{j=1}^{n_{\lambda}} N_{\lambda,1}^j(\xi) \hat{\lambda}_{\varepsilon,1}^j.
\end{split}
\end{align}
This procedure eliminates the additional nodal unknowns $\boldsymbol{\hat{\lambda}}$. However, the constraint of vanishing nodal gaps $\hat{g}^j$ will not be exactly fulfilled anymore. 
The only difference of the discretized weak form~\eqref{line_pen_discreteweakform}, i.e. the one resulting from a space-continuous penalty potential, and 
\eqref{line_lm_discreteweakformpenalty}, i.e the one resulting from a discretized Lagrange multiplier potential and a subsequent penalty regularization, lies in the definition of the scalar contact forces 
$\lambda_{\varepsilon}(\xi)$ and $\varepsilon \langle g(\xi) \rangle$, respectively.

%-------------------------------------------------------------------------------
%
\subsubsection{Comparison of the two penalty approaches}
\label{sec:line_comparison}
%
%-------------------------------------------------------------------------------

\begin{figure}[ht]
 \centering
  \subfigure[Problem setup and geometry]
   {
    \includegraphics[height=0.18\textwidth]{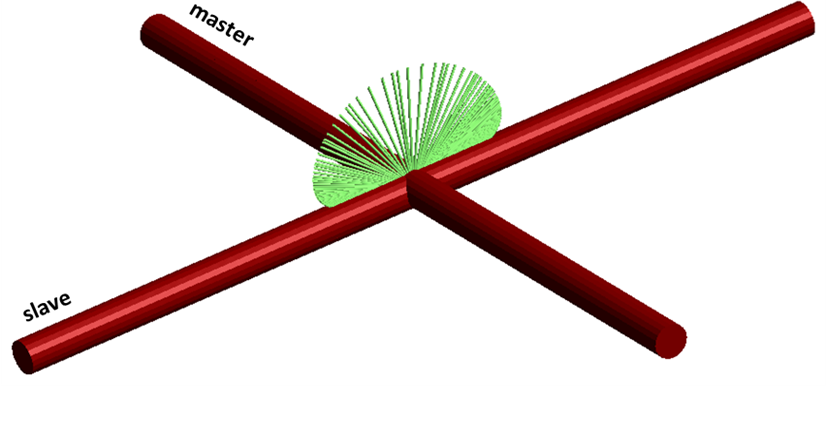}
    \label{fig:2straightlargeangleeles_setup}
   }
   \subfigure[Evolution of gap function]
   {
    \includegraphics[height=0.19\textwidth]{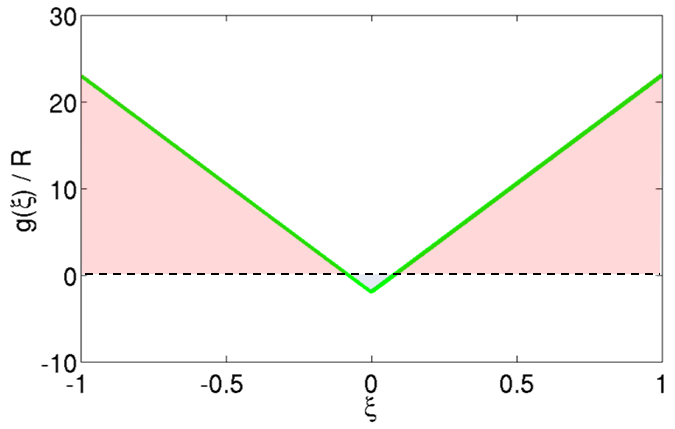}
    \label{fig:2straightlargeangleeles_gap}
   }
   \subfigure[Evolution of contact force]
   {
    \includegraphics[height=0.19\textwidth]{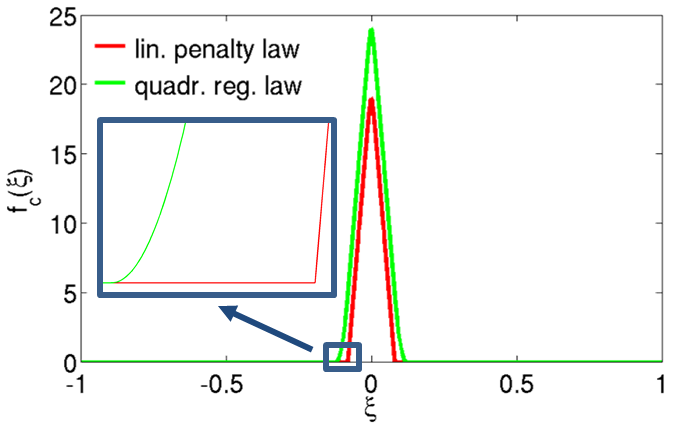}
    \label{fig:2straightlargeangleeles_force}
   }
  \caption{Two straight beam elements with large penetration and large contact angle}
  \label{fig:2straightlargeangleeles}
\end{figure}

The main advantage of the formulation presented in Section \ref{sec:line_lmdiscretization} is that it results from a consistent Lagrange multiplier discretization. As long as the trial and 
weighting spaces $\mathcal{S}_{h}, \mathcal{V}_{h}, \mathcal{S}_{\lambda h}$ and $\mathcal{V}_{\lambda h}$ are chosen such that a proper discrete inf-sup-stability condition is satisfied, 
no contact-related locking effects have to be expected, even for large values of the penalty parameter. This does in general not hold for the 
formulation presented in Section~\ref{sec:line_weakform}, where contact-related locking might occur for very high penalty parameters.
% \begin{minipage}{15.0 cm}
% \textbf{Remark:} In this context, the formulation according to \eqref{line_lm_discreteweakformpenalty} shows strong similarities to finite element technologies (e.g. EAS, ANS etc.) based on 
% mixed formulations and/or strain re-interpolation which are often applied in order to avoid/relieve certain locking phenomena.\\
% \end{minipage}
When considering highly slender beams, moderate values of the penalty parameter are often sufficient in order to satisfy the contact constraint 
with the desired accuracy. In Section~\ref{sec:numerical_examples}, it will be verified numerically that within this range of penalty parameters the spatial convergence behavior 
is not deteriorated by contact-related locking effects when applying the contact formulation according to Section~\ref{sec:line_weakform}. A crucial advantage of the latter formulation lies in its 
efficiency and its straight-forward implementation. On the one hand, the numerical implementation of the variant presented in Section~\ref{sec:line_lmdiscretization} requires an additional 
element evaluation loop in order to determine the nodal gaps according to \eqref{line_lm_nodalgaps}, or, in other words, the penalty-based elimination of the Lagrange multipliers 
cannot exclusively be conducted on element level. On the other hand, in combination with the standard gap function definition according to~\eqref{line_constraint}, this variant requires 
a very fine finite element discretization when applied to contact problems involving highly slender beams.
\begin{figure}[ht]
 \centering
  \subfigure[Problem setup and geometry]
   {
    \includegraphics[height=0.18\textwidth]{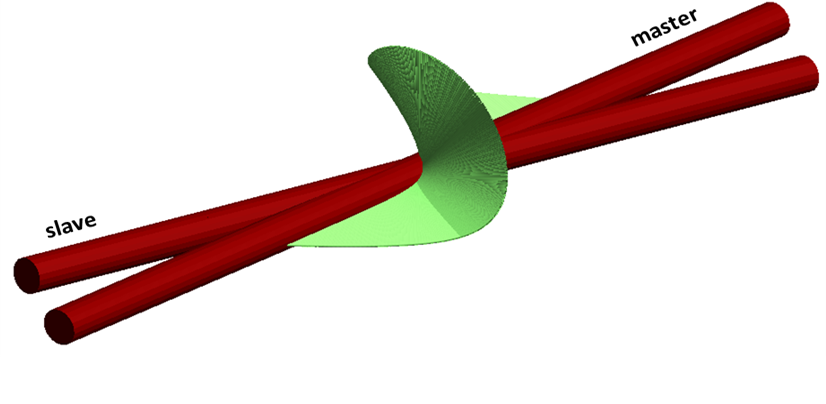}
    \label{fig:2straightsmallangleeles_setup}
   }
   \subfigure[Evolution of gap function]
   {
    \includegraphics[height=0.19\textwidth]{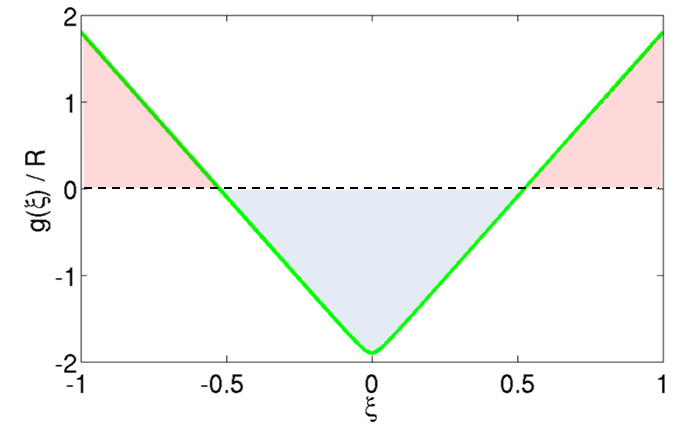}
    \label{fig:2straightsmallangleeles_gap}
   }
   \subfigure[Evolution of contact force]
   {
    \includegraphics[height=0.19\textwidth]{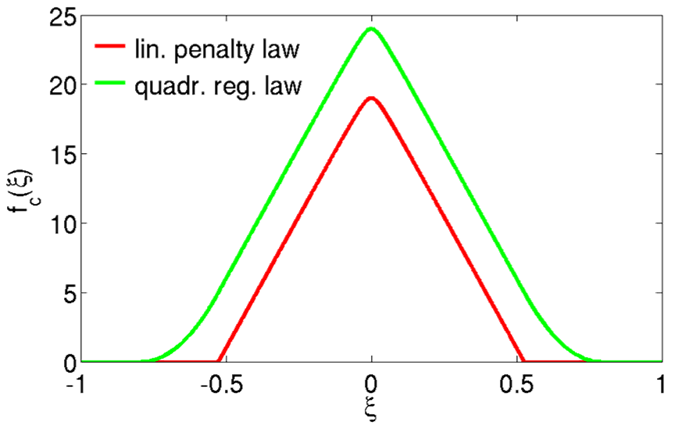}
    \label{fig:2straightsmallangleeles_force}
   }
  \caption{Two straight beam elements with large penetration and small contact angle}
  \label{fig:2straightsmallangleeles}
\end{figure}
This fact will be illustrated in the following by means of Figures \ref{fig:2straightsmallangleeles} and \ref{fig:2straightlargeangleeles}. In Figure \ref{fig:2straightsmallangleeles_setup}, 
two straight beam elements with cross-section radii $R_1=R_2=R$ characterized by a comparatively small contact angle and a large penetration of almost $g(\xi_c)\approx 2R$ are depicted. The resulting 
contact line force vector field according to \eqref{line_pen_contactforce} is illustrated in green color. Furthermore, in Figure \ref{fig:2straightsmallangleeles_gap}, the evolution of the gap function 
is plotted over the length of the slave beam element. With increasing penalty parameter, the formulation according to Section~\ref{sec:line_lmdiscretization} forces the nodal 
gaps in~\eqref{line_lm_nodalgaps} to vanish. Roughly speaking, this means that the areas enclosed by positive gaps and the areas enclosed 
by negative gaps, as indicated with red and blue color in Figure \ref{fig:2straightsmallangleeles_gap}, must balance each other.
For small contact angles and reasonable spatial discretizations, this is possible. However, when looking at the gap-function evolution resulting from two almost perpendicular beams as illustrated 
in Figure \ref{fig:2straightlargeangleeles_gap}, such a balancing can only be achieved if the beam element length is reduced drastically. This need for 
a sufficiently fine spatial discretization increases the numerical effort of this method. Alternatively, one might modify the definition of the gap function $g(\xi)$, such that negative/positive gap 
contributions are weighted stronger/weaker. Since such an extra effort is not necessary for the procedure proposed in Section~\ref{sec:line_weakform}, we want to focus on this variant.

%-------------------------------------------------------------------------------
%
\subsubsection{Penalty method vs. Lagrange multiplier method}
\label{sec:line_penaltyvslagrange}
%
%-------------------------------------------------------------------------------

Constraint enforcement by means of Lagrange multipliers is common practice in the field of computational contact mechanics for solids, especially in combination with mortar methods 
(see Section~\ref{sec:line_lmdiscretization}), due to some advantageous properties, for example concerning the accuracy of contact resolution. Even though the 
application of the Lagrange multiplier method for constraint enforcement in beam-to-beam contact scenarios has already been investigated in \cite{litewka2005}, 
the vast majority of publications in this field is based on regularized constraint enforcement via the penalty method.
This fact can be justified by a couple of reasons: When considering discretizations based on structural models the ratio of surface degrees of freedom to all degrees of freedom 
(=1 for beams) is much larger than for solid discretizations based on a 3D continuum theory. Consequently, also the ratio of additional Lagrange multiplier degrees of freedom 
to displacement degrees of freedom would be comparatively high when enforcing, e.g., beam-to-beam line contact constraints (see Section~\ref{sec:line_lmdiscretization}) by means of 
Lagrange multipliers. Furthermore, when modeling slender structures by means of mechanical beam models, which are often based on the assumption of rigid cross-sections, 
computational efficiency is one of the key aspects whereas the resolution of exact contact pressure distributions and other mechanical effects on the length scale of the 
cross-section, which is typically by orders of magnitude smaller than the length dimension of the beam, is not of primary interest. If one is primarily interested 
in the global system behavior, even penetrations on the order of magnitude of the cross-section radius are often tolerable. Typically, penalty parameters required to limit 
the penetrations to such values decrease with the beam thickness. Often, the required values are proportional to the beam bending stiffness and therefore the penalty contributions do not 
significantly deteriorate the conditioning of the system matrix which is usually dominated by high axial and shear stiffness terms.\\

Besides the arguments above, there is one further crucial point, which makes the penalty method not only preferable to constraint enforcement via Lagrange multipliers, but which 
even prohibits the use of the latter method. Many of the perhaps most efficient and elegant beam models available in the literature (see e.g. the comparison of 
ANS beams and geometrically exact beams in \cite{romero2008}), are based on the assumption of rigid cross-sections. Especially when considering very thin beams, 
this assumption is well-justified and the properties of the resulting beam formulations are desirable from a numerical point of view. However, combining the assumption 
of rigid cross-sections and contact constraint enforcement via Lagrange multipliers leads to the following dilemma when considering, e.g., the dynamic collision of two 
beams: In the range of large contact angles, the initial kinetic energy will be transformed into elastic bending energy and back to kinetic energy during the impact. However, 
with decreasing contact angle the elastic bending deformation decreases and in the limit of two matching, exactly parallel beams the amount of elastic deformation during the 
collision drops to zero, since the cross-sections are rigid. The accelerations and contact forces resulting from such a scenario are unbounded and the resulting numerical 
problem become singular. Thus, undoubtedly, a certain amount of cross-section flexibility is indispensable when modeling such a scenario. This cross-section flexibility can be provided by a penalty force 
law such as the one in Section~\ref{sec:line_weakform}, which already has the structure of a typical hyper-elastic strain energy function and models the accumulated stiffness 
of the cross-sections of the two contacting beams. Of course, this idea can be refined by deriving more sophisticated penalty laws in form of reduced models based on a 
continuum mechanical analysis of the cross-section deformation and stiffness. However, since our primary intention is the regularization of parallel-impact scenarios 
and not the resolution of local deformations on the cross-section scale, we will keep the simple and convenient force law according to \eqref{penlaw_linposquad} in the following.
Nevertheless, the adaption of the presented theory to more general penalty laws is straightforward. Furthermore, with these considerations in mind, the penalty parameter 
in the context of rigid-cross-section beam contact is no longer a pure mathematical tool of constraint enforcement, but it rather has a physical meaning: it serves as 
mechanical model of the beam cross-section stiffness. This interpretation simplifies the determination of a proper penalty parameter.

%-------------------------------------------------------------------------------
%
\section{Endpoint-to-line and endpoint-to-endpoint contact contributions}
\label{sec:endpoint}
%
%-------------------------------------------------------------------------------
\begin{figure}[ht]
 \centering
   \subfigure[Element parameter space]
   {
    \includegraphics[width=0.21\textwidth]{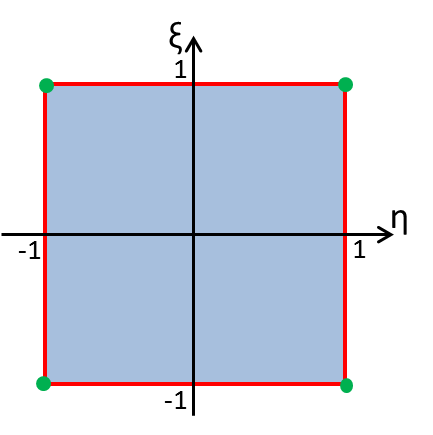}
    \label{fig:endpoints1}
   }
   \hspace{0.08\textwidth}
   \subfigure[Endpoint-to-line]
   {
    \includegraphics[width=0.20\textwidth]{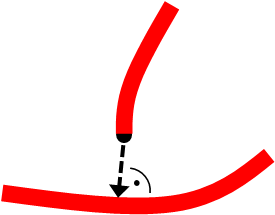}
    \label{fig:endpoints2}
   }
   \hspace{0.08\textwidth}
   \subfigure[Endpoint-to-endpoint]
   {
    \includegraphics[width=0.25\textwidth]{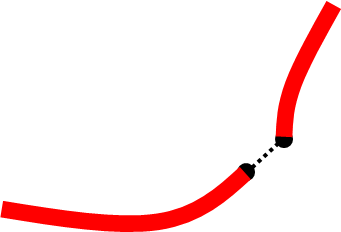}
    \label{fig:endpoints3}
   }
  \caption{Possible contact configurations involving the interior as well as the endpoints of the beams.}
  \label{fig:endpoints}
\end{figure}
The contact formulations presented in the last two sections have only considered solutions of the minimal distance problem within the element parameter domain $\xi,\eta \in [-1;1]$ 
as represented by the blue area in Figure~\ref{fig:endpoints1}. Due to the $C^1$-continuity of our discrete centerline representation, also solutions coinciding with the element nodes are 
found by this procedure. However, a minimal distance solution can also occur in form of a boundary minimum at the physical endpoints of the contacting beams. The boundary solutions indicated by the 
four red lines in Figure~\ref{fig:endpoints1} represent solutions with one parameter taking on the value $-1$ or $1$ and the other parameter being arbitary. Mechanically, these 
solutions can be interpreted as the minimal distance appearing between a physical beam endpoint and an arbitrary beam segment as indicated in Figure~\ref{fig:endpoints2}. Additionally, a minimum 
can also occur in form of the distance between the physical endpoints of both beams (see Figure~\ref{fig:endpoints3}), which corresponds to the four green corner points in Figure~\ref{fig:endpoints1}.
Neglecting these boundary minima can lead to impermissibly large penetrations and even to an entirely undetected crossing 
of the beams. At first view, these contact configurations seem to be comparatively rare for thin beams and the mechanical influence of these contact contributions seems to be limited. However, 
practical simulations have shown that neglecting these contributions does not only lead to a slight inconsistency of the mechanical model itself but also 
to a drastically reduced robustness of the nonlinear solution scheme, since initially undetected large penetrations can lead to considerable jumps in the contact forces during the iterations of a nonlinear 
solution scheme. 
While for the endpoint-to-endpoint case, the contact point coordinates are already given, the endpoint-to-line case requires a unilateral closest-point-projection similar to the one in~\eqref{line_mindist}. 
Depending on which beams endpoint is given, this unilateral closest-point-projection either searches for the closest point $\eta_c$ to a given point $\xi \in \{-1,1\}$ or for the closest 
point $\xi_c$ to a given point $\eta \in \{-1,1\}$. As soon as the contact point coordinates are known, one can directly apply the residual contribution 
of the point-to-point contact formulation according to~\eqref{point_discreteweakform}. From a geometrical point of view, applying this model means that the beam endpoints are 
approximated by hemispherical surfaces.
Again, it is justified to only consider the variation contribution with fixed $\xi$ and fixed $\eta$ for $\delta g$ according to~\eqref{point_weakform}, since either the considered parameter 
coordinate is indeed fixed (if representing a physical endpoint) or the corresponding tangent vector is perpendicular to the contact normal (if representing the projection onto a segment).
Nevertheless, one has to distinguish between the cases endpoint-to-endpoint and endpoint-to-line contact in order to correctly include the increments $\Delta \xi$ and $\Delta \eta$ in the 
linearizations of the contact residuals (see~\ref{anhang:linearizationendpoint} for details).

%-------------------------------------------------------------------------------
%
\section{Numerical examples}
\label{sec:numerical_examples}
%
%-------------------------------------------------------------------------------
In this section, we want to verify the robustness and accuracy of our new line-to-line contact formulation presented in Section~\ref{sec:line}. For all examples, a standard Newton-Raphson scheme is 
applied in order to solve the nonlinear system of equations $\mb{R}_{tot}$ resulting from the discretized weak form~\eqref{global_system}. 
As convergence criteria we check the Euclidean norms of the displacement increment vector $\Delta \mb{D}^k$ and of the residual vector $\mb{R}^k_{tot}$ at Newton iteration $k$. 
For convergence, these norms have to fall 
below prescribed tolerances $\delta_{\mb{R}}$ and $\delta_{\mb{D}}$, i.e. $||\mb{R}^k_{tot}||<\delta_{\mb{R}}$ and $||\Delta \mb{D}^k||<\delta_{\mb{D}}$. If nothing to the contrary is mentioned, 
these tolerances are chosen according to the following standard values $\delta_{\mb{R}}=\delta_{\mb{D}}=1.0 \cdot 10^{-7}$.

%-------------------------------------------------------------------------------
%
\subsection{Example 1: Patch test}
\label{sec:examples_example1}
%
%-------------------------------------------------------------------------------
\begin{figure}[ht]
 \centering
 \includegraphics[width=0.6\textwidth]{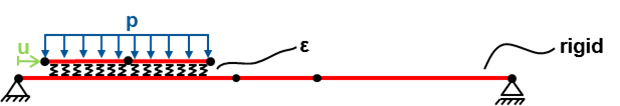}
 \caption{Static patch pest: Initial configuration}
 \label{fig:patchtest_initialconfig}
\end{figure}
\begin{figure}[ht]
 \centering
   \subfigure[Integration without segmentation at the beam endpoints]
   {
    \includegraphics[width=0.485\textwidth]{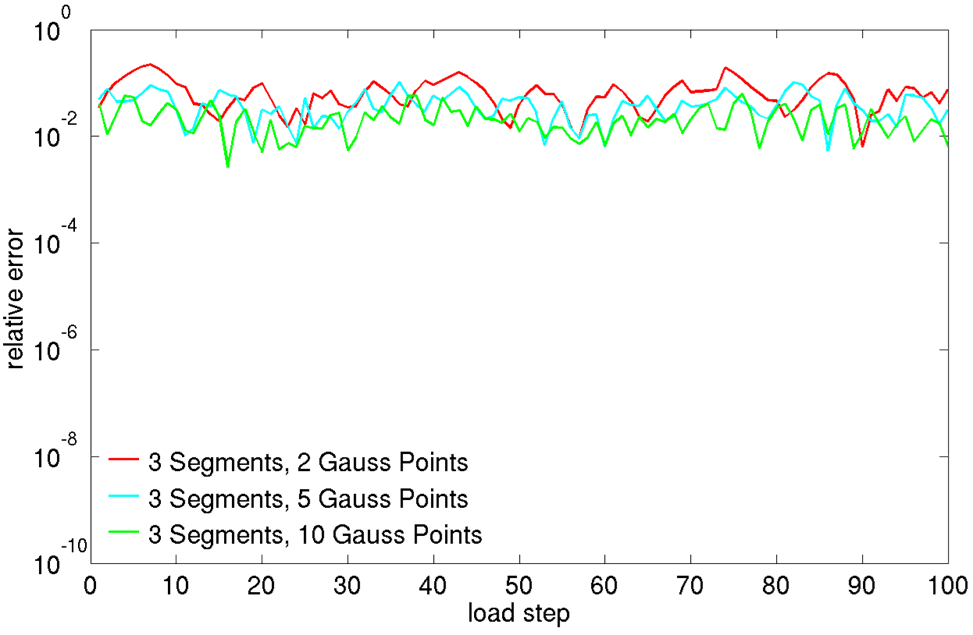}
    \label{fig:line_patchtest_errorsingap_NES}
   }
   \subfigure[Integration with segmentation at the beam endpoints]
   {
    \includegraphics[width=0.485\textwidth]{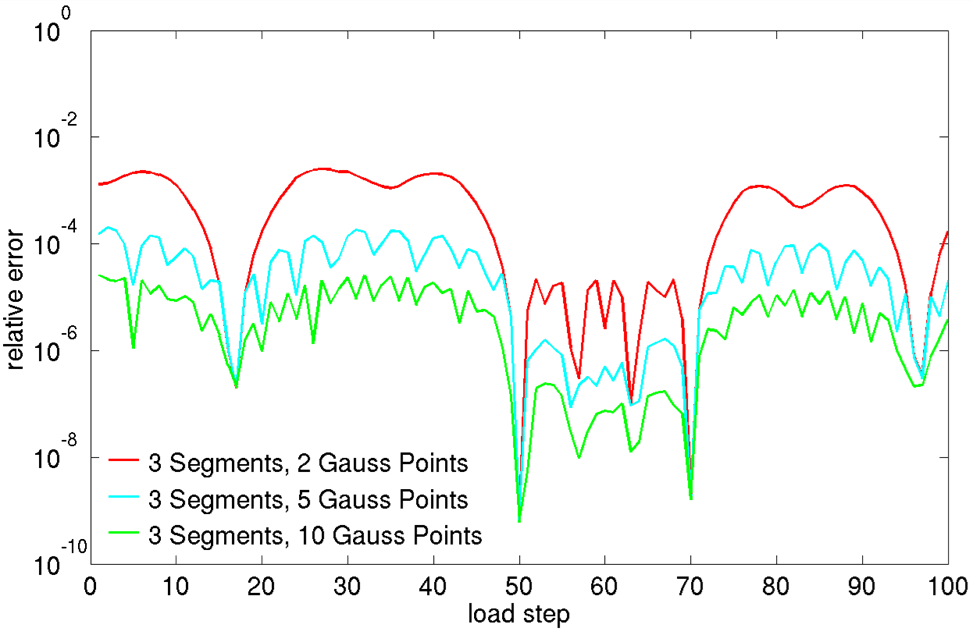}
    \label{fig:line_patchtest_errorsingap_ES}
   }
  \caption{Static patch pest: Average error of the gap at the Gauss points for different Gauss rules}
  \label{fig:line_patchtest_errorsingap}
\end{figure}

The first example is a simple static patch test that should verify the effectiveness of the integration interval segmentation introduced in Section~\ref{sec:line_integrationsegments}. 
As illustrated in Figure~\ref{fig:patchtest_initialconfig}, the example consists of one completely fixed, rigid beam discretized with three beam elements (different element lengths) and a second, deformable 
beam discretized by two beam elements with cross-section radii $R_1=R_2=0.005$, Young´s moduli $E_1=E_2=1.0 \cdot 10^9$, length of the first beam $l_1=2.0$ and length of the second 
beam $l_2=0.8$. The second beam is loaded by a constant transverse line load $p=1.0$ and its left endpoint is exposed to a Dirichlet-displacement of $\Delta u=1.001$ within $100$ 
equidistant load steps. Furthermore, contact interaction between the two beams is modeled by the linear penalty law according to \eqref{penlaw_lin} with a penalty parameter $\varepsilon=500$.
As a consequence of the constant transverse line load and the chosen penalty parameter, there exists a trivial analytical solution with a constant gap $g_{ref}=-p/\varepsilon=-0.002$ along the entire upper beam. 
In order to verify the working principle of the integration interval segmentation in the presence of strong discontinuities, we have chosen the first (rigid) beam as slave beam. In Figure 
\ref{fig:line_patchtest_errorsingap}, the average relative error
\begin{align*}
e_{rel}=\sum \limits_{i=1}^{n_{GP,tot}}\frac{g_i-g_{ref}}{n_{GP,tot} \cdot g_{ref}}
\end{align*}
of the gaps $g_i$ at the active Gauss points is plotted over the number of load steps for the formulations with and without integration interval segmentation at the 
beam endpoints in combination with different numbers of Gauss points $n_{GP,tot}$.
In all cases, three integration intervals per slave element have been applied. From Figure \ref{fig:line_patchtest_errorsingap_NES}, one observes that the strong discontinuity of the contact force 
$\varepsilon \langle g(\xi_{ij}) \rangle$ occurring in the integrand of \eqref{line_pen_discreteweakform_multipleIS} leads to a considerable integration error that only gradually decreases when increasing 
the number of Gauss points. As expected, 
the formulation with integration interval segmentation (see Figure \ref{fig:line_patchtest_errorsingap_ES}) yields a significantly lower integration error level and a faster decline in the error with increasing number 
of Gauss points. Yet, even this formulation does not allow for an exact integration, in general, since the test functions $\mb{N}_{1}$ and $\mb{N}_{2}$ 
in~\eqref{line_pen_discreteweakform_multipleIS} have no closed-form polynomial representation across the element boundaries. However, it will be shown in the next examples that the 
corresponding integration error is typically lower than the overall discretization error and therefore of no practical relevance. Furthermore, compared to a formulation with integration interval segmentation 
at all master beam element nodes, which would then allow for exact numerical integration, the proposed segmentation strategy is considerably less computationally expensive. 

%-------------------------------------------------------------------------------
%
\subsection{Example 2: Twisting of two beams}
\label{sec:examples_example2}
%
%-------------------------------------------------------------------------------
\begin{figure}[ht]
 \centering
   \subfigure[Initial and deformed configuration of contacting beams]
   {
    \includegraphics[height=0.25\textwidth]{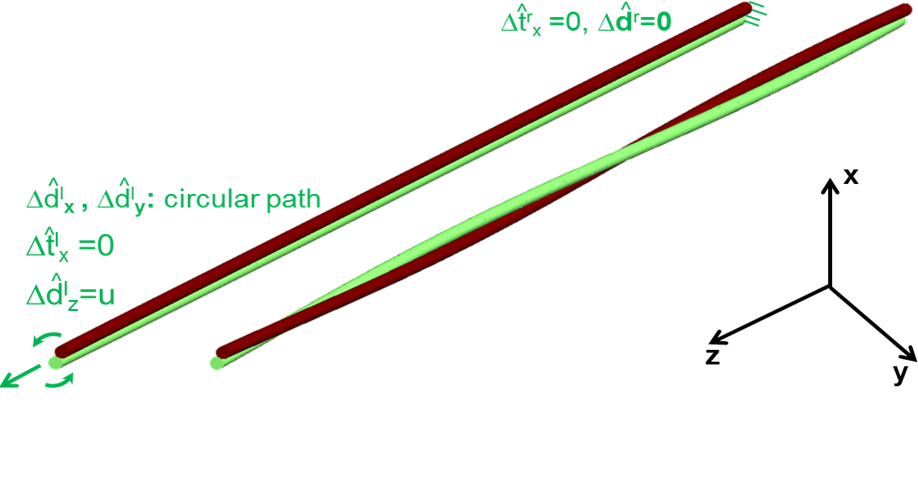}
    \label{fig:line_twobeamstwisting_config}
   }
   \hspace{0.1\textwidth}
   \subfigure[Relative $L^2$-error over element length]
   {
    \includegraphics[height=0.27\textwidth]{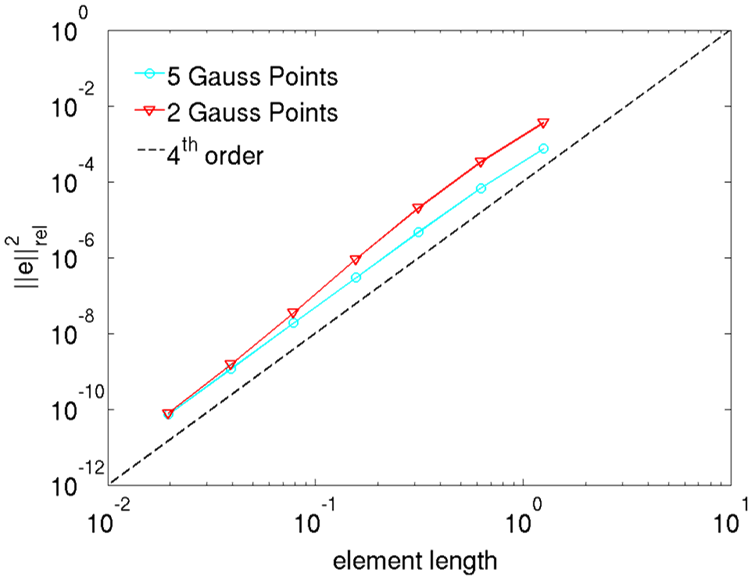}
    \label{fig:line_twobeamstwisting_L2}
   }
  \caption{Two initially straight and parallel beams in contact: Dirichlet-controlled twisting into a double-helical shape}
  \label{fig:line_twobeamstwisting}
\end{figure}
The second example aims at verifying the accuracy and consistency of the line-to-line contact formulation by investigating the spatial convergence behavior. Thereto, we consider two initially straight 
and parallel beams with circular cross-sections and radii $R_1=R_2=R=0.01$, initial lengths $l_1=l_2=l=5$ and Youngs moduli $E_1=E_2=E=1.0 \cdot 10^9$ as illustrated in 
Figure~\ref{fig:line_twobeamstwisting_config}. The initial geometries of the two beams $k=1,2$ are given by the analytical expression:
\begin{align}
\label{line_example2_initialgeometry}
  \mb{r}_{0k}(s)=
  \left(\begin{array}{lll}
         (-1)^{k-1}r \\
         0 \\
         s
         \end{array}\right),\,\, 
         s \in [0;l], \,\,
         r\!=\!R\!-\!\frac{|g_0|}{2}, \,\,
         k=1,2.
\end{align}
The distance of the two beams is chosen such that they exhibit an initial gap of $g_0=-0.1R$. We clamp the beams at one end and move the beam 
cross-sections in a Dirichlet-controlled manner at the other end such that the corresponding 
cross-section center points move on a circular path. By this procedure, the two beams get twisted into a double-helical shape as illustrated in Figure \ref{fig:line_twobeamstwisting_config}. 
We try to adapt the system parameters in a way such that the analytical solution for the deformed beams is exactly represented by a helix with constant slope 
according to
\begin{align}
\label{line_example2_helix}
  \mb{r}_k(\varphi)\!=\!
  \left(\begin{array}{lll}
         r \cos \left[\varphi+(k-1)\pi \right] \\
         r \sin \left[ \varphi+(k-1) \pi \right] \\
         h \varphi
         \end{array}\right),\,\, 
         \varphi \! \in \! [0;2\pi], \,\,
         r\!=\!R\!-\!\frac{|g_0|}{2}, \,\,
         h\!=\!\!\sqrt{\left( \left(\frac{(1.0 + \epsilon) l}{2\pi}\right)^2\!-r^2 \right)}, \,\,
         \epsilon\!=\!0.01, \,\,
         k\!=\!1,2.
\end{align}
In the following, we only present the corresponding results, while the derivation based on the projected ODEs representing the strong form of the Kirchhoff 
theory (see \cite{meier2015}) is summarized in~\ref{anhang:analyticalsolution_twistingoftwobeams}. Before the actual twisting process starts, the two beams are pre-stressed by an axial displacement 
at the left endpoints (superscript ``l'') 
\begin{align}
\label{line_example2_axialdisp}
\Delta \hat{d}_{1,z}^l = \Delta \hat{d}_{2,z}^l = u=2\pi h-l \approx 4.9647 \cdot 10^{-2}
\end{align}
within one load step. Then, these points are moved on a circular path with radius $r=R-|g_0|/2=0.0095$, i.e.
\begin{align}
\label{line_example2_circularpath}
  \Delta \hat{d}_{1,x}^l \!=\!-r\left[1 \!-\! \cos \left(\dfrac{k 2\pi}{n_l}\right) \right], \,\, \Delta \hat{d}_{1,y}^l\!=\!r \sin\left(\dfrac{k 2\pi}{n_l}\right), \,\,
  \Delta \hat{d}_{2,x}^l\!=\!r\left[1 \!-\! \cos\left(\dfrac{k 2\pi}{n_l}\right) \right], \,\, \Delta \hat{d}_{2,y}^l\!=\!-r \sin\left(\dfrac{k 2\pi}{n_l}\right), \,\, k=1,...,n_l,
\end{align}
within $n_l=8$ further load steps in order to end up with one full twist rotation. The translational displacements at the right endpoints of the right beams (superscript ``r'') are set to zero, i.e.
\begin{align}
\label{line_example2_rightendpoint}
  \Delta \hat{d}_{1,x}^r=\Delta \hat{d}_{2,x}^r=\Delta \hat{d}_{1,y}^r=\Delta \hat{d}_{2,y}^r=\Delta \hat{d}_{1,z}^r=\Delta \hat{d}_{2,z}^r=0.
\end{align}
Furthermore, the $x-$components of all tangential degrees of freedom (see also Section \ref{sec:beamformulation}) are set to 
zero, i.e.
\begin{align}
\label{line_example2_fixedtangentdofs}
  \Delta \hat{t}_{1,x}^l=\Delta \hat{t}_{2,x}^l=\Delta \hat{t}_{1,x}^r=\Delta \hat{t}_{2,x}^r=0,
\end{align}
whereas the $y-$ and $z-$components of these nodal tangents are not prescribed but part of the numerical solution. As shown in~\ref{anhang:analyticalsolution_twistingoftwobeams}, these boundary 
conditions for the tangential degrees of freedom are sufficient in order to impose the necessary boundary moments at the endpoints. If, finally, the penalty parameter is chosen according to
\begin{align}
\label{line_example2_penalty}
  \varepsilon=-\frac{(1+\epsilon)r}{(r^2+h^2)g_0}\left( EA\epsilon + \frac{EI(1+\epsilon)h^2}{(r^2+h^2)^2} \right),
\end{align}
the resulting analytical solution obeys the analytical representation of \eqref{line_example2_helix}, thus showing a gap of $g_0$ between the two beams that is constant along the beam lengths. As already 
mentioned earlier, the penalty parameter and the resulting gap between the two beams occurring in the analytical solution \eqref{line_example2_helix} can be interpreted as a mechanical 
model for the contact-surface/cross-section flexibility of the considered beams. Furthermore, the derived analytical solution corresponds to a mechanical state consisting of constant axial 
tension~$\epsilon$, constant bending curvature $\kappa=\frac{(1+\epsilon)r}{r^2+h^2}$ and vanishing torsion along both beams. In Figure \ref{fig:line_twobeamstwisting_L2}, the relative $L^2$-error of 
the FE solution for beam 1 is plotted with respect to the analytical solution over the element length for discretizations 
with $4, 8, 16, 32, 64, 128$ and $256$ elements per beam. For all convergence plots in this work, the following definition of the relative $L^2$-error has been applied:
\begin{align}
\label{line_defL2error}
||e||^2_{rel} = \frac{1}{u_{max}} \sqrt{ \frac{1}{l}\int_0^l ||\mb{r}_h-\mb{r}_{ref}||^2 ds}.
\end{align}
Herein, $\mb{r}_h$ denotes the numerical solution of the beam centerline position for a certain discretization. For all examples without analytical
solution, the standard choice for the reference solution $\mb{r}_{ref}$ is a numerical solution using a spatial discretization
that is by a factor of four finer than the finest discretization shown in the corresponding convergence plot. The normalization with the element length $l$
makes the error independent of the length of the considered beam. The second normalization leads to a more convenient relative error measure, which relates the $L^2$-error
to the maximal displacement $u_{max}$ occurring for the investigated load case.\\

In order to investigate the influence of the applied Gauss rule, we compare the cases of a $5$-point and a $2$-point Gauss rule with one integration interval per element in both cases. 
According to Figure \ref{fig:line_twobeamstwisting_L2}, the $5$-point-variant converges towards the analytical solution up to machine precision with the optimal order $\mathcal{O}(h^{4})$ as expected for the 
applied third-order beam elements. Throughout this work, this $5$-point-rule will be the default value if nothing to the contrary is mentioned. Reducing the number of Gauss integration points to 
a value of $2$ leads to slight increase of the $L^2$-error in the range of comparatively rough spatial discretizations. 
However, for finer discretizations the $2$-point curve converges towards the $5$-point curve. When looking at the upper right data point in Figure \ref{fig:line_twobeamstwisting_L2}, one observes 
the remarkable result that a total of $8$ contact evaluation points per beam ($4$ elements per beam with $2$ Gauss points per element) is sufficient in order to end up with a relative error 
that is far below $1\%$.\\

In Section~\ref{sec:point_limitationspoint}, we have derived a lower bound $\alpha_{min}$ for the contact angle, above which a unique bilateral closest point projection exists. 
In the following, we briefly want to verify the corresponding result \eqref{limitationspoint_requirementsecondderiv8} by means of a slightly modified version of the considered twisting example. 
Thereto, we assume that the maximal admissible ratio of cross-section to curvature radius supported by the beam theory is $1\%$, i.e. $\mu_{max}=0.01$. For simplicity, we additionally assume 
that the helix radius given in \eqref{line_example2_helix} equals the beam cross-section radius, i.e. $r=R$ and consequently $g_0=0$. With $\mu_{max}=0.01$, the minimal 
admissible slope for a helix with constant slope similar to \eqref{line_example2_helix} can be calculated as:
\begin{align}
\mu_{max}=\bar{\kappa}R=\frac{R^2}{R^2+h_{min}^2}=0.01 \quad \rightarrow \quad h_{min}^2=99R^2.
\end{align}
Furthermore, after some geometrical considerations, one can calculate for the case $h=h_{min}$ the actual contact angle enclosed by two corresponding tangents, 
which is a constant angle in case of helical beams similar to \eqref{line_example2_helix}: 
\begin{align}
\label{limitationspoint_examplehelix}
\alpha=\arccos \left( \frac{\mb{r}_{1,\varphi}^T(\varphi) \mb{r}_{2,\varphi}(\varphi)}{||\mb{r}_{1,\varphi}(\varphi)||\cdot ||\mb{r}_{2,\varphi}(\varphi)||} \right)=\arccos \left( \frac{h_{min}^2-R^2}{h_{min}^2+R^2} \right)=\arccos \left( 0.98 \right) \approx 11.5^{\circ}.
\end{align}
This is exactly the same result that we would obtain for the lower bound $\alpha_{min}$ by inserting $\mu_{max}=0.01$ into \eqref{limitationspoint_requirementsecondderiv8}. This means that the helix geometry according to
example $2$ represents an extreme case, where all worst-case assumptions made in the derivation \eqref{limitationspoint_requirementsecondderiv8a} become true and where, for a given admissible radius 
ratio~$\mu_{max}=0.01$, a non-unique closest point solution appears exactly at the contact angle $\alpha_{min}$ predicted as lower bound by equation \eqref{limitationspoint_requirementsecondderiv8}. 
On the other hand, this example shows that \eqref{limitationspoint_requirementsecondderiv8} provides the best possible lower bound, since it actually occurs 
in a practical example. Furthermore, it can be concluded that the considered twisting example, leading to a constant gap function along both beams, can of course 
not be modeled by means of the standard point-to-point contact formulation.

%-------------------------------------------------------------------------------
%
\subsection{Example 3: General contact of two beams}
\label{sec:examples_example3}
\begin{figure}[ht]
 \centering
   \subfigure[Example 3: Contact force distribution]
   {
    \includegraphics[height=0.28\textwidth]{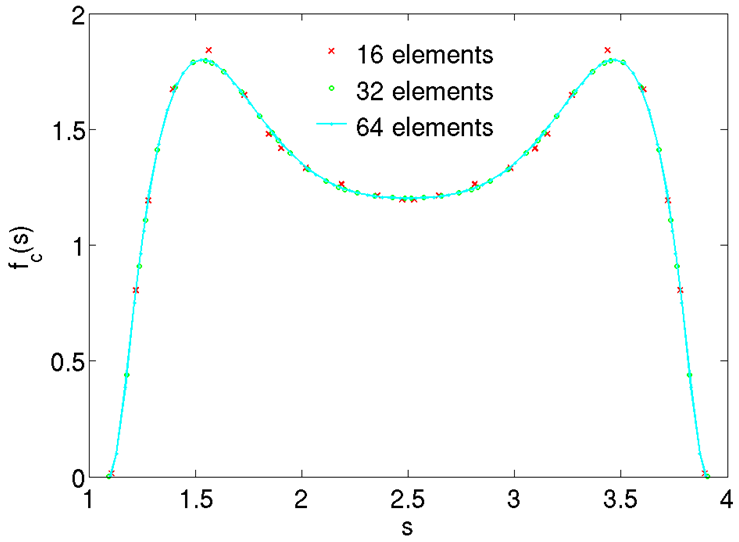}
    \label{fig:line_twobeamstwisting_largerdist_force}
   }   
   \hspace{0.1\textwidth}
   \subfigure[Example 4: Contact force distribution]
   {
    \includegraphics[height=0.28\textwidth]{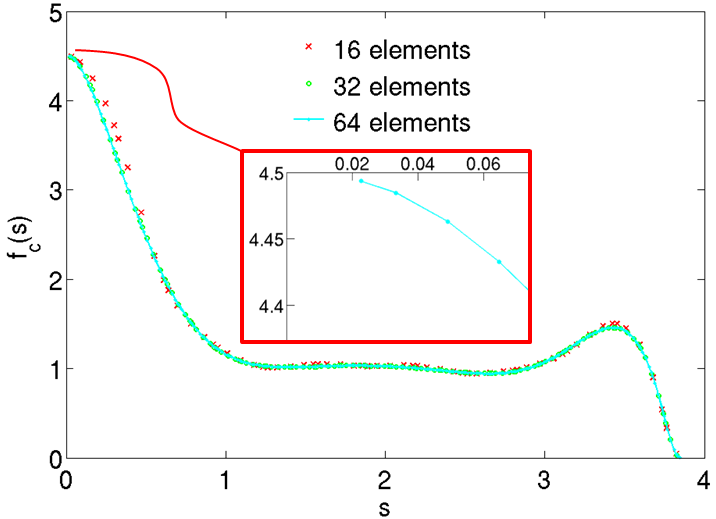}
    \label{fig:line_twobeamstwisting_nonmatchingclose_force}
   }
  \caption{Comparison of Example 3 and Example 4: Contact force distribution for different spatial discretizations}
  \label{fig:line_twobeamstwisting_largerdist_and_nonmatchingclose}
\end{figure}

%
%-------------------------------------------------------------------------------
So far, we have only considered scenarios with a constant gap function along the beam length. By means of the following examples, the more general case of non-constant gaps, 
and especially the case of a change in sign in the gap evolution along the beam, will be investigated. At positions with a change in sign in the gap function, the contact force according to the 
standard law in~\eqref{penlaw_lin} drops to zero. As illustrated in Figure \ref{fig:2straightsmallangleeles_force}, this leads to a kink in the force evolution at this point, which becomes more and more 
pronounced with increasing contact angle (see Figure \ref{fig:2straightlargeangleeles_force}). This weak discontinuity in the integrand may in general increase the numerical integration error and 
can be avoided by replacing the standard linear force law by the smoothed force law in~\eqref{penlaw_linposquad} (see again 
Figures~\ref{fig:2straightsmallangleeles_force} and \ref{fig:2straightlargeangleeles_force}).
\begin{figure}[ht]
 \centering
   \subfigure[Final geometry]
   {
    \includegraphics[height=0.27\textwidth]{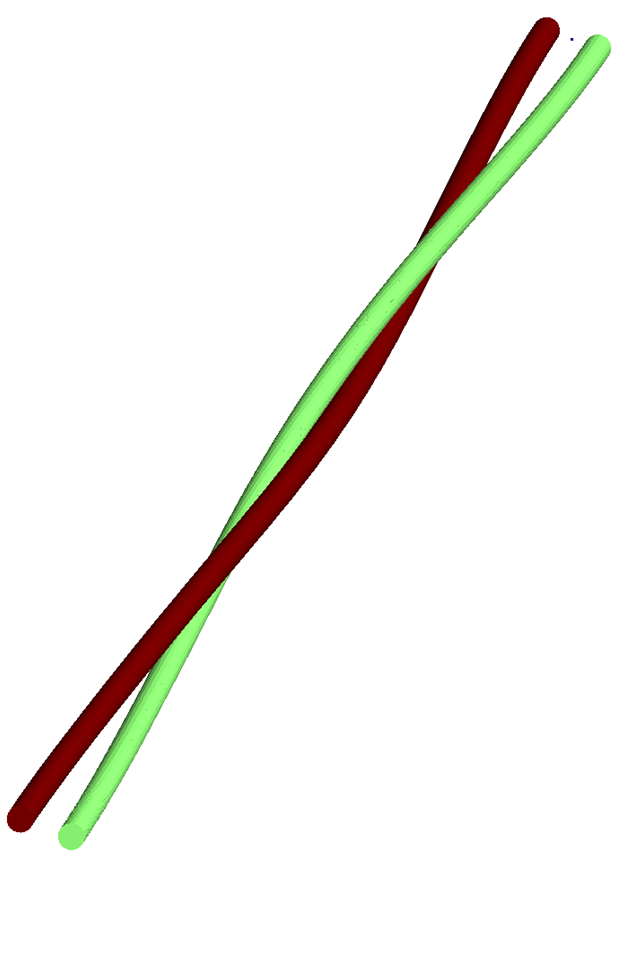}
    \label{fig:line_twobeamstwisting_largerdist_finalconfig}
   }
   \hspace{0.02\textwidth}
   \subfigure[$L^2$-error for linear penalty law]
   {
    \includegraphics[height=0.27\textwidth]{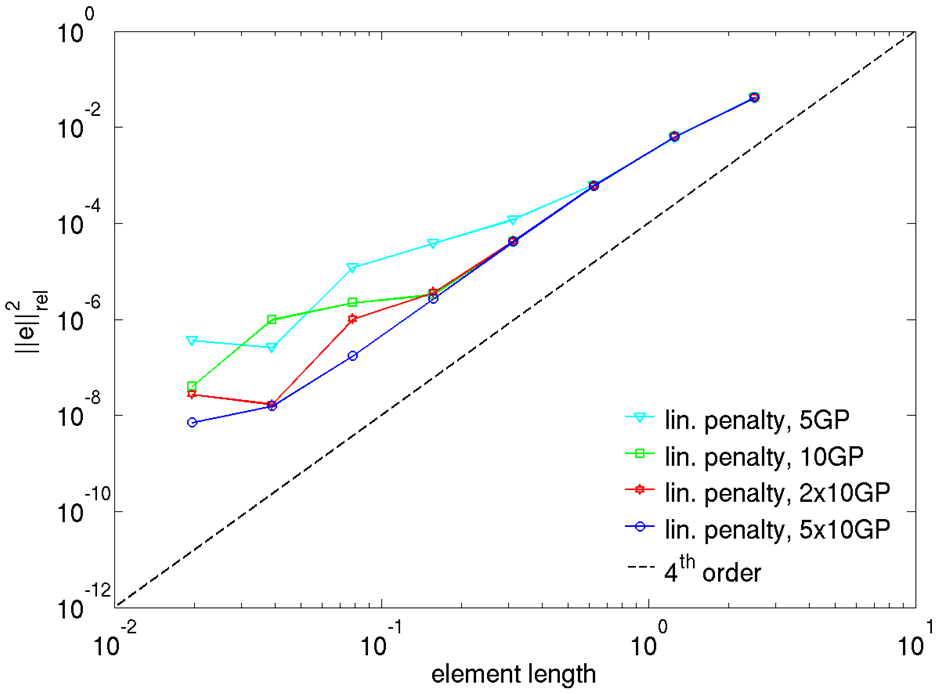}
    \label{fig:line_twobeamstwisting_largerdist_L2_a}
   }
   \subfigure[$L^2$-error for quadr. regularized penalty law]
   {
    \includegraphics[height=0.27\textwidth]{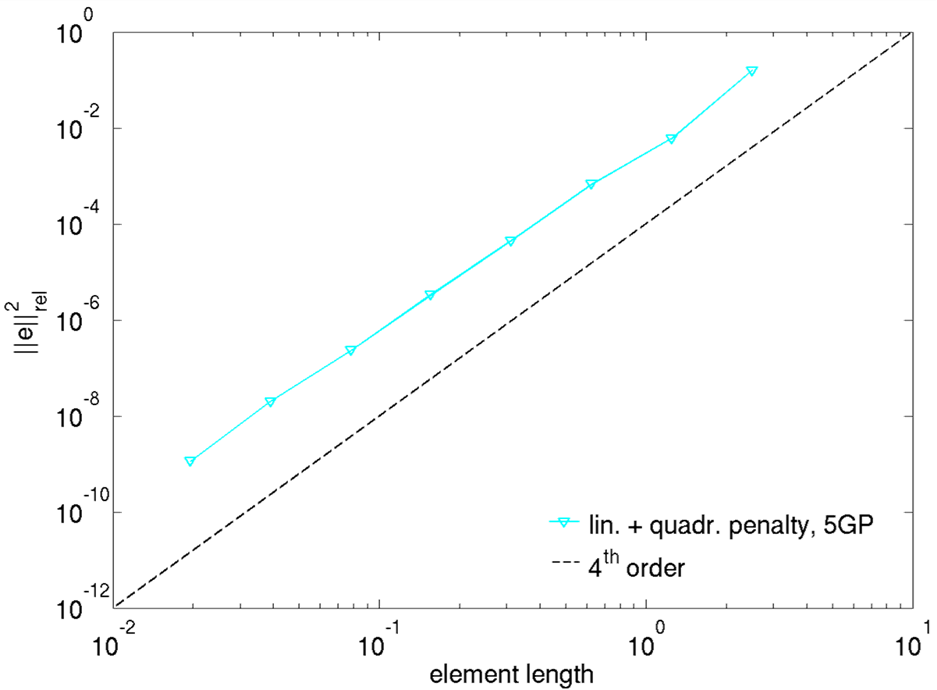}
    \label{fig:line_twobeamstwisting_largerdist_L2_b}
   }
  \caption{Two initially straight and parallel beams with larger initial distance: Dirichlet-controlled twisting}
  \label{fig:line_twobeamstwisting_largerdist}
\end{figure}
The influence of these two different force laws on the integration error and eventually on the spatial convergence behavior will be investigated by means of the following example: We consider beam 
geometries and material parameters identical to the last example. The penalty parameter is decreased to $\varepsilon=1000$. Also, the initial configuration is similar to the one illustrated in 
Figure \ref{fig:line_twobeamstwisting_config} of the last example. However, this time the initial distance between the beams is increased to a value of $2r=4R=0.04$. The Dirichlet boundary conditions 
of the tangential degrees of freedom are slightly changed in order to completely avoid any cross-section rotation at the boundaries. Correspondingly we have:
\begin{align}
\label{line_example3_fixedtangentdofs}
  \Delta \hat{t}_{1,x}^l=\Delta \hat{t}_{2,x}^l=\Delta \hat{t}_{1,x}^r=\Delta \hat{t}_{2,x}^r=\Delta \hat{t}_{1,y}^l=\Delta \hat{t}_{2,y}^l=\Delta \hat{t}_{1,y}^r=\Delta \hat{t}_{2,y}^r=0.
\end{align}
Thus, this time the tangents are completely clamped at both ends. Furthermore, no axial pre-stressing is applied, i.e.
\begin{align}
\label{line_example3_axialdisp}
\Delta \hat{d}_{1,z}^l = \Delta \hat{d}_{2,z}^l = 0.
\end{align}
The remaining Dirichlet conditions are similar to the last example, see~\eqref{line_example2_circularpath} and \eqref{line_example2_rightendpoint}.
The resulting deformed configuration is illustrated in Figure \ref{fig:line_twobeamstwisting_largerdist_finalconfig}.
Due to the larger separation of the beams, the gap function increases from negative values to positive values when approaching the beam endpoints. The corresponding contact force evolutions 
resulting from different spatial discretizations are illustrated in Figure \ref{fig:line_twobeamstwisting_largerdist_force}. In 
Figures~\ref{fig:line_twobeamstwisting_largerdist_L2_a} and \ref{fig:line_twobeamstwisting_largerdist_L2_b}, the relative $L^2$-error with respect to a numerical reference solution 
is plotted for the formulation based on a linear penalty law and the formulation based on the quadratically regularized force law (regularization parameter $\bar{g}=0.1R=0.001$). 
In case of 
the simple linear penalty law, the number of Gauss points 
has to be enhanced by a factor of $10$ as compared to the standard $5$-point rule in order to ensure $\mathcal{O}(h^{4})$ convergence within the considered range of spatial discretizations 
(see Figure~\ref{fig:line_twobeamstwisting_largerdist_L2_a}). Thus, obviously, the increased integration error resulting from the kink in the penalty force law dominates the spatial 
discretization error if the standard $5$-point Gauss rule is applied. Only an increase in the number of Gauss points, and therefore an increase in the numerical effort, reduces this integration error. 
An elimination of this kink by means of a smoothed penalty law enables the same accuracy and the optimal convergence order $\mathcal{O}(h^{4})$ already with the standard $5$-point Gauss rule 
(see Figure~\ref{fig:line_twobeamstwisting_largerdist_L2_b}) and consequently reduces the numerical effort drastically.
%-------------------------------------------------------------------------------
%
\subsection{Example 4: Influence of integration interval segmentation on convergence behavior}
\label{sec:examples_example4}
\begin{figure}[ht]
 \centering
   \subfigure[Final geometry]
   {
    \includegraphics[height=0.28\textwidth]{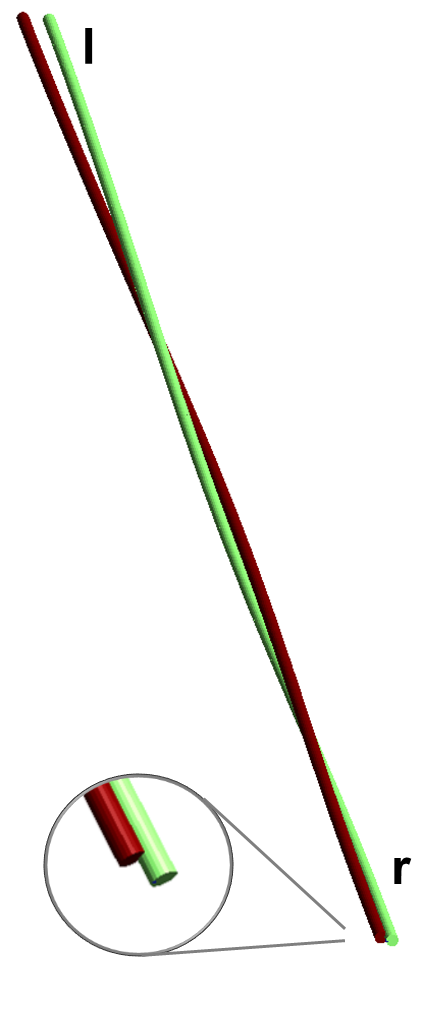}
    \label{fig:line_twobeamstwisting_nonmatchingclose_finalconfig}
    
   }
   \subfigure[Smooth force law with integration segmentation]
   {
    \includegraphics[height=0.29\textwidth]{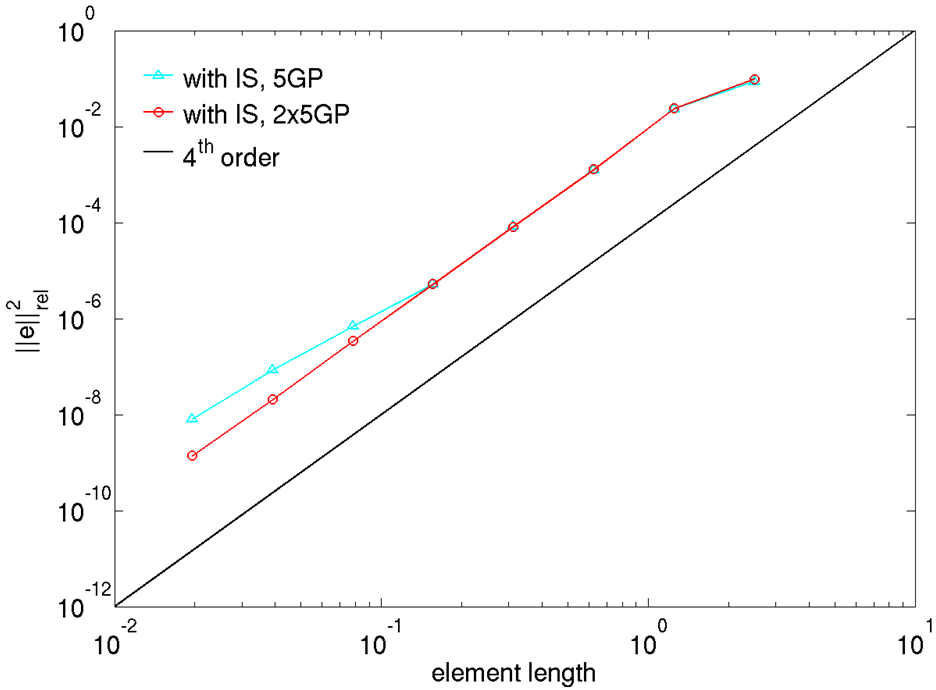}
    \label{fig:line_twobeamstwisting_nonmatchingclose_L2_a}
   }
   \subfigure[Smooth force law without integration segmentation]
   {
    \includegraphics[height=0.29\textwidth]{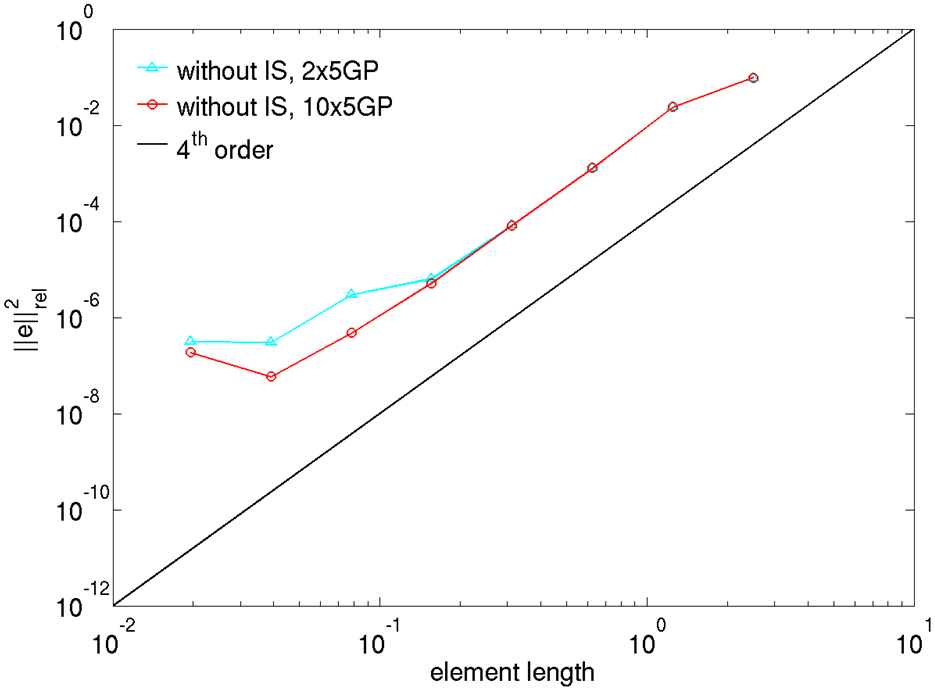}
    \label{fig:line_twobeamstwisting_nonmatchingclose_L2_b}
   }
  \caption{Two initially straight and parallel beams with non-matching endpoints: Dirichlet-controlled twisting}
  \label{fig:line_twobeamstwisting_nonmatchingclose}
\end{figure}

%
%-------------------------------------------------------------------------------
In the first example, we have already illustrated how the integration error can be reduced by means of an integration interval segmentation at the beam endpoints. 
Now, we want to investigate the influence of this method on the spatial convergence behavior. Again, we consider beam geometry, material parameters as well as the penalty parameter to be 
identical to the last example. In order to enforce an integration across the beam endpoints, the initial geometry of one of the beams is shifted by a value of $r=2R=0.02$ along the positive 
$z$-axis leading to the representation:
\begin{align}
\label{line_example4_initialgeometry}
  \mb{r}_{k0}(s)=
  \left(\begin{array}{lll}
         (-1)^{k-1}r \\
         0 \\
         s+(k-1)r
         \end{array}\right),\,\, 
         s \in [0;l], \,\,
         r\!=\!2R, \,\,
         k=1,2.
\end{align}
For this example, we apply the following Dirichlet boundary conditions at the endpoints of the two considered beams:
\begin{align}
\label{line_example4_alldirichs}
\begin{split}
  \Delta \hat{d}_{1,x}^r&=-0.12,\,\, \Delta \hat{d}_{2,x}^r=0.12,\\
  \Delta \hat{d}_{1,y}^r&=\Delta \hat{d}_{2,y}^r=\Delta \hat{d}_{1,z}^r=\Delta \hat{d}_{2,z}^r=0,\\
  \Delta \hat{d}_{1,x}^l \!&=\!-r\left[1 \!-\! \cos \left(\dfrac{k 2\pi}{n_l}\right) \right], \,\, \Delta \hat{d}_{1,y}^l\!=\!r \sin\left(\dfrac{k 2\pi}{n_l}\right), \,\,
  \Delta \hat{d}_{2,x}^l\!=\!r\left[1 \!-\! \cos\left(\dfrac{k 2\pi}{n_l}\right) \right], \,\, \Delta \hat{d}_{2,y}^l\!=\!-r \sin\left(\dfrac{k 2\pi}{n_l}\right), \,\, k=1,...,n_l, \\
  \Delta \hat{d}_{1,z}^l&=\Delta \hat{d}_{2,z}^l=0, \\
  \Delta \hat{t}_{1,x}^l&=\Delta \hat{t}_{2,x}^l=\Delta \hat{t}_{1,x}^r=\Delta \hat{t}_{2,x}^r=\Delta \hat{t}_{1,y}^l=\Delta \hat{t}_{2,y}^l=\Delta \hat{t}_{1,y}^r=\Delta \hat{t}_{2,y}^r=0.
\end{split}
\end{align}
The two additional conditions in the first line of \eqref{line_example4_alldirichs} enforce a negative gap and consequently active contact forces at the (non-matching) right endpoints of the beams. 
By this means, we enforce an integration across a contact force jump at these endpoints which is sensible in order to investigate the effectiveness of the integration interval segmentation. 
All the remaining Dirichlet conditions appearing in \eqref{line_example4_alldirichs} are similar to Section~\ref{sec:examples_example3}. The deformed geometry resulting from these 
boundary conditions is illustrated 
in Figure \ref{fig:line_twobeamstwisting_nonmatchingclose_finalconfig}. Furthermore, the contact force evolutions corresponding to different finite element meshes are presented in 
Figure~\ref{fig:line_twobeamstwisting_nonmatchingclose_force}. The contact force evolution shows the expected jump from $f_c(s\!=\!0.02^-)\!=\!0$ to $f_c(s\!=\!0.02^+)\! \approx \!4.5$ at 
position~$s\!=\!0.02$ (see also the detail view in Figure~\ref{fig:line_twobeamstwisting_nonmatchingclose_force}).
In Figures~\ref{fig:line_twobeamstwisting_nonmatchingclose_L2_a} and \ref{fig:line_twobeamstwisting_nonmatchingclose_L2_b}, the relative $L^2$-error with respect to a numerical reference solution 
is plotted for the formulation based on the quadratically regularized force law (regularization parameter $\bar{g}=0.1R=0.001$), once with integration interval segmentation 
(Figure~\ref{fig:line_twobeamstwisting_nonmatchingclose_L2_a}) and once without a corresponding segmentation (Figure~\ref{fig:line_twobeamstwisting_nonmatchingclose_L2_b}).\\

According to Figure \ref{fig:line_twobeamstwisting_nonmatchingclose_L2_a}, the remaining integration error of the formulation with interval segmentation and a $5$-point Gauss rule slightly 
deteriorates the spatial convergence behavior. However, by applying two instead of one $5$-point Gauss integration intervals per element, this influence of the integration error vanishes 
and we observe the optimal convergence order $\mathcal{O}(h^{4})$. On the contrary, the convergence behavior of the formulation without integration interval segmentation 
(see Figure \ref{fig:line_twobeamstwisting_nonmatchingclose_L2_b}) is still deteriorated by the integration error for a $5$-point Gauss rule even with two intervals per element. 
Even if the number of intervals is increased to $10$, i.e. an increase of the number of Gauss points by a factor of $5$, this negative influence is still visible in the range of fine discretizations. 
Furthermore, it is worth mentioning 
that this effect is expected to become even more pronounced in practical applications, where the displacements are not Dirichlet-controlled in the direct neighborhood of the 
strong discontinuity. All in all, it seems that the integration interval segmentation solely applied at the beam endpoints represents a sensible compromise of integration accuracy and 
computational efficiency. Additionally, in dynamic simulations, this strategy prevents from force and energy jumps in scenarios where active Gauss points of standard integration schemes based 
on fixed, non-segmented integration intervals would slide across master beam endpoints.

%-------------------------------------------------------------------------------
%
\subsection{Example 5: Simulation of a biopolymer network}
\label{sec:examples_example5}
%
%-------------------------------------------------------------------------------

In a first practically relevant example, we apply the presented simulation framework in order to investigate the influence of mechanical contact interaction on the three-dimensional Brownian 
motion of filaments in 
biopolymer networks. Biopolymer networks are tight meshes of highly slender polymer filaments (e.g. Actin filaments) embedded in a liquid phase, often interconnected by means of a second molecule species 
(so-called cross-linkers). These networks can for example be found in biological cells. There, they crucially determine the mechanical properties of cells and biologically highly relevant processes such as 
cell-migration or cell-division.
\begin{figure}[ht!]
 \centering
   \subfigure[Undeformed initial configuration]
   {
    \includegraphics[width=0.31\textwidth]{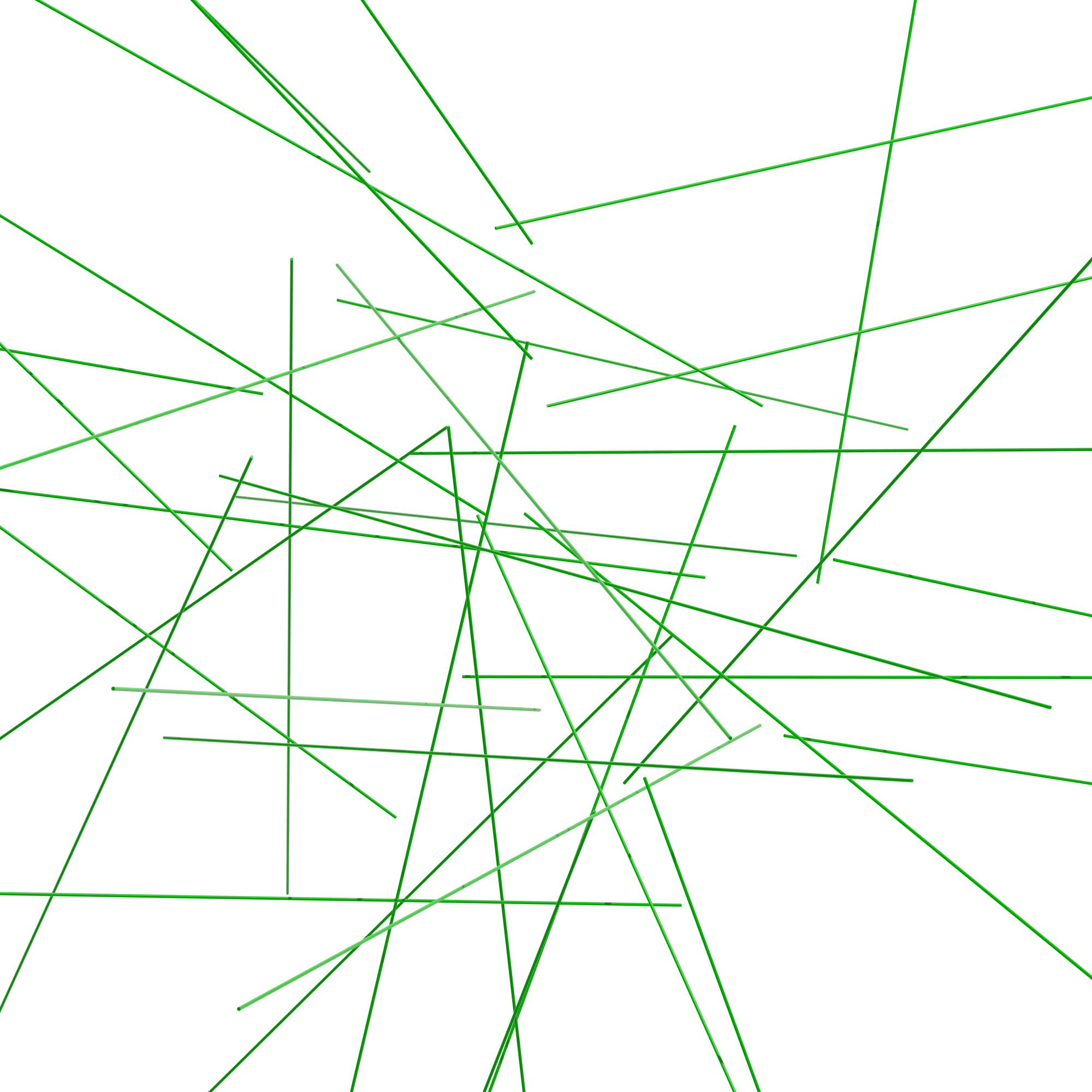}
    \label{fig:statmech_37filaments_step000}
   }
   \subfigure[Deformed configuration at step 500]
   {
    \includegraphics[width=0.31\textwidth]{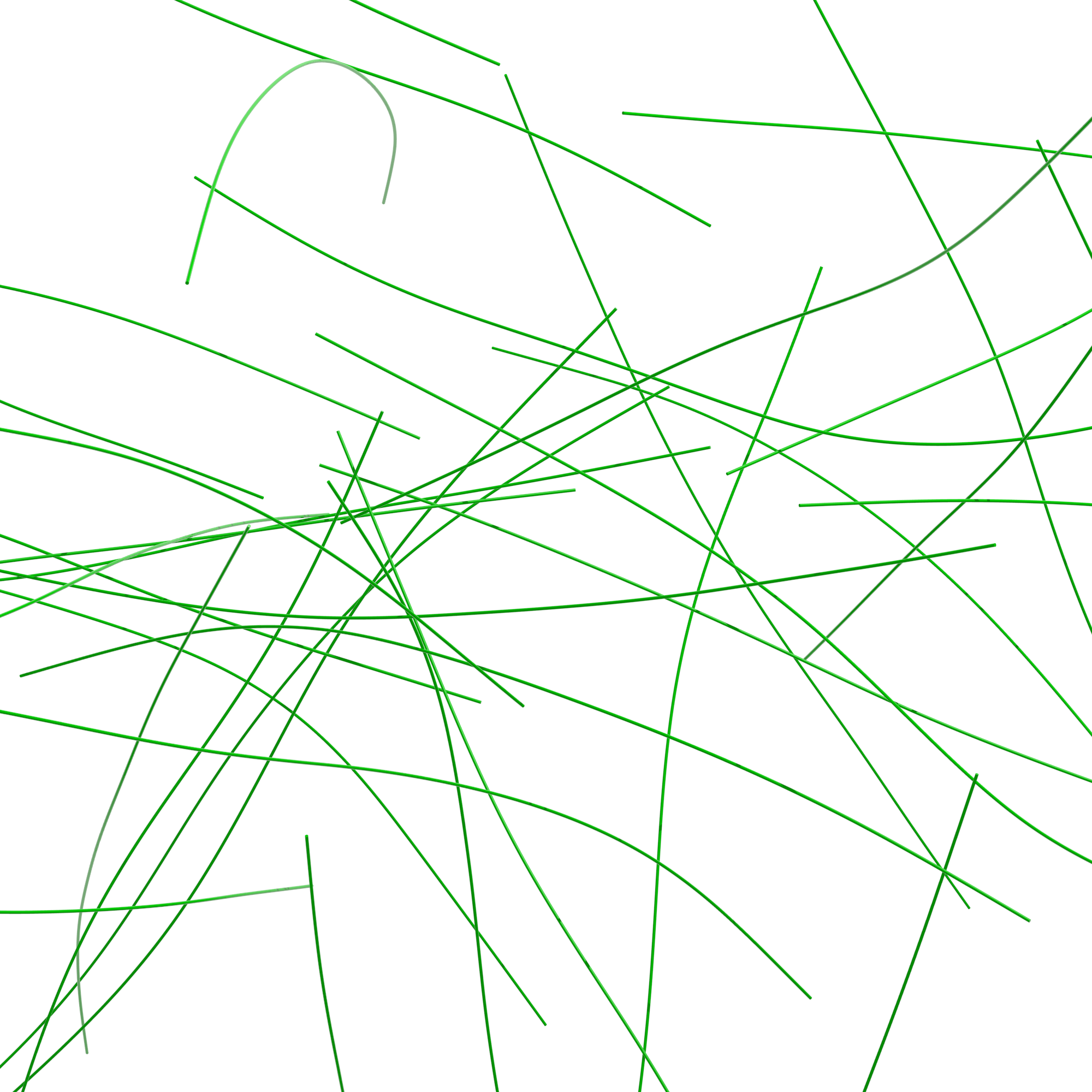}
    \label{fig:statmech_37filaments_step500}
   }
   \subfigure[Deformed configuration: zoom-factor 3]
   {
    \includegraphics[width=0.31\textwidth]{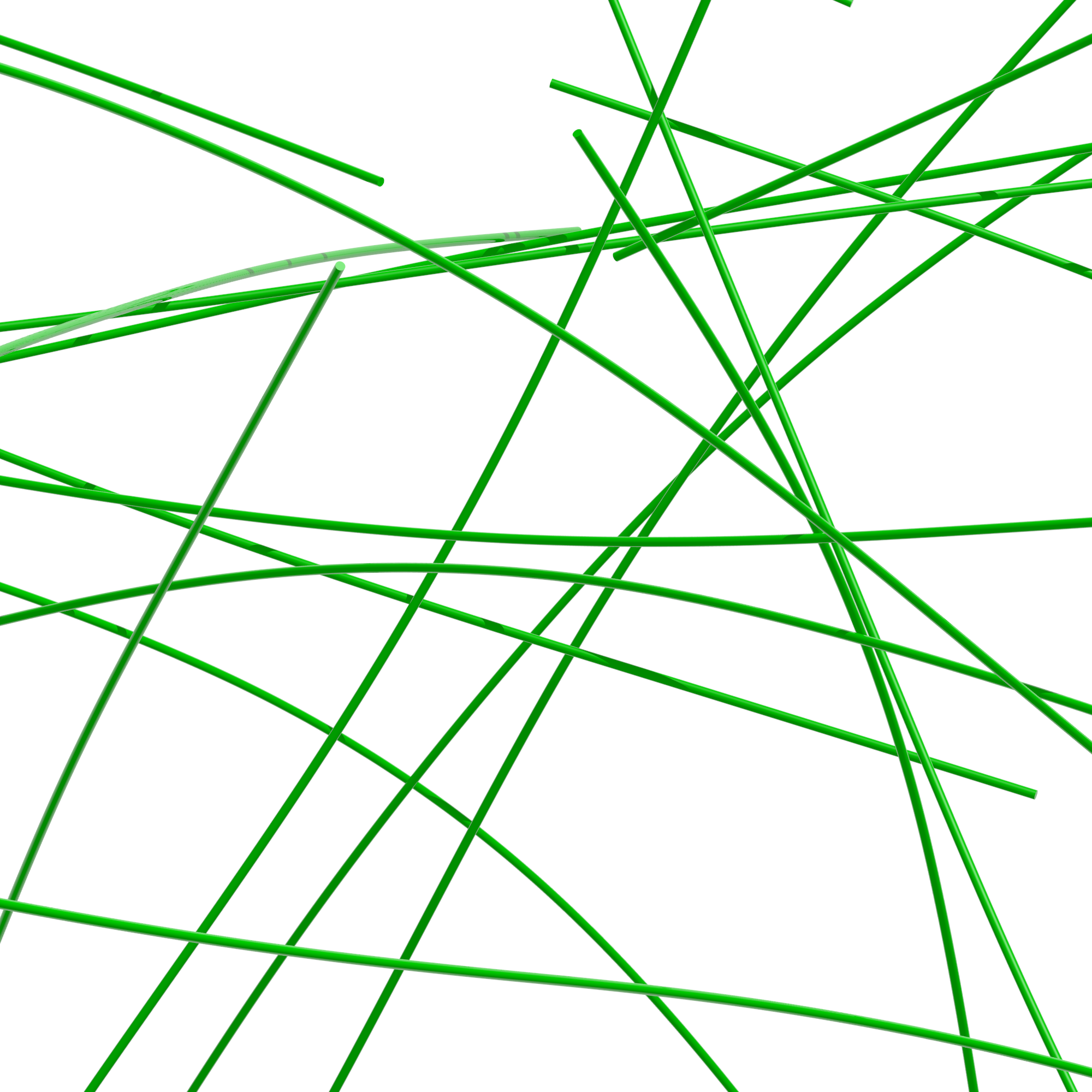}
    \label{fig:statmech_37filaments_step500_zoom}
   }
  \caption{Brownian dynamics simulation of the free diffusion of Actin filaments: Deformed configurations at different time steps}
  \label{fig:statmech_37filaments_configs}
\end{figure}
\begin{figure}[ht]
 \centering
   \subfigure[Active line contacts]
   {
    \includegraphics[height=0.29\textwidth]{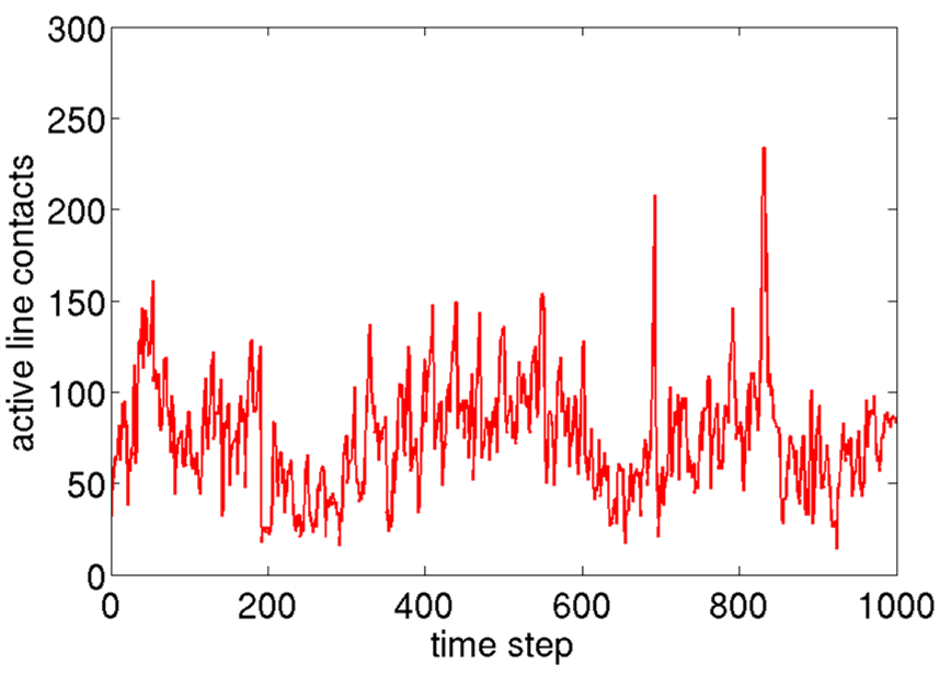}
    \label{fig:statmech_37filaments_pureline_activelines}
   }
   \subfigure[Active endpoint contacts]
   {
    \includegraphics[height=0.29\textwidth]{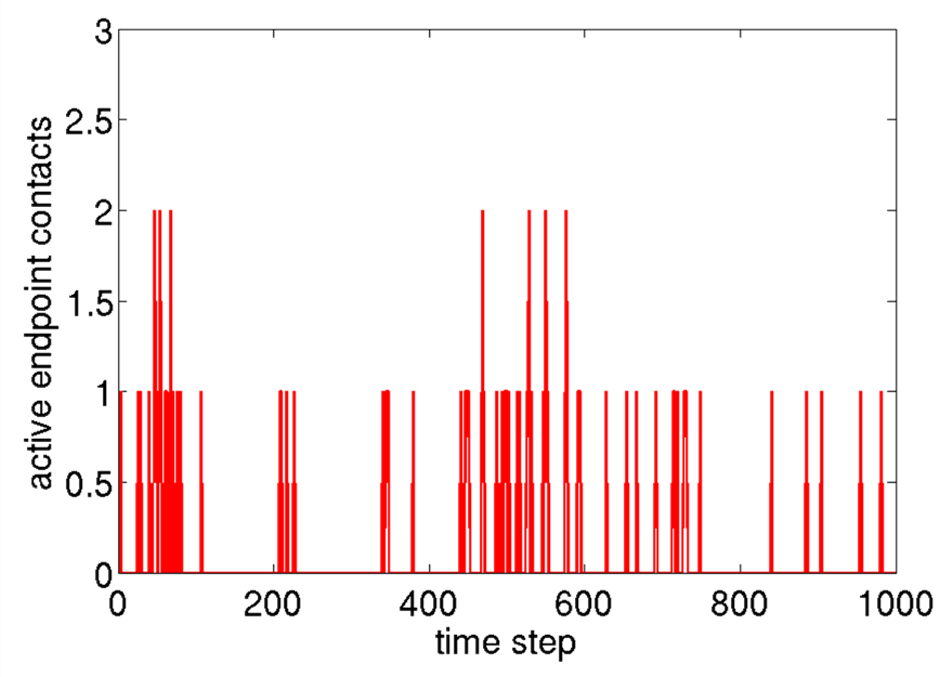}
    \label{fig:statmech_37filaments_pureline_activeendpoints}
   }
   \subfigure[Modeling error]
   {
    \includegraphics[height=0.29\textwidth]{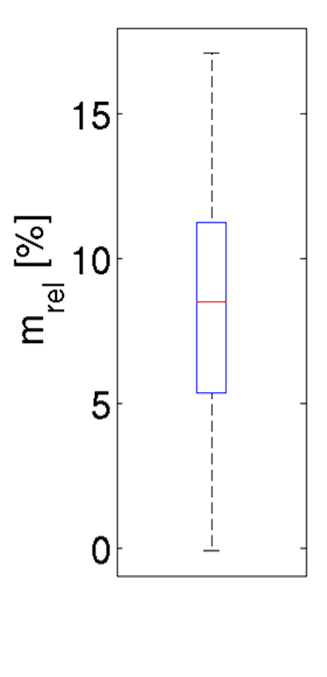}
    \label{fig:statmech_37filaments_boxplot}
   }
  \caption{Brownian dynamics simulation of the free diffusion of Actin filaments}
  \label{fig:statmech_37filaments_pureline}
\end{figure}
In a recent contribution \cite{cyron2012}, a finite element model for the Brownian motion of these filaments in the absence of mechanical 
contact interaction has been proposed. Accordingly, these slender filaments are described by means of a geometrically nonlinear beam theory. The mechanical interaction of the filaments with 
the surrounding fluid is modeled by means of external distributed line loads consisting of velocity-proportional 
viscous drag forces and thermal excitation forces. The latter are modeled as stochastic forces characterized by a mean value of zero, a variance determined by the absolute temperature and the properties of the 
surrounding fluid and finally by spatial and temporal correlation lengths which are assumed to be zero (see \cite{cyron2012} for details). Due to the physical length scales relevant for such systems, inertia forces can 
often be neglected, thus leading to a system of first-order stochastic partial differential equations (SPDEs). 
While spatial discretization is performed via the finite element method in combination with a geometrically exact beam formulation (Section~\ref{sec:beamformulation}), 
a Backward Euler scheme is applied in order to discretize the resulting semi-discrete problem in time.
Here, we combine the Brownian dynamics simulation framework presented in \cite{cyron2012} with the contact algorithm proposed in the sections before in order to simulate 
the free diffusion of Actin filaments (without consideration of cross-linker molecules). Thereto, we consider a system of $37$ initially straight and randomly distributed Actin filaments with circular 
cross-section of radius $R=2.45 \cdot 10^{-3}$, length $l=2$ and Youngs modulus $E=1.3 \cdot 10^{9}$ (all quantities given in the units $mg$, $\mu m$, and $s$) as illustrated in 
Figure~\ref{fig:statmech_37filaments_step000}. All further physical system parameters describing the viscous and stochastic forces are identical to those applied in \cite{mueller2014} 
and can be found therein.\\

The simulation was performed based on a spatial discretization with $8$ beam elements per filament, a time step size of $\Delta t=1.0 \cdot 10^{-4}$ and a total simulation time 
of $t_{end}=1.0 \cdot 10^{-1}$. Furthermore, the contact parameters have been chosen as $\varepsilon=5.0 \cdot 10^4$ and $\bar{g}=2.0 \cdot 10^{-3}$ in combination with $50$ integration intervals 
per slave element based on a $5$-point Gauss rule, respectively. The spatial configurations at times $t=0.0$ and $t_{end}=0.1$ as well as a corresponding detail view at $t_{end}=0.1$ are illustrated in 
Figure~\ref{fig:statmech_37filaments_configs}. 
Due to the stochastic forces, the velocity field of these filaments is strongly fluctuating in space and in time, thus leading to drastic and frequent changes in the active contact sets. This property in 
combination with the high filament slenderness ratio of approximately $800$ makes this example very demanding concerning the robustness of the 
proposed contact algorithm. The Newton-Raphson convergence tolerances are set to $\delta_{\mb{R}}\!=\!\delta_{\mb{D}}\!=\!10^{-6}$. 
In this example, where dynamic collisions at all possible filament-to-filament orientations can occur, the significance of the endpoint contact contributions introduced in Section~\ref{sec:endpoint} 
becomes apparent. In order to underpin this statement, the corresponding total 
numbers of active line contact Gauss points and active beam endpoint contacts have been plotted over the simulation time in Figures~\ref{fig:statmech_37filaments_pureline_activelines} 
and \ref{fig:statmech_37filaments_pureline_activeendpoints}. Accordingly, even for this comparatively small example, the endpoint contact 
contributions occur with significant frequency. Neglecting these endpoint contact forces would not only allow for nonphysically large penetrations, it would also lead to non-convergence of the 
Newton-Raphson scheme in many time steps. The contact angles measured for this example during the simulation time lie within the range $\alpha \in [4^{\circ};90^{\circ}]$, thus covering almost the entire possible scope. 
Considering the maximal curvature $\bar{\kappa}_{max}\approx 2.0$ measured during 
the simulation time and the beam cross-section radius $R=2.45 \cdot 10^{-3}$, the lower bound for the contact angle that would allow for a point-to-point contact formulation can 
be calculated as $\alpha_{min} \approx 8^{\circ}$. This means that even this example, which is dominated by rather large contact angle configurations, cannot be completely covered by a standard 
point-to-point contact formulation.
Finally, we conclude this section by an exemplary statistical analysis of a physically relevant quantity. Concretely, the influence of the mechanical contact interaction 
on the filament diffusion measured by the mean square displacement per time step is evaluated. This quantity is defined as:
\begin{align}
\label{meansquare}
  \langle \Delta r^P \rangle := \frac{1}{n_{step}} \sum \limits_{i=1}^{n_{step}} \Delta r_i^P \quad \text{with} \quad \Delta r_i^P=||\Delta \mb{r}_i^P||.
\end{align}
In \eqref{meansquare}, $n_{step}=10^{3}$ denotes the number of time steps of the simulation and $\Delta \mb{r}_i^P$ the displacement increment of a material filament point $P$, here 
chosen as the midpoint of a filament located close to the center of the considered network in the initial configuration, at time step $i$. In order to enable a statistical 
analysis, we have performed $100$ realizations of the underlying Gaussian process by generating $100$ different sets of random numbers representing the space-time distribution 
of the external stochastic line loads. Having determined the mean square displacement $\langle \Delta r^P \rangle_{c}^l$ of the case where contact is considered and 
$\langle \Delta r^P \rangle_{nc}^l$ of the case where contact is neglected, where the superscript $l=1,...,100$ represents the stochastic realization, we can 
define the modeling error $m^l_{rel}$ of realization $l$:
\begin{align}
\label{meansquare2}
  m^l_{rel}:=\frac{\langle \Delta r^P \rangle_{nc}^l-\langle \Delta r^P \rangle_{c}^l}{\langle \Delta r^P \rangle_{c}^l}.
\end{align}
Statistical evaluation of the measured modeling errors finally yield a mean value of $8.5\%$ and a variance of $2.5\%$ (see Figure~\ref{fig:statmech_37filaments_boxplot}). 
In other words, for the considered example, 
the mean square displacement per time step is overestimated by $8.5\%$ in average when neglecting mechanical contact interaction. Of course, this analysis only has an exemplary 
character, since system parameters such as fluid and filament properties, magnitude of stochastic forces, considered simulation time and/or type of chosen (periodic) boundary 
conditions (not considered here) might drastically change the influence of mechanical contact interaction on the filament diffusion behavior. Nevertheless, this result represents 
a valuable first indication that mechanical contact may decrease diffusivity noticeably. Beyond this example, there are 
many questions of interest in this field of application, e.g. the influence of mechanical contact interaction on the development of thermodynamically stable or unstable phases 
in cross-linked biopolymer networks (see e.g. \cite{mueller2014}), where a robust contact simulation framework such as the one proposed in this contribution is of 
essential importance.

%-------------------------------------------------------------------------------
%
\subsection{Example 6: Simulation of the static twisting process of a rope}
\label{sec:examples_example6}
\begin{figure}[h]
 \centering
   \subfigure[Undeformed initial configuration]
   {
    \includegraphics[width=0.31\textwidth]{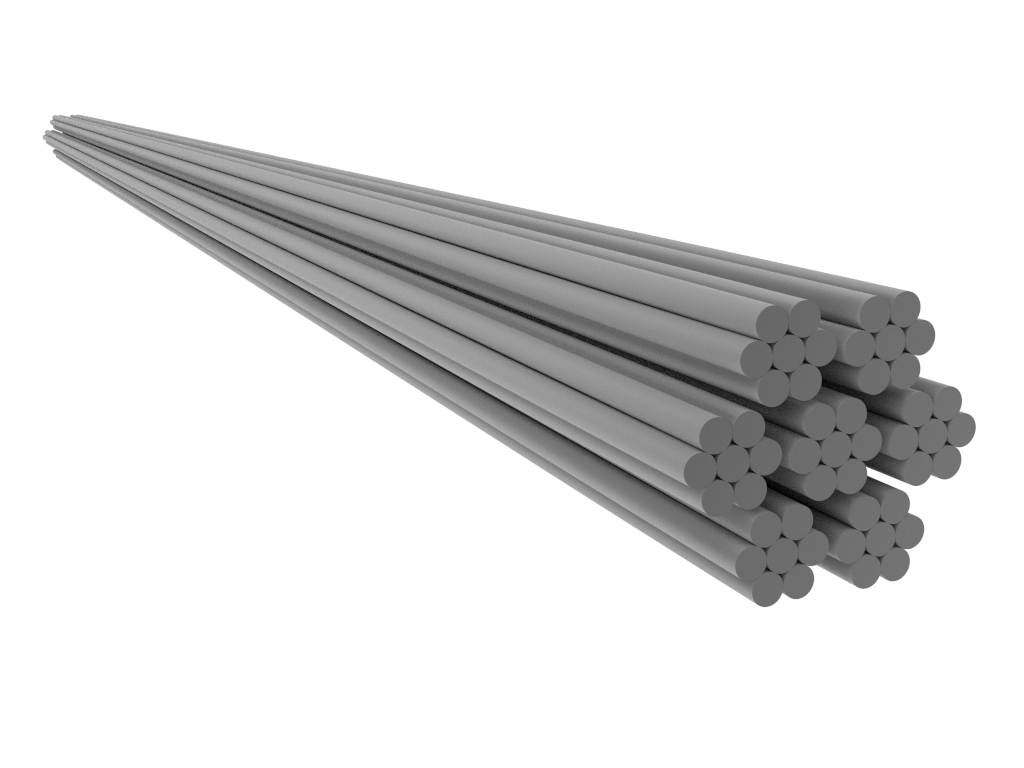}
    \label{fig:manybeams_twisting_step000}
   }
      \subfigure[Deformed configuration at load step 20]
   {
    \includegraphics[width=0.31\textwidth]{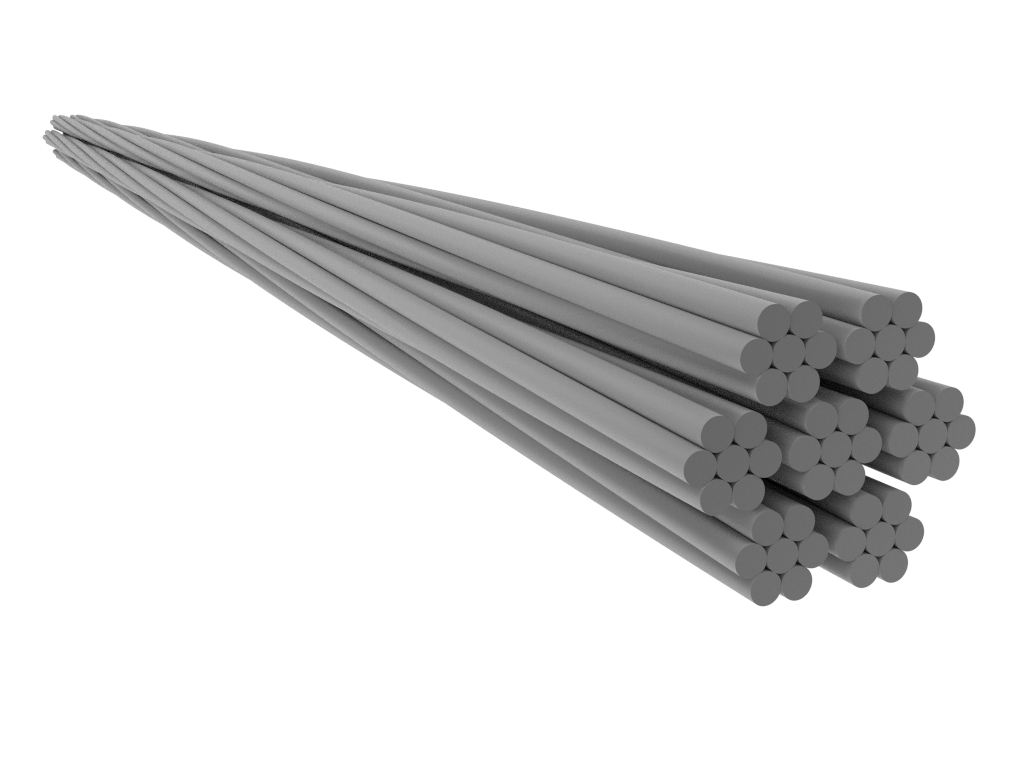}
    \label{fig:manybeams_twisting_step020}
   }
      \subfigure[Deformed configuration at load step 40]
   {
    \includegraphics[width=0.31\textwidth]{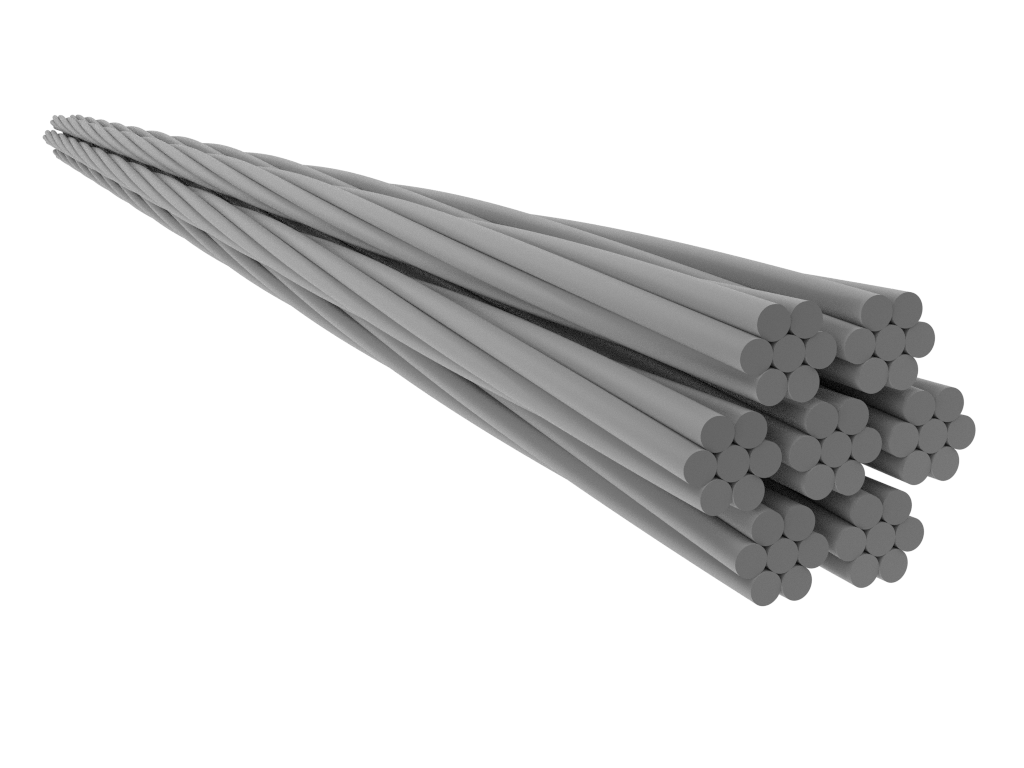}
    \label{fig:manybeams_twisting_step030}
   }
      \subfigure[Deformed configuration at load step 60]
   {
    \includegraphics[width=0.31\textwidth]{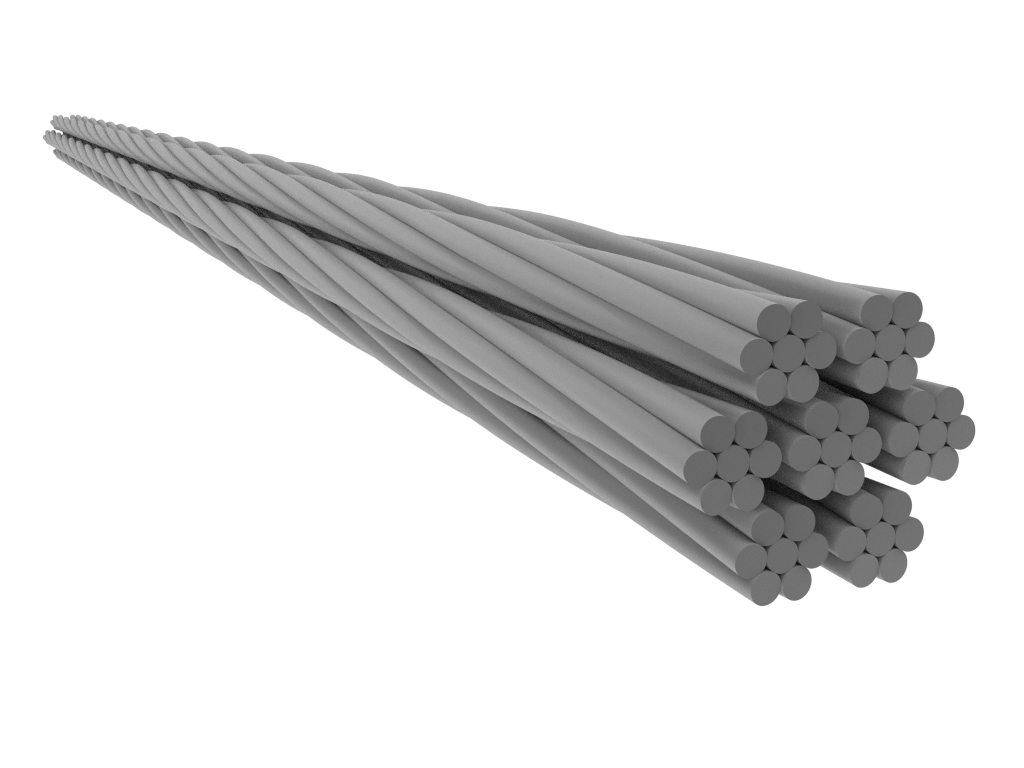}
    \label{fig:manybeams_twisting_step060}
   }
      \subfigure[Deformed configuration at load step 80]
   {
    \includegraphics[width=0.31\textwidth]{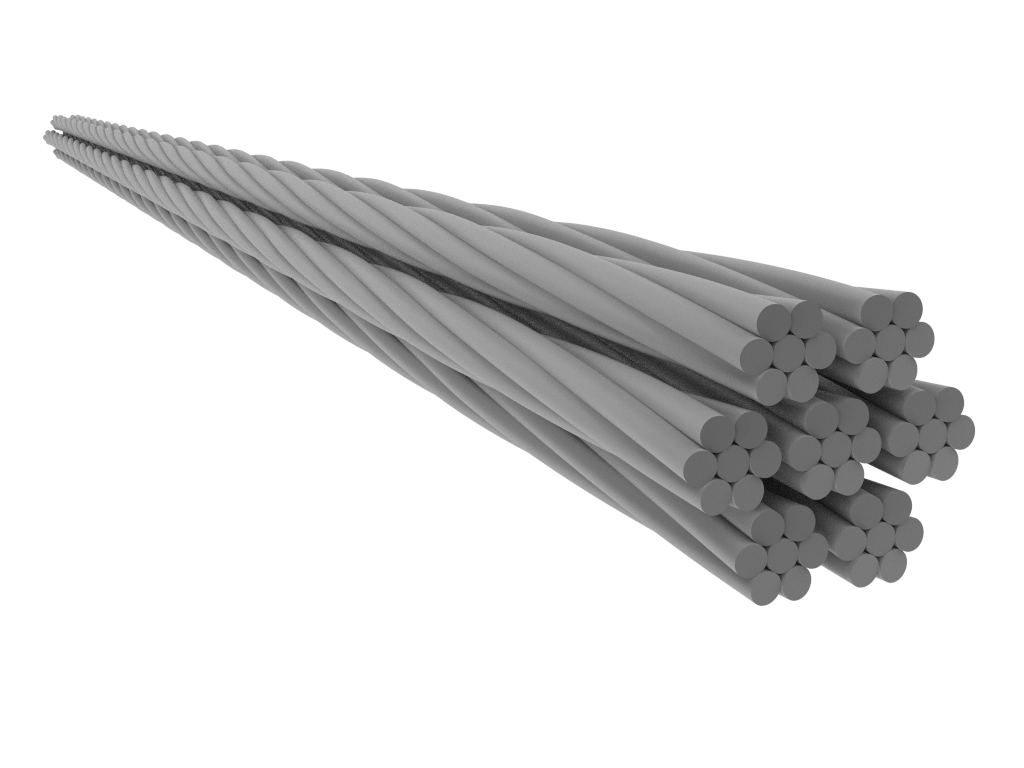}
    \label{fig:manybeams_twisting_step080}
   }
      \subfigure[Deformed configuration at load step 100]
   {
    \includegraphics[width=0.31\textwidth]{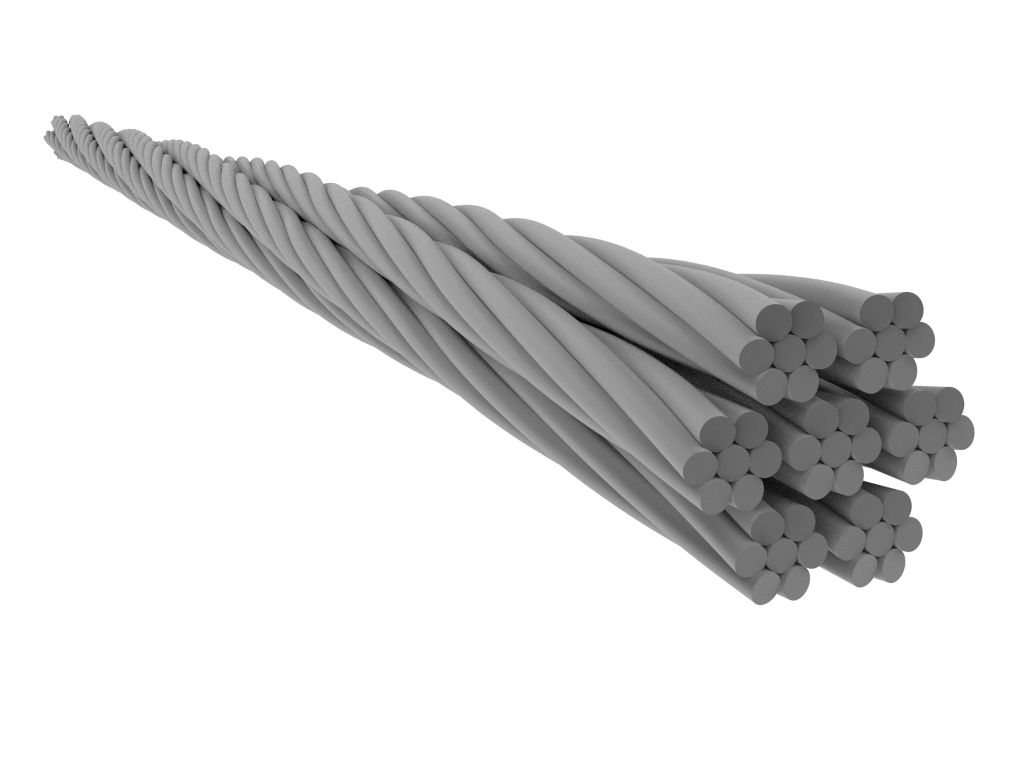}
    \label{fig:manybeams_twisting_step100}
   }
  \caption{Static simulation of the twisting process of a rope consisting of $7\times 7$ fibers: Deformed configurations at different load steps}
  \label{fig:manybeams_twisting}
\end{figure}

%
%-------------------------------------------------------------------------------
In this last example, the static twisting process of a rope will be investigated. 
The considered rope is built from $7 \times 7$ individual fibers with length $l=5$, circular cross-section of radius $R=0.01$ and Young´s modulus $E=10^9$.
The arrangement of the initially straight fibers in seven sub-bundles with seven fibers per sub-bundle is illustrated in Figure~\ref{fig:manybeams_twisting_step000}.
For spatial discretization, we use $10$ beam elements per fiber. The contact parameters have been chosen as 
$\varepsilon=5.0 \cdot 10^5, \bar{g}=0.1R=0.001$ in combination with seven $5$-point integration intervals per element.
In the first stage of the twisting process, each of the seven sub-bundles is twisted by four full rotations within $80$ static load steps.
The twisting process is performed in a Dirichlet-controlled manner, such that the cross-section center points at one end of the sub-bundles (front side in Figure~\ref{fig:manybeams_twisting}) 
are moving on a circular path (see also Example 2 of Section~\ref{sec:examples_example2}) with respect to the individual sub-bundle center points, while the corresponding points at the other end of the 
sub-bundles (back side in Figure~\ref{fig:manybeams_twisting}) remain fixed. The deformed configurations at characteristic load steps after one, two, three and four full rotations are illustrated 
in Figures~\ref{fig:manybeams_twisting_step020}-\ref{fig:manybeams_twisting_step080}. In the second stage of the twisting process, all seven sub-bundles together are twisted by one 
further rotation within $20$ additional static load steps. This time, the cross-section center points are moving on a circular path with respect to the center point of the entire $7 \times 7$-rope. 
The deformed configuration at the end of this twisting process is illustrated in Figure~\ref{fig:manybeams_twisting_step100}. While the cross-section center points of all fiber endpoints at one 
end of the rope (front side in Figure~\ref{fig:manybeams_twisting}) are fixed in axial direction, the cross-section center points of all fiber endpoints at the other end of the rope 
(back side in Figure~\ref{fig:manybeams_twisting}) are free to move in axial direction.\\

\begin{figure}[h]
 \centering
   \subfigure[Axial reaction torque during static twisting process]
   {
    \includegraphics[height=0.3\textwidth]{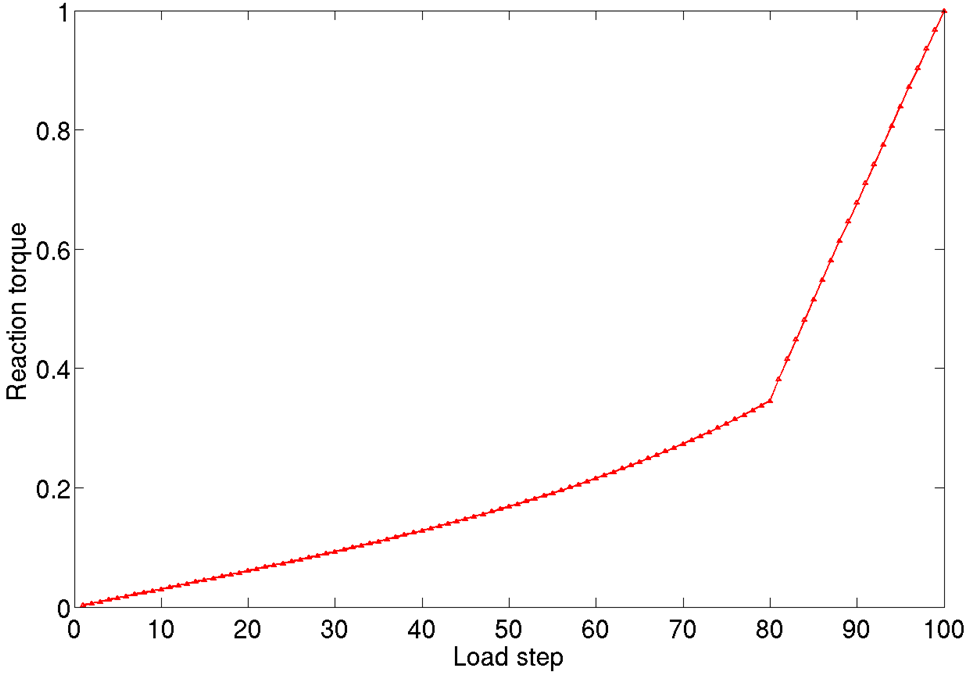}
    \label{fig:rope_static_torque}
   }
   \hspace{0.02\textwidth}
   \subfigure[Minimal contact angle $\min{(\alpha)}$ and lower bound $\alpha_{min}$]
   {
    \includegraphics[height=0.3\textwidth]{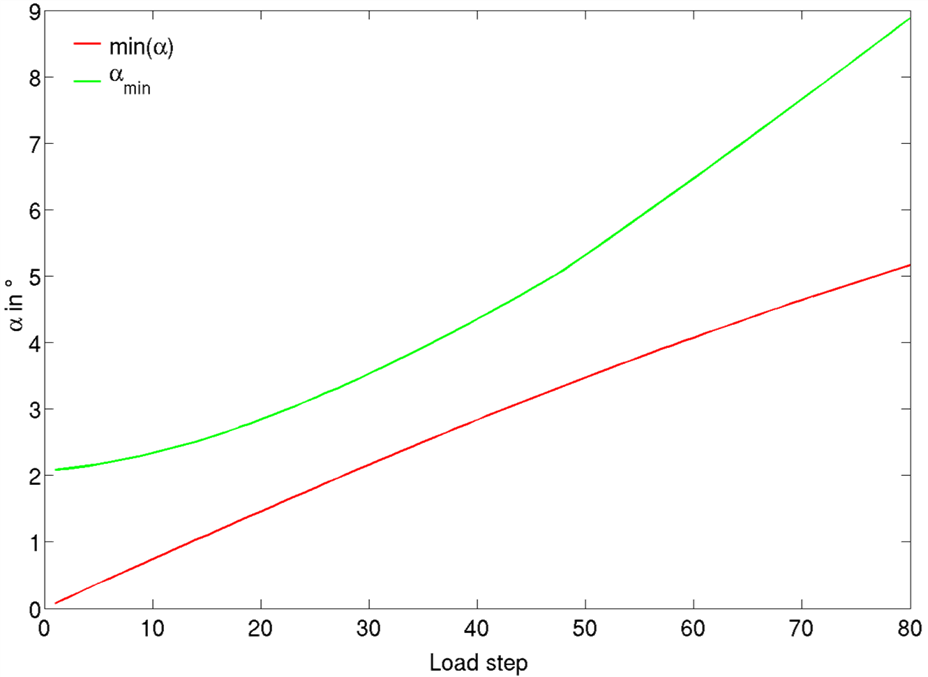}
    \label{fig:rope_static_angles}
   }
  \caption{Mechanically relevant quantities in the numerical simulation of a rope: Reaction torque and contact angle}
  \label{fig:rope_mechanical_quantities}
\end{figure}

Additionally, a constant axial tensile force $\bar{f}_{ax}=1000$ acting on each of these axially freely 
movable fiber endpoints provides axial pre-stressing during the entire twisting process. In contrary to Section~\ref{sec:examples_example2}, the fiber endpoints are simply supported but not 
clamped. Consequently, Dirichlet conditions are only applied to the positional degrees of freedom $\mb{\hat{d}}^i$ at the endpoints but not to the tangential degrees of freedom $\mb{\hat{t}}^i$.
As already mentioned in Section~\ref{sec:examples_example2}, each individual fiber is free of mechanical torsion at the end of this twisting process, since only external 
(contact and reaction) forces, but no external torsional moments are acting on the fibers. Nevertheless, of course, an overall external axial torque resulting from the moment contributions of 
the reaction forces at the beam endpoints with respect to the centerline of the rope is necessary in order to guarantee for static equilibrium of the twisted rope at different load steps. 
The corresponding evolution of this external axial torque during the deformation process normalized by the maximal torque occurring at load step $100$ is 
plotted in Figure~\ref{fig:rope_static_torque}. Interestingly, the evolution of the twisting torque over the twisting angle is almost linear within the two stages of deformation, i.e. 
the behavior of the rope is similar to the twisting response of a slender continuum. The higher slope in the second twisting stage, where all sub-bundles are twisted with respect to the centerline 
of the rope, results from the increased overall elastic resistance. The external work which is required in order to perform the considered twisting process in a quasi-static manner is proportional 
to the area enclosed by the graph of the twisting torque evolution and the horizontal axis of Figure~\ref{fig:rope_static_torque}.
From a purely mechanical point of view, it is quite obvious that the presented example, which is dominated by line-to-line contact interaction along the entire length of the rope, should 
better be modeled by a line-to-line than a point-to-point type contact formulation. However, we also want to motivate this choice from a mathematical point of view.
Thereto, the minimal contact angle $\min{(\alpha)}$ occurring within the entire rope at a specific load step has been plotted over the first 80 load steps in Figure~\ref{fig:rope_static_angles}. 
As expected, the value of this 
minimal angle increases with increasing twisting angle/load step. In order to mathematically evaluate the possibility of applying the point-to-point contact formulation to this example, 
we will compare this minimal contact angle with the lower bound $\alpha_{min}$ according to~\eqref{limitationspoint_requirementsecondderiv8}, above which a unique bilateral closest point 
solution can be guaranteed. To be able to do so, we have additionally plotted the evolution $\alpha_{min}=\arccos{(1-2R\bar{\kappa}_{max})}$ over the first 80 loads steps. Here, $R=0.01$ represents the 
cross-section radius and $\bar{\kappa}_{max}$ is the maximal curvature value occurring in the entire rope for the considered load step. As can be seen from Figure~\ref{fig:rope_static_angles}, 
the curve representing the actual minimal angle $\min{(\alpha)}$ lies entirely below the curve representing the minimal admissible angle $\alpha_{min}$. Thus, a unique bilateral closest point solution 
cannot be guaranteed and hence, as expected, the point-to-point contact formulations is not suitable for this example.
%-------------------------------------------------------------------------------

\section{Conclusion}
\label{sec:conclusion}

%-------------------------------------------------------------------------------
Within this contribution, a new finite element formulation describing the contact behavior of slender beams in 
complex 3D contact configurations involving arbitrary beam-to-beam orientations has been proposed. It has been shown by means of a mathematically concise 
investigation of standard point-to-point beam contact models that these formulations fail to describe 
a considerable range of practically relevant contact configurations as consequence of a non-unique bilateral closest point projection.  
In contrary, the proposed line-to-line formulation models contact interaction of slender continua by means of 
distributed line forces. It has been shown analytically that the corresponding unilateral closest point projection relevant for this line contact formulation always possesses 
a unique solution and thus is applicable for any geometrical contact configuration. By means of theoretical and numerical investigations, different contact discretizations, i.e. 
Gauss-point-to-segment or mortar type formulations, as well as different constraint enforcement strategies, based on penalty regularization or Lagrange multipliers, have been evaluated.\\ 

On the basis of these comparisons, it has been concluded that a penalty-based Gauss-point-to-segment 
formulation is most suitable for the considered range of beam-to-beam contact applications. On the one hand, the penalty regularization of the contact constraint, which can be interpreted 
as mechanical model of the cross-section stiffness, has been shown to be indispensable when employing beam models based on the assumption of rigid cross-sections. On the other hand, as 
compared to mortar-type approaches, the proposed Gauss-point-to-segment contact discretization is favorable in terms of computational efficiency and implementation effort. Additionally, 
it has been argued and verified numerically that the required range of penalty parameters does typically not induce contact-related locking phenomena when considering thin beams.
The proposed formulation is supplemented by a consistently linearized integration interval segmentation that avoids numerical integration across strong discontinuities at beam endpoints. 
In combination with a smoothed contact force law and the applied $C^1$-continuous beam element formulation, this procedure drastically reduces the numerical integration error. 
It has been verified numerically that this reduction of the integration error is an essential prerequisite in order to obtain optimal spatial convergence rates. Moreover, the resulting line-to-line 
contact algorithm has been supplemented by the contact contributions of the beam endpoints. It has been shown, that these contributions are likely to occur in systems with arbitrarily 
distributed and oriented fibers.\\

Finally, it has been verified by means of several numerical examples that all the described individual model components are necessary in order to obtain an accurate, 
consistent and robust contact algorithm that can model complex systems of slender fibers with arbitrary contact configurations. Amongst others, a new numerical test case suitable for 
line-to-line contact scenarios has been designed and a corresponding analytical solution based on the Kirchhoff theory of thin rods has been derived. 
This test case in combination with the associated analytical solution can serve as valuable benchmark for the proposed model but also for future beam-to-beam contact formulations.

%-------------------------------------------------------------------------------

\appendix

%-------------------------------------------------------------------------------

%-------------------------------------------------------------------------------
%
\section{Residual contributions and linearization of applied beam element formulation}
\label{anhang:reslin_beamelement}
%
%-------------------------------------------------------------------------------
After some reformulations of~\eqref{weakform}, the residual contributions $\mb{r}_{int}, \mb{r}_{kin}$ and $\mb{r}_{ext}$ of one torsion-free beam element according to Section~\ref{sec:beamformulation}
due to internal, inertia and external forces can be derived:
\begin{align}
\label{res_beamelement}
\begin{split}
   \mb{r}_{int}&=\int \limits_{-1}^{1} \left[ \mb{N}^{\prime T} \! \! \left(EA \mb{t_1}+EI\mb{t_2} \right)
   + \mb{N}^{\prime \prime T} \! EI\mb{t_3}  \right]\frac{l_{ele}}{2}d\xi,  \quad
  \mb{r}_{kin}= \int \limits_{-1}^{1} \! \mb{N}^T \! \rho A \ddot{\boldsymbol{r}} \frac{l_{ele}}{2}d\xi, \\
  \mb{r}_{ext}&= - \int \limits_{-1}^{1} \left[\mb{N}^T \mb{\tilde{f}} + \mb{N}^{\prime T} \! \left( \mb{\tilde{m} \times \mb{t_4}} \right) \right]\frac{l_{ele}}{2}d\xi
  -  \Bigg[\mb{N}^T \bar{\mb{f}} + \mb{N}^{\prime T} \! \left(\bar{\mb{m}} \times \mb{t_4} \right) \Bigg]_{\varGamma_{\sigma}} = 0.
\end{split}
\end{align}
Here, we have introduced the following additional abbreviations:
\begin{align}
\mb{t_1} := \frac{\mb{r}^{\prime}}{||\mb{r}^{\prime}||} \left( ||\mb{r}^{\prime}||-1 \right), \, \, \,
\mb{t_2} := \frac{2\mb{r}^{\prime}(\mb{r}^{\prime T} \mb{r}^{\prime \prime})^2}{||\mb{r}^{\prime}||^6}-\frac{\mb{r}^{\prime}(\mb{r}^{\prime \prime T}
\mb{r}^{\prime \prime })
+\mb{r}^{\prime \prime }(\mb{r}^{\prime T} \mb{r}^{\prime \prime})}{||\mb{r}^{\prime}||^4}, \, \, \,
\mb{t_3} := \frac{\mb{r}^{\prime \prime}}{||\mb{r}^{\prime}||^2}-\frac{\mb{r}^{\prime}(\mb{r}^{\prime T} \mb{r}^{\prime \prime})}{||\mb{r}'||^4}, \,\,\,
\mb{t_4} := \frac{\mb{r}^{\prime}}{||\mb{r}^{\prime}||^2}.
\end{align}
Consequently, we obtain the following expressions for the corresponding linearizations $\mb{k}_{int}, \mb{k}_{kin}$ and $\mb{k}_{ext}$:
\begin{align}
\label{linres_beamelement}
\begin{split}
   \mb{k}_{int}&=\frac{\partial \mb{r}_{int}}{\partial \mb{d}}=\int \limits_{-1}^{1} \left[ \mb{N}^{\prime T} \! \! \left(EA \frac{\partial \mb{t_1}}{\partial \mb{d}}
   +EI \frac{\partial \mb{t_2}}{\partial \mb{d}} \right)
   + \mb{N}^{\prime \prime T} \! EI \frac{\partial \mb{t_3}}{\partial \mb{d}}  \right]\frac{l_{ele}}{2}d\xi,  \quad
     \mb{k}_{kin}=\frac{\partial \mb{r}_{kin}}{\partial \mb{d}}= \rho A \ddot{d}_{,d} \int \limits_{-1}^{1} \! \mb{N}^T \mb{N} \frac{l_{ele}}{2}d\xi  = \text{const.}, \\
  \mb{k}_{ext}&=\frac{\partial \mb{r}_{ext}}{\partial \mb{d}}= - \int \limits_{-1}^{1} \left[\mb{N}^T \mb{\tilde{f}} + \mb{N}^{\prime T} \! \left( \mb{S}(\mb{\tilde{m}) \frac{\partial \mb{t_4}}{\partial \mb{d}}} \right) \right]\frac{l_{ele}}{2}d\xi
  -  \Bigg[\mb{N}^T \bar{\mb{f}} + \mb{N}^{\prime T} \! \left( \mb{S}(\bar{\mb{m}}) \frac{\partial \mb{t_4}}{\partial \mb{d}} \right) \Bigg]_{\varGamma_{\sigma}} = 0,
\end{split}
\end{align}
where $\ddot{d}_{,d}$ is typically a constant factor depending on the applied time integration scheme and $\mb{S}(.)$ is a skew-symmetric matrix that represents the 
cross-product, i.e. $\mb{S}(\mb{a})\mb{b}=\mb{a}\times \mb{b} \, \forall \, \mb{a}, \mb{b} \in \Re^3$. Additionally, we have:
\begin{align}
\begin{split}
 \frac{\partial \mb{t_1}}{\partial \mb{d}} & = \left[ \frac{\left(||\mb{r}^{\prime}||-1\right)}{||\mb{r}^{\prime}|| }  \mb{I}_3
 +\frac{1}{||\mb{r}^{\prime}||^3} \left(\mb{r}^{\prime} \otimes \mb{r}^{\prime T} \right)
 \right]\mb{N}^{\prime}, \\
 %---------------------------------------------------------------------------------------------------------------------------------
 \frac{\partial \mb{t_2}}{\partial \mb{d}}
   & = \Bigg[ 
   \left\{ \frac{2(\mb{r}^{\prime T} \mb{r}^{\prime \prime})^2}{||\mb{r}^{\prime}||^6}-\frac{(\mb{r}^{\prime \prime T} \mb{r}^{\prime \prime})}{||\mb{r}^{\prime}||^4}  \right\} \mb{I}_3
  +\left\{ \frac{-12(\mb{r}^{\prime T} \mb{r}^{\prime \prime})^2}{||\mb{r}^{\prime}||^8}+\frac{4(\mb{r}^{\prime \prime T} \mb{r}^{\prime \prime})}{||\mb{r}^{\prime}||^6}  \right\} 
   \left(\mb{r}^{\prime} \otimes \mb{r}^{\prime T} \right)
  +\frac{4(\mb{r}^{\prime T} \mb{r}^{\prime \prime})}{||\mb{r}^{\prime}||^6} \left(\mb{r}^{\prime} \otimes \mb{r}^{\prime \prime T} \right) \\
  & \,\,\,\,\,\,\,\, +\frac{4(\mb{r}^{\prime T} \mb{r}^{\prime \prime})}{||\mb{r}^{\prime}||^6} \left(\mb{r}^{\prime \prime} \otimes \mb{r}^{\prime T} \right)
  -\frac{1}{||\mb{r}^{\prime}||^4} \left(\mb{r}^{\prime \prime} \otimes \mb{r}^{\prime \prime T} \right)
   \Bigg]\mb{N}^{\prime} \\
  & + \left[ -\frac{(\mb{r}^{\prime T} \mb{r}^{\prime \prime})}{||\mb{r}^{\prime}||^4}  \mb{I}_3
  +\frac{4(\mb{r}^{\prime T} \mb{r}^{\prime \prime})}{||\mb{r}^{\prime}||^6} \left(\mb{r}^{\prime} \otimes \mb{r}^{\prime T} \right)
  -\frac{2}{||\mb{r}^{\prime}||^4} \left(\mb{r}^{\prime} \otimes \mb{r}^{\prime \prime T} \right)
  -\frac{1}{||\mb{r}^{\prime}||^4} \left(\mb{r}^{\prime \prime} \otimes \mb{r}^{\prime T} \right)
  \right]\mb{N}^{\prime \prime}, \\
 %---------------------------------------------------------------------------------------------------------------------------------
 \frac{\partial \mb{t_3}}{\partial \mb{d}} & = \!
  \left[ -\frac{(\mb{r}^{\prime T} \mb{r}^{\prime \prime})}{||\mb{r}^{\prime}||^4}  \mb{I}_3
 +\frac{4(\mb{r}^{\prime T} \mb{r}^{\prime \prime})}{||\mb{r}^{\prime}||^6} \left(\mb{r}^{\prime} \otimes \mb{r}^{\prime T} \right)
 -\frac{2}{||\mb{r}^{\prime}||^4} \left(\mb{r}^{\prime \prime} \otimes \mb{r}^{\prime T} \right)
 -\frac{1}{||\mb{r}^{\prime}||^4} \left(\mb{r}^{\prime} \otimes \mb{r}^{\prime \prime T} \right)
 \right]\mb{N}^{\prime} \\
  & + \left[ \frac{1}{||\mb{r}^{\prime}||^2}  \mb{I}_3
 -\frac{1}{||\mb{r}^{\prime}||^4} \left(\mb{r}^{\prime} \otimes \mb{r}^{\prime T} \right)
 \right]\mb{N}^{\prime \prime}, \\
 %--------------------------------------------------------------------------------------------------------------------------------
  \frac{\partial \mb{t_4}}{\partial \mb{d}} & = \! \left[ \frac{1}{||\mb{r}^{\prime}||^2 }  \mb{I}_3
 -\frac{2}{||\mb{r}^{\prime}||^4}  \left(\mb{r}^{\prime} \otimes \mb{r}^{\prime T} \right)
 \right]\mb{N}^{\prime}
\end{split}
\end{align}

It can easily be shown that in the absence of external moments, i.e. $\mb{\tilde{m}}\!=\!\mb{\bar{m}}\!=\!\mb{0}$, the overall stiffness matrix is symmetric. 
If the MCS method is applied, it is sensible to slightly reformulate the element residual contribution due to axial tension. 
Eventually, the original contribution $\mb{r}_{int,EA}$ and the alternative MCS-contribution read:
\begin{align}
\label{res_beamelementmcs}
\begin{split}
   \mb{r}_{int,EA}&=EA \! \int \limits_{-1}^{1} \mb{N}^{\prime T}  \mb{t_1} \frac{l_{ele}}{2}d\xi = 
   EA \int \limits_{-1}^{1} \left(\frac{\partial \epsilon (\xi)}{\partial \mb{d}}\right)^T \! \! \epsilon(\xi) \, \frac{l_{ele}}{2}d\xi \quad \text{with} \quad
   \epsilon=||\mb{r}^{\prime}||-1, \quad 
   \frac{\partial \epsilon}{\partial \mb{d}}=\frac{\mb{r}^{\prime T}\mb{N}^{\prime}}{||\mb{r}^{\prime}||},\\
   \bar{\mb{r}}_{int,EA}&=
   EA \int \limits_{-1}^{1} \left(\frac{\partial \epsilon(\xi^i)}{\partial \mb{d}}\right)^T \! \!  L^i(\xi) \, L^j(\xi) \, \epsilon(\xi^j) \, \frac{l_{ele}}{2}d\xi 
   \quad \text{with} \quad i,j=1,2,3; \quad \xi^1\!\!=\!-1,\,\xi^2\!\!=\!0,\,\xi^3\!\!=\!1.
\end{split}
\end{align}
Accordingly, the corresponding contributions of the axial tension terms to the element stiffness matrix yield:
\begin{align}
\label{linres_beamelementmcs}
\begin{split}
   \mb{k}_{int,EA}&=
   EA \int \limits_{-1}^{1} \left[ \left(\frac{\partial^2 \epsilon (\xi)}{\partial \mb{d}^2}\right)^T \! \! \epsilon(\xi) 
    +\left(\frac{\partial \epsilon (\xi)}{\partial \mb{d}}\right)^T \left(\frac{\partial \epsilon (\xi)}{\partial \mb{d}}\right) 
\right] \frac{l_{ele}}{2} d\xi \quad \text{with} \quad
    \frac{\partial^2 \epsilon (\xi)}{\partial \mb{d}^2}=
    \frac{\mb{N}^{\prime T}}{||\mb{r}^{\prime}||}\left( \mb{I}_3 - \frac{\mb{r}^{\prime} \otimes \mb{r}^{\prime T}}{||\mb{r}^{\prime}||^2} \right)\mb{N}^{\prime},\\
       \bar{\mb{k}}_{int,EA}&=
   EA \int \limits_{-1}^{1} \left[ \left(\frac{\partial^2 \epsilon(\xi^i)}{\partial \mb{d}^2}\right)^T \! \!  L^i(\xi) \, L^j(\xi) \, \epsilon(\xi^j) 
   + \left(\frac{\partial \epsilon(\xi^i)}{\partial \mb{d}}\right)^T \! \!  L^i(\xi) \, L^j(\xi) \, \left(\frac{\partial \epsilon(\xi^i)}{\partial \mb{d}}\right) \right ]\frac{l_{ele}}{2}d\xi.
\end{split}
\end{align}
In equations \eqref{res_beamelementmcs} and \eqref{linres_beamelementmcs} applies the summation convention over the repeated indices $i$ and $j$.

%-------------------------------------------------------------------------------
%
\section{Linearization of point-to-point, endpoint-to-line and endpoint-to-endpoint contact contributions}
\label{anhang:linearizationendpoint}
%
%-------------------------------------------------------------------------------

Since the endpoint contact contributions can be regarded as a special case of the point contact formulation, we start with the linearization 
of this formulation. The linearization of~\eqref{point_discreteweakform} has the following general form: 
\begin{align}
\label{point_general_lin}
\mb{k}_{con,l}=\dfrac{d \mb{r}_{con,l}}{d \mb{d}_{12} }=
  \dfrac{\partial \mb{r}_{con,l}}{\partial \mb{d}_{12} } + \dfrac{\partial \mb{r}_{con,l}}{\partial \xi_c}\dfrac{d \xi_c}{d \mb{d}_{12} }
+ \dfrac{\partial \mb{r}_{con,l}}{\partial \eta_c}\dfrac{d \eta_c}{d \mb{d}_{12} }
  \quad \text{for} \quad l=1,2.
\end{align}
Here, the derivatives $d \xi_c/d \mb{d}_{12}$ and $d \eta_c/d \mb{d}_{12}$ stem from a linearization of the orthogonality conditions~\eqref{point_orthocond}:
\begin{align}
\begin{split}
\label{point_linorthogonality}
         \mb{A} (\xi_c,\eta_c) \cdot
         \left(
         \frac{d \xi_c}{d \mb{d}_{12}}^T,
         \frac{d \eta_c}{d \mb{d}_{12}}^T
         \right)^T
         & =  -\mb{B} (\xi_c,\eta_c), \\    
         \text{with} \quad \mb{A} & 
         =\left(
         \begin{matrix}
         p_{1,\xi} & p_{1,\eta} \\
         p_{2,\xi} & p_{2,\eta} \\     
         \end{matrix}
         \right)
         =\left(
         \begin{matrix}
         \mb{r}_{1,\xi}^T \mb{r}_{1,\xi} +(\mb{r}_1-\mb{r}_2)^T \mb{r}_{1,\xi \xi} & -\mb{r}_{1,\xi}^T \mb{r}_{2,\eta} \\
         \mb{r}_{1,\xi}^T \mb{r}_{2,\eta} & -\mb{r}_{2,\eta}^T \mb{r}_{2,\eta} +(\mb{r}_1-\mb{r}_2)^T \mb{r}_{2,\eta \eta}       
         \end{matrix}
         \right),\\
         \text{and} \quad \mb{B} & 
         =\left(
         \begin{matrix}
         p_{1,\mb{d}_{12}} \\
         p_{2,\mb{d}_{12}}      
         \end{matrix}
         \right)
         \hspace{0.058\textwidth}=\left(
         \begin{matrix}
         (\mb{r}_1-\mb{r}_2)^T \mb{N}_{1,\xi} + \mb{r}_{1,\xi}^T \mb{N}_{1} & -\mb{r}_{1,\xi}^T \mb{N}_{2} \\
         \mb{r}_{2,\eta}^T \mb{N}_{1} & (\mb{r}_1-\mb{r}_2)^T \mb{N}_{2,\eta} - \mb{r}_{2,\eta}^T \mb{N}_{2}       
         \end{matrix}
         \right) .
\end{split}
\end{align}
Here, the terms $p_{1,\xi}, p_{1,\eta}, p_{2,\xi}$ and $p_{2,\eta}$, which are collected in matrix $\mb{A}$, can be used for an iterative 
solution of the orthogonality conditions~\eqref{point_orthocond} for the unknown closest point coordinates $\xi_c$ and $\eta_c$ by means of 
a local Newton-Raphson scheme. The partial derivatives of the residual vectors with respect to $\mb{d}_{12}$
as occurring in~\eqref{point_general_lin} are given by:
\begin{align}
\label{point_discreteweakformapp2}
\begin{split}
  \frac{\partial \mb{r}_{con,1}}{\partial \mb{d}_{12}} & =  
  \varepsilon \left( 
  \mb{N}_{1}^T \mb{n} \frac{\partial g}{\partial \mb{d}_{12}}
  + g \mb{N}_{1}^T  \frac{\partial \mb{n}}{\partial \mb{d}_{12}}
  \right), \quad
    \frac{\partial \mb{r}_{con,2}}{\partial \mb{d}_{12}} =  
  \varepsilon \left( 
  \mb{N}_{2}^T \mb{n} \frac{\partial g}{\partial \mb{d}_{12}}
  + g \mb{N}_{2}^T  \frac{\partial \mb{n}}{\partial \mb{d}_{12}}
  \right), \\
  \frac{\partial g}{\partial \mb{d}_{12}} & = \mb{n}^T \left[ \mb{N}_{1}, -\mb{N}_{2} \right], \quad
  \frac{\partial \mb{n}}{\partial \mb{d}_{12}} = \frac{\mb{I}_3-\mb{n} \otimes \mb{n}^T}{||\mb{r}_1-\mb{r}_2||} \left[ \mb{N}_{1}, -\mb{N}_{2} \right],
\end{split}
\end{align}
Correspondingly, the partial derivatives with respect to the closest point coordinates $\xi_c$ and $\eta_c$ take the following form:
\begin{align}
\label{point_discreteweakformapp3}
\begin{split}
  \frac{\partial \mb{r}_{con,1}}{\partial \xi_c} & =  
  \varepsilon \left( 
  \mb{N}_{1}^T \mb{n} g_{,\xi}
  + g \mb{N}_{1,\xi}^T \mb{n}
  + g \mb{N}_{1}^T  \mb{n}_{,\xi}
  \right)\big|_{(\xi_c,\eta_c)}, \quad
  \frac{\partial \mb{r}_{con,2}}{\partial \xi_c}  =  
  \varepsilon \left( 
  \mb{N}_{2}^T \mb{n} g_{,\xi}
  + g \mb{N}_{2}^T  \mb{n}_{,\xi}
  \right)\big|_{(\xi_c,\eta_c)}, \\
   \frac{\partial \mb{r}_{con,1}}{\partial \eta_c} & =  
  \varepsilon \left( 
  \mb{N}_{1}^T \mb{n} g_{,\eta}
  + g \mb{N}_{1}^T  \mb{n}_{,\eta}
  \right)\big|_{(\xi_c,\eta_c)}, \quad
  \frac{\partial \mb{r}_{con,2}}{\partial \eta_c}  =  
  \varepsilon \left( 
  \mb{N}_{2}^T \mb{n} g_{,\eta}
  + g \mb{N}_{2,\eta}^T \mb{n}
  + g \mb{N}_{2}^T  \mb{n}_{,\eta}
  \right)\big|_{(\xi_c,\eta_c)}, \\
  g_{,\xi} &= \mb{n}^T \mb{r}_{1,\xi}, \quad
  g_{,\eta}= -\mb{n}^T \mb{r}_{2,\eta}, \quad
  % \mb{n}_{,\xi}=\frac{\mb{r}_{1,\xi}}{||\mb{r}_1-\mb{r}_2||}, \quad
  %\mb{n}_{,\eta}=-\frac{\mb{r}_{2,\eta}}{||\mb{r}_1-\mb{r}_2||}.
  \mb{n}_{,\xi}=\frac{\mb{I}_3-\mb{n} \otimes \mb{n}^T}{||\mb{r}_1-\mb{r}_2||}\mb{r}_{1,\xi}, \quad
  \mb{n}_{,\eta}=-\frac{\mb{I}_3-\mb{n} \otimes \mb{n}^T}{||\mb{r}_1-\mb{r}_2||}\mb{r}_{2,\eta}.
\end{split}
\end{align}
Depending on the case (point-, line- or endpoint-contact), \eqref{point_discreteweakformapp3} can be simplified due to $\mb{n}^T\mb{r}_{1,\xi}=0$ and/or $\mb{n}^T\mb{r}_{2,\eta}=0$.
In case of endpoint contact, only the partial derivatives $d \xi_c / d\mb{d}_{12}$ and $d \eta_c / d \mb{d}_{12}$ have to be adapted, while 
all other terms remain unchanged. In case of contact between an endpoint of beam 1, i.e. $\xi_c=-1$ or $\xi_c=1$, with a segment $\eta_c \in [-1;1]$ on beam 2, we consider the second 
line of~\eqref{point_linorthogonality} in order to determine $d \eta_c / d\mb{d}_{12}$, while $d \xi_c / d \mb{d}_{12}$ vanishes:
\begin{align}
\frac{d \xi_c}{d \mb{d}_{12}} = \mb{0} \quad \text{and} \quad
         \frac{d \eta_c}{d \mb{d}_{12}} = -\frac{p_{2,\mb{d}_{12}}}{p_{2,\eta}}.
\end{align}
Correspondingly, the condition $p_{2}(\eta_c)=0$ and the derivative $p_{2,\eta}$ can be used for an iterative determination of $\eta_c$. 
In case of contact between an endpoint of beam 2, i.e. $\eta_c=-1$ or $\eta_c=1$, with a curve segment $\xi_c \in [-1;1]$ on beam 1, we have to consider the first line of~\eqref{point_linorthogonality} 
in order to determine $d \xi_c / d \mb{d}_{12}$, while $d \eta_c / d \mb{d}_{12}$ vanishes:
\begin{align}
\frac{d \xi_c}{d \mb{d}_{12}} = -\frac{p_{1,\mb{d}_{12}}}{p_{1,\xi}} \quad \text{and} \quad
\frac{d \eta_c}{d \mb{d}_{12}} = \mb{0}.     
\end{align}
In this case, the condition $p_{1}(\xi_c)=0$ and the derivative $p_{1,\xi}$ can be used for an iterative determination of $\xi_c$.
When the contact between two endpoints is considered, $\xi_c=-1$ or $\xi_c=1$ and $\eta_c=-1$ or $\eta_c=1$, we have:
\begin{align}
\frac{d \xi_c}{d \mb{d}_{12}}  = \frac{d \eta_c}{d \mb{d}_{12}} = \mb{0}.     
\end{align}

%-------------------------------------------------------------------------------
%
\section{Linearization of the line-to-line contact formulation}
\label{anhang:linearizationlinecontact}
%
%-------------------------------------------------------------------------------
We briefly want to repeat the linearization~\eqref{line_linearization} of the contributions $\mb{r}_{con,1}^{ij}$ and $\mb{r}_{con,2}^{ij}$ of one individual Gauss point:
\begin{align}
\label{line_linearization_app}
\begin{split}
\mb{k}_{con,l}^{ij} =\dfrac{d \mb{r}_{con,l}^{ij}}{d \mb{d}_{12}} & =
  \dfrac{\partial \mb{r}_{con,l}^{ij}}{\partial \mb{d}_{12}} 
 +\dfrac{\partial \mb{r}_{con,l}^{ij}}{\partial \xi_{ij}}\dfrac{d \xi_{ij}}{d \mb{d}_{12}} 
 +\dfrac{\partial \mb{r}_{con,l}^{ij}}{\partial \eta_c}\dfrac{d \eta_c}{d \mb{d}_{12}}
 +\dfrac{\partial \mb{r}_{con,l}^{ij}}{\partial \xi_{1,i}}\dfrac{d \xi_{1,i}}{d \mb{d}_{12}}
 +\dfrac{\partial \mb{r}_{con,l}^{ij}}{\partial \xi_{2,i}}\dfrac{d \xi_{2,i}}{d \mb{d}_{12}}, \quad l=1,2, \\
  \text{with} \quad \dfrac{d \xi_{ij}}{d \mb{d}_{12}} & = \dfrac{\partial \xi_{ij}}{\partial \xi_{1,i}}\dfrac{d \xi_{1,i}}{d \mb{d}_{12}}+\dfrac{\partial \xi_{ij}}{\partial \xi_{2,i}}\dfrac{d \xi_{2,i}}{d \mb{d}_{12}}, \\
  \text{and} \quad \dfrac{d \eta_c}{d \mb{d}_{12}} & = \dfrac{\partial \eta_{c}}{\partial \xi_{ij}}\dfrac{d \xi_{ij}}{d \mb{d}_{12}}+\dfrac{\partial \eta_{c}}{\partial \mb{d}_{12}}.
\end{split}
\end{align}
We focus on the most general case with an integration interval segmentation being applied on both sides of the slave element.
In the line contact case, the orthogonality condition $p_2$ on beam 2 is relevant. Its linearization reads:
\begin{align}
\label{anhang_linf2}
 p_{2,\xi} \frac{d \xi}{d \mb{d}_{12}} + p_{2,\eta_c} \frac{d \eta_c}{d \mb{d}_{12}} = -p_{2,\mb{d}_{12}} \quad \rightarrow \quad
  \frac{d \eta_c}{d \mb{d}_{12}} = \Bigg( 
 \underbrace{\frac{-p_{2,\xi_{ij}}}{p_{2,\eta}}}_{=\frac{\partial \eta_c}{\partial \xi_{ij}}} 
 \cdot \frac{d \xi_{ij}}{d \mb{d}_{12}} 
 +\underbrace{\frac{-1}{p_{2,\eta}}p_{2,\mb{d}_{12}}}_{=\frac{\partial \eta_c}{\partial \mb{d}_{12}}} 
 \Bigg) \Bigg|_{\,(\xi_{ij},\eta_c(\xi_{ij}))}.
\end{align}
With the help of \eqref{line_integrationparamvalues}, the linearization $d \xi_{ij} / d \mb{d}_{12}$ of the evaluation points on the slave beam follows as
\begin{align}
\frac{d \xi_{ij}}{d \mb{d}_{12}} = \dfrac{\partial \xi_{ij}}{\partial \xi_{1,i}}\dfrac{d \xi_{1,i}}{d \mb{d}_{12}}+\dfrac{\partial \xi_{ij}}{\partial \xi_{2,i}}\dfrac{d \xi_{2,i}}{d \mb{d}_{12}}
\quad \text{with} \quad \dfrac{\partial \xi_{ij}}{\partial \xi_{1,i}} = \frac{1.0-\bar{\xi}_j}{2}
\quad \text{and} \quad \dfrac{\partial \xi_{ij}}{\partial \xi_{2,i}} = \frac{1.0+\bar{\xi}_j}{2},
\end{align}
where $\bar{\xi}_j$ are constant Gauss point coordinates. Since $\eta$ is fixed at the master beam endpoints, one obtains from \eqref{anhang_linf2}:
\begin{align}
 \frac{d \xi_{1,i}}{d \mb{d}_{12}} = \Bigg(\underbrace{\frac{-1}{p_{2,\xi}}p_{2,\mb{d}_{12}}}_{=\frac{\partial \xi_{B1}}{\partial \mb{d}_{12}}}\Bigg) \Bigg|_{\,(\xi_{B1}(\eta_{EP}),\eta_{EP})}
 \quad \text{and} \quad
 \frac{d \xi_{2,i}}{d \mb{d}_{12}} = \Bigg(\underbrace{\frac{-1}{p_{2,\xi}}p_{2,\mb{d}_{12}}}_{=\frac{\partial \xi_{B2}}{\partial \mb{d}_{12}}}\Bigg) \Bigg|_{\,(\xi_{B2}(\eta_{EP}),\eta_{EP})}.
\end{align}
Since the linearizations  $\partial \mb{r}_{con,l}^{ij} / \partial \xi_{1,i}$ and $\partial \mb{r}_{con,l}^{ij} / \partial \xi_{2,i}$ solely stem from the explicit dependence of the 
total Jacobian $J(\xi_{ij},\xi_{1,i},\xi_{2,i})$ on the boundary coordinates $\xi_{1,i}$ and $\xi_{2,i}$, these linearizations can be rewritten as follows:
\begin{align}
\label{anhang_linearizationxiB}
\dfrac{\partial \mb{r}_{con,l}^{ij}}{\partial \xi_{1,i}} = \dfrac{\mb{r}_{con,l}^{ij}}{J(\xi_{ij},\xi_{1,i},\xi_{2,i})}\cdot J_{,\xi_{1,i}}(\xi_{ij},\xi_{1,i},\xi_{2,i}), \quad
\dfrac{\partial \mb{r}_{con,l}^{ij}}{\partial \xi_{2,i}} = \dfrac{\mb{r}_{con,l}^{ij}}{J(\xi_{ij},\xi_{1,i},\xi_{2,i})}\cdot J_{,\xi_{2,i}}(\xi_{ij},\xi_{1,i},\xi_{2,i}) \quad \text{with} \quad l=1,2.
\end{align}
The linearizations of the Jacobian occurring in \eqref{anhang_linearizationxiB} follow directly from their definition in equation \eqref{line_totaljacobian}:
\begin{align}
J_{,\xi_{1,i}}(\xi_{ij},\xi_{1,i},\xi_{2,i}) = -\frac{J_{ele}(\xi(\bar{\xi}_i))}{2}, \quad
J_{,\xi_{2,i}}(\xi_{ij},\xi_{1,i},\xi_{2,i}) =  \frac{J_{ele}(\xi(\bar{\xi}_i))}{2}.
\end{align}
The derivative $\partial \mb{r}_{con,l}^{ij}/\partial \mb{d}_{12}$ with respect to $\mb{d}_{12}$ shows strong similarities to the corresponding terms in~\ref{anhang:linearizationendpoint}:
\begin{align}
\begin{split}
  \dfrac{\partial \mb{r}_{con,1}^{ij}}{\partial \mb{d}_{12}}  \! &= \! w_j J(\xi_{ij},\xi_{1,i},\xi_{2,i}) \varepsilon \frac{\partial g(\xi_{ij})}{\partial \mb{d}_{12}}  \mb{N}_{1}^T(\xi_{ij}) \mb{n}(\xi_{ij})
  + w_j J(\xi_{ij},\xi_{1,i},\xi_{2,i}) \varepsilon  g(\xi_{ij}) \mb{N}_{1}^T(\xi_{ij}) \frac{\partial \mb{n}(\xi_{ij})}{\partial \mb{d}_{12}},\\
  \dfrac{\partial \mb{r}_{con,2}^{ij}}{\partial \mb{d}_{12}}  \! &= \!  -w_j J(\xi_{ij},\xi_{1,i},\xi_{2,i}) \varepsilon  \varepsilon \frac{\partial g(\xi_{ij})}{\partial \mb{d}_{12}}  \mb{N}_{2}^T(\eta_{c}(\xi_{ij})) \mb{n}(\xi_{ij})
  -w_j J(\xi_{ij},\xi_{1,i},\xi_{2,i}) \varepsilon  g(\xi_{ij})  \mb{N}_{2}^T(\eta_{c}(\xi_{ij})) \frac{\partial \mb{n}(\xi_{ij})}{\partial \mb{d}_{12}}.
\end{split}
\end{align}
The terms $\partial g /\partial \mb{d}_{12}$ and $\partial \mb{n}/\partial \mb{d}_{12}$ are identical to the ones presented in~\eqref{point_discreteweakformapp2}. The partial derivatives 
of the residual contributions $\mb{r}_{con,1}^{ij}$ and $\mb{r}_{con,2}^{ij}$ with respect to the evaluation points $\xi_{ij}$ and $\eta_c$ have the following form:
\begin{align}
\begin{split}
  \dfrac{\partial \mb{r}_{con,1}^{ij}}{\partial \xi_{ij}}  \! &= \! w_j J_{,\xi_{ij}}(\xi_{ij},\xi_{1,i},\xi_{2,i}) \varepsilon g(\xi_{ij})  \mb{N}_{1}^T(\xi_{ij}) \mb{n}(\xi_{ij})
  + w_j J(\xi_{ij},\xi_{1,i},\xi_{2,i}) \varepsilon g_{,\xi_{ij}}(\xi_{ij})  \mb{N}_{1}^T(\xi_{ij}) \mb{n}(\xi_{ij})\\
  & + w_j J(\xi_{ij},\xi_{1,i},\xi_{2,i}) \varepsilon g(\xi_{ij})  \mb{N}_{1,\xi_{ij}}^T(\xi_{ij}) \mb{n}(\xi_{ij})
  + w_j J(\xi_{ij},\xi_{1,i},\xi_{2,i}) \varepsilon g(\xi_{ij})  \mb{N}_{1}^T(\xi_{ij}) \mb{n}_{,\xi_{ij}}(\xi_{ij}),\\
  \dfrac{\partial \mb{r}_{con,2}^{ij}}{\partial \xi_{ij}}  \! &= - w_j J_{,\xi_{ij}}(\xi_{ij},\xi_{1,i},\xi_{2,i}) \varepsilon g(\xi_{ij})  \mb{N}_{2}^T(\xi_{ij}) \mb{n}(\xi_{ij})
  - w_j J(\xi_{ij},\xi_{1,i},\xi_{2,i}) \varepsilon g_{,\xi_{ij}}(\xi_{ij})  \mb{N}_{2}^T(\xi_{ij}) \mb{n}(\xi_{ij})\\
  & \,\, \, \,\,\, - w_j J(\xi_{ij},\xi_{1,i},\xi_{2,i}) \varepsilon g(\xi_{ij})  \mb{N}_{2}^T(\xi_{ij}) \mb{n}_{,\xi_{ij}}(\xi_{ij}),\\  
   \dfrac{\partial \mb{r}_{con,1}^{ij}}{\partial \eta_c}  \! &= \!
    w_j J(\xi_{ij},\xi_{1,i},\xi_{2,i}) \varepsilon g_{,\eta_c}(\xi_{ij})  \mb{N}_{1}^T(\xi_{ij}) \mb{n}(\xi_{ij})
   + w_j J(\xi_{ij},\xi_{1,i},\xi_{2,i}) \varepsilon g(\xi_{ij})  \mb{N}_{1}^T(\xi_{ij}) \mb{n}_{,\eta_c}(\xi_{ij}),\\
   \dfrac{\partial \mb{r}_{con,2}^{ij}}{\partial \eta_c}  \! &=
   - w_j J(\xi_{ij},\xi_{1,i},\xi_{2,i}) \varepsilon g_{,\eta_c}(\xi_{ij})  \mb{N}_{2}^T(\xi_{ij}) \mb{n}(\xi_{ij})
   - w_j J(\xi_{ij},\xi_{1,i},\xi_{2,i}) \varepsilon g(\xi_{ij})  \mb{N}_{2,\eta_c}^T(\xi_{ij}) \mb{n}(\xi_{ij})\\
   & \,\, \, \,\,\, - w_j J(\xi_{ij},\xi_{1,i},\xi_{2,i}) \varepsilon g(\xi_{ij})  \mb{N}_{2}^T(\xi_{ij}) \mb{n}_{,\eta_c}(\xi_{ij}).
\end{split}
\end{align}
The partial derivatives of $g$ and $\mb{n}$ are again identical to the ones presented in~\eqref{point_discreteweakformapp3}. The partial derivative 
$J_{,\xi_{ij}}=J_{ele,\xi_{ij}}(\xi_{2,i}-\xi_{1,i})/2$ of the total Jacobian is only relevant in case of a non-constant element Jacobian $J_{ele}$.
It should again be emphasized that this most general linearization in~\eqref{line_linearization_app} is only necessary for slave elements with valid master beam 
endpoint projections according to~\eqref{line_segmentprojection}. In practical simulations, for the vast majority of contact element pairs this is not the case, 
i.e. $d \xi_{1,i} / d \mb{d}_{12} = \mb{0}$ and $d \xi_{2,i} / d \mb{d}_{12} = \mb{0}$, and the linearization in~\eqref{line_linearization_compact} is sufficient.

%-------------------------------------------------------------------------------
%
\section{Derivation of an analytical solution for the example ``Twisting of two beams''}
\label{anhang:analyticalsolution_twistingoftwobeams}
%
%-------------------------------------------------------------------------------

For completeness, we briefly repeat the strong form of the projected Kirchhoff equilibrium equations:
\begin{align}
\label{anhang_FSDGL}
\begin{split}
  f_{\parallel}^{\prime} + \dfrac{\kappa}{1+\epsilon} \left(\tau m_{n} + m_{b}^{\prime} + \tilde{m}_{b} \right) + \tilde{f}_{\parallel}&=0, \\
  -\left( \dfrac{\tau m_{n} + m_{b}^{\prime} + \tilde{m}_{b}}{1+\epsilon}\right)^{\prime} - \dfrac{\tau}{1+\epsilon} \left( \kappa m_{\parallel} + m_{n}^{\prime} - \tau m_{b} + \tilde{m}_{n} \right) 
 +\kappa f_{\parallel} + \tilde{f}_{n} &=0, \\
  \left(\dfrac{-\tau m_{b} + m_{n}^{\prime} + \kappa m_{\parallel} + \tilde{m}_{n}}{1+\epsilon}\right)^{\prime} - \dfrac{\tau}{1+\epsilon} \left( \tau m_{n} + m_{b}^{\prime} + \tilde{m}_{b} \right) 
 +\tilde{f}_{b} &=0, \\
  m_{\parallel}^{\prime} - \kappa m_{n} + \tilde{m}_{\parallel} &= 0,
\end{split}
\end{align}
and the constitutive equations for the case of an initially straight beam with circular cross-sections:
\begin{align}
\label{anhang_FSConstitutive}
\begin{split}
  f_{\parallel} &= EA \epsilon, \\
  m_{\parallel} &= GI_T \left( \tau + \varphi^{\prime}\right), \\
  m_n &= 0, \\
  m_b &= EI \kappa,
\end{split}
\end{align}
derived in \cite{meier2015}. Here, the indices $\parallel$, $n$ and $b$ denote the components of vector-valued quantities into the directions of the unit tangent vector $\mb{t}_{FS}$, the unit normal vector $\mb{n}_{FS}$ and the unit binormal vector $\mb{b}_{FS}$ of the Frenet-Serret frame aligned to the 
considered space curve representing the beam centerline. For further definitions of the quantities occurring in \eqref{anhang_FSDGL} as well as \eqref{anhang_FSConstitutive} and for the derivation of the projected equilibrium equations, the interested reader is referred 
to Section 2.4 of \cite{meier2015}. In the following, we investigate the possibility of finding a parameter choice for the example ``Twisting of two beams'' that leads 
to a solution in form of a helix with constant slope according to \eqref{line_example2_helix} for both considered beams. Per definition, such a helix with radius $r$ and slope $h$ exhibits the following constant 
expressions for the mathematical curvature $\bar{\kappa}$ and torsion $\bar{\tau}$ along the beams length:
\begin{align}
\label{anhang_mathematical_curvature_and_torsion}
  \bar{\kappa} = \frac{r}{h^2+r^2} = \text{const.} \quad\text{and} \quad \bar{\tau} = \frac{h}{h^2+r^2} =\text{const.}
\end{align}
Since the mathematical curvature $\bar{\kappa}$ and torsion $\bar{\tau}$ are defined as angle increments per (current) arc-length increment, and the mechanically relevant quantities 
$\kappa=(1+\epsilon)\bar{\kappa}$ and $\tau=(1+\epsilon)\bar{\tau}$ are defined as angle increments per initial/undeformed arc-length, we finally get the following expressions for the kinematic 
quantities in \eqref{anhang_FSConstitutive}:
\begin{align}
\label{anhang_curvature_and_torsion}
  \kappa = \frac{r(1+\epsilon)}{h^2+r^2} = \text{const.} \quad \rightarrow \quad m_b^{\prime}=0 \quad \text{and} \quad \tau = \frac{h(1+\epsilon)}{h^2+r^2} =\text{const.}
\end{align}
The external load for a beam in static equilibrium according to 
Figure~\ref{fig:line_twobeamstwisting_config} consists of discrete point forces and moments at the left and right endpoints of the beams due to the applied Dirichlet conditions and a line 
load~$\tilde{f}_{n}$ in $n_{FS}$-direction stemming from the contact interaction. In case of a prescribed constant gap $g_0<0$ in the deformed equilibrium 
configuration, this contact line load obeys the following relation:
\begin{align}
\label{anhang_contact_lineload}
\tilde{f}_{n} = \varepsilon g_0.
\end{align}
All remaining distributed external loads vanish. Concretely, this means that:
\begin{align}
\label{anhang_external_loads}
\tilde{f}_{\parallel} = \tilde{f}_{b} = \tilde{m}_{\parallel} = \tilde{m}_{n} = \tilde{m}_{b} = 0.
\end{align}
Furthermore, we try to find the most simple solution of this kind with a prescribed constant axial tension $\epsilon=0.01$ and a constant mechanical torsion $\tau + \varphi^{\prime}=\text{const.}$ 
Together with \eqref{anhang_FSConstitutive} and \eqref{anhang_curvature_and_torsion}, this requirement leads to:
\begin{align}
\label{anhang_tension_and_mechanicaltorsion}
  f_{\parallel}^{\prime}=m_{\parallel}^{\prime}=0.
\end{align}
Inserting equations \eqref{anhang_FSConstitutive}-\eqref{anhang_tension_and_mechanicaltorsion} into the equilibrium equations \eqref{anhang_FSDGL} leads to only one remaining relation
\begin{align}
\label{anhang_FSDGL2}
   - \dfrac{\tau}{1+\epsilon} \left( \kappa m_{\parallel} - \tau m_{b} \right)  + \kappa f_{\parallel} + \tilde{f}_{n} = 0,
\end{align}
that has to be satisfied by the system parameters, while the other three equilibrium equations of \eqref{anhang_FSDGL} are satisfied automatically. From the family of solutions in~\eqref{anhang_FSDGL2}, 
we restrict ourselves to one with vanishing mechanical torsion:
\begin{align}
\label{anhang_vanishingtorsion}
   m_{\parallel} = GI_T \left( \tau + \varphi^{\prime}\right) = 0 \quad \rightarrow \quad \varphi^{\prime}=-\tau=-\frac{h}{h^2+r^2}.
\end{align}
Altogether, equations \eqref{anhang_FSDGL2} and \eqref{anhang_vanishingtorsion} postulate the following requirement for the penalty parameter:
\begin{align}
\label{anhang_finalpenalty}
   \dfrac{h^2(1+\epsilon)^2}{(h^2+r^2)^2} \cdot \frac{EIr}{h^2+r^2}  + \frac{r(1+\epsilon)EA \epsilon}{h^2+r^2} +  \varepsilon g_0 = 0 \quad \rightarrow \quad 
   \varepsilon=-\frac{(1+\epsilon)r}{(r^2+h^2)g_0}\left( EA\epsilon + \frac{EI(1+\epsilon)h^2}{(r^2+h^2)^2} \right).
\end{align}
In a next step, the Dirichlet boundary conditions have to be determined. The relation $r=R-|g_0|/2$ for the helix radius appearing in \eqref{line_example2_helix} stems from the simple observation 
that the distance between the two helix centerlines has to satisfy $2r=2R-|g_0|$ in order to generate the required gap $g_0$. With $r$ being defined this way, the 
derivation of the conditions~\eqref{line_example2_circularpath} and \eqref{line_example2_rightendpoint} is trivial in order to end up with a helix with radius $r$. However, the 
condition~\eqref{line_example2_axialdisp} for the axial displacement requires some further calculations. Thereto, we have to express the required constant 
axial tension $\epsilon=0.01$ of the helix as a function of the total length $l_c$ of the deformed helix in order to determine the helix slope $h$:
\begin{align}
\label{anhang_helixslope}
   \epsilon=\frac{l_c-l}{l}=\frac{1}{l}\int \limits_{\varphi=0}^{2\pi} \left|\left| \frac{d\mb{r}_k(\varphi)}{d \varphi} \right|\right| \, d\varphi -1  = \frac{2\pi \sqrt{r^2+h^2}}{l}-1 \quad \rightarrow \quad 
   h\!=\!\!\sqrt{\left( \left(\frac{(1.0 + \epsilon) l}{2\pi}\right)^2\!-r^2 \right)}.
\end{align}
Equation \eqref{anhang_helixslope} yields the required helix slope in case of a given helix radius $r$ and a prescribed axial tension $\epsilon$. The required axial displacement $u$ of 
the right endpoint follows from \eqref{line_example2_helix} and its initial position according to \eqref{line_example2_initialgeometry} as
\begin{align}
\label{anhang_axialdisplacement}
  \Delta d_{1,z}^l = \Delta d_{2,z}^l = u =2\pi h -l.
\end{align}
Finally, the corresponding Dirichlet-conditions for the tangential degrees of freedom have to be determined. The constant bending moment $m_b$ along the beam has to be considered by means of proper moment boundary conditions at the beam endpoints. According to 
\cite{meier2015} (see e.g. Table $1$), the virtual work contribution of an external moment vector $m_b^j \mb{b}_{FS}^j$ in $b$-direction at the left/right boundary node $j=l,r$ leads to a residual entry
\begin{align}
  m_b^j \mb{b}_{FS}^{jT} \left(\delta \alpha^j \mb{t}_{FS}^j + \frac{\mb{t}_{FS}^j \times \delta \mb{t}^j}{||\mb{t}^j||}\right) 
  = \frac{m_b^j}{||\mb{t}^j||} \delta \mb{t}^{jT}\mb{n}_{FS}^{j} = \frac{m_b^j}{||\mb{t}^j||} \delta t_{n}^{j} \quad \text{with} \quad \mb{t}_{FS}^j=\frac{\mb{t}^j}{||\mb{t}^j||}, \,\, 
  \mb{b}_{FS}^{jT} \mb{t}_{FS}^j = 0, \,\, \mb{b}_{FS}^{j} \times \mb{t}_{FS}^j = \mb{n}_{FS}^{j}
\end{align}
into the $n$-component of the corresponding nodal tangential degrees of freedom. Since the local $n$-directions coincide with the global $x$-directions at the beam endpoints here, it is sufficient 
to prescribe the $x$-components of the nodal tangents via Dirichlet constraints in order to enable proper reaction moments. According to the analytical solution in~\eqref{line_example2_helix}, these 
$x$-components have to vanish:
\begin{align}
  \Delta t_{1,x}^l=\Delta t_{2,x}^l=\Delta t_{1,x}^r=\Delta t_{2,x}^r=0.
\end{align}
Since we have chosen the system parameters in a way that leads to vanishing mechanical torsion, \textit{no} additional torsional external moments have to be applied at the beam endpoints, 
i.e. $m_{\parallel}^l \mb{t}_{FS}^l=m_{\parallel}^r \mb{t}_{FS}^r=\mb{0}$. This is the reason why the application of the torsion-free Kirchhoff beam element presented in 
Section~\ref{sec:beamformulation} is justified for this example and leads to the correct mechanical solution. As explained in \cite{meier2014} and \cite{meier2015}, the axial components of the 
nodal tangents represent the axial tension ($\epsilon^j=t_{\parallel}^j-1$) at the nodes and cannot be prescribed, but are a part of the FE solution.

%%---------------------------------------------------------------------------
%
\bibliographystyle{plain}
\bibliography{allgemein.bib}

\begin{thebibliography}{10}

\bibitem{chamekh2009}
M.~Chamekh, S.~Mani-Aouadi, and M.~Moakher.
\newblock {Modeling and numerical treatment of elastic rods with frictionless
  self-contact}.
\newblock {\em Computer Methods in Applied Mechanics and Engineering},
  198(47-48):3751--3764, 2009.

\bibitem{chamekh2014}
M.~Chamekh, S.~Mani-Aouadi, and M.~Moakher.
\newblock {Stability of elastic rods with self-contact}.
\newblock {\em Computer Methods in Applied Mechanics and Engineering},
  279:227--246, 2014.

\bibitem{crisfield1999}
M.~A. Crisfield and G.~Jelenic.
\newblock {Objectivity of strain measures in the geometrically exact
  three-dimensional beam theory and its finite-element implementation}.
\newblock {\em Proceedings of the Royal Society of London. Series A:
  Mathematical, Physical and Engineering Sciences}, 455(1983):1125--1147, 1999.

\bibitem{cyron2012}
C.J. Cyron and W.A. Wall.
\newblock {Numerical method for the simulation of the Brownian dynamics of
  rod-like microstructures with three-dimensional nonlinear beam elements}.
\newblock {\em International Journal for Numerical Methods in Engineering},
  90(8):955--987, 2012.

\bibitem{durville2004}
D.~Durville.
\newblock {Modelling of contact­friction interactions in entangled fibrous
  materials}.
\newblock In {\em VI World Wide Congress on Computational Mechanics, Beijing},
  2004.

\bibitem{durville2007}
D.~Durville.
\newblock {Finite Element Simulation of Textile Materials at Mesoscopic Scale}.
\newblock In {\em Finite element modelling of textiles and textile composites,
  Saint-Petersbourg : Russian Federation}, 2007.

\bibitem{durville2010}
D.~Durville.
\newblock Simulation of the mechanical behaviour of woven fabrics at the scale
  of fibers.
\newblock {\em International Journal of Material Forming}, 3(2):1241--1251,
  2010.

\bibitem{durville2012}
D.~Durville.
\newblock Contact-friction modeling within elastic beam assemblies: an
  application to knot tightening.
\newblock {\em Computational Mechanics}, 49(6):687--707, 2012.

\bibitem{eugster2013}
S.~R. Eugster, C.~Hesch, P.~Betsch, and Ch. Glocker.
\newblock Director-based beam finite elements relying on the geometrically
  exact beam theory formulated in skew coordinates.
\newblock {\em International Journal for Numerical Methods in Engineering},
  97(2):111--129, 2014.

\bibitem{neto2015}
A.~Gay~Neto, P.~M. Pimenta, and P.~Wriggers.
\newblock {Self-contact modeling on beams experiencing loop formation}.
\newblock {\em Computational Mechanics}, 55(1):193--208, 2015.

\bibitem{jelenic1999}
G.~Jelenic and M.~A. Crisfield.
\newblock {Geometrically exact 3D beam theory: implementation of a
  strain-invariant finite element for statics and dynamics}.
\newblock {\em Computer Methods in Applied Mechanics and Engineering},
  171(1--2):141--171, 1999.

\bibitem{konjukhov2008}
A.~Konyukhov and K.~Schweizerhof.
\newblock {On the solvability of closest point projection procedures in contact
  analysis: Analysis and solution strategy for surfaces of arbitrary geometry}.
\newblock {\em Computer Methods in Applied Mechanics and Engineering},
  197(33-40):3045--3056, 2008.

\bibitem{konjukhov2010}
A.~Konyukhov and K.~Schweizerhof.
\newblock {Geometrically exact covariant approach for contact between curves}.
\newblock {\em Computer Methods in Applied Mechanics and Engineering},
  199(37-40):2510--2531, 2010.

\bibitem{kulachenko2012}
A.~Kulachenko and T.~Uesaka.
\newblock {Direct simulations of fiber network deformation and failure}.
\newblock {\em Mechanics of Materials}, 51:1--14, 2012.

\bibitem{laursen2002}
T.~A. Laursen.
\newblock {\em {Computational contact and impact mechanics}}.
\newblock Springer-Verlag Berlin Heidelberg, 2002.

\bibitem{litewka2005}
P.~Litewka.
\newblock {The penalty and Lagrange multiplier methods in the frictional 3d
  beam-to-beam contact problem}.
\newblock {\em Civil and Environmental Engineering Reports}, 1:189--207, 2005.

\bibitem{litewka2007}
P.~Litewka.
\newblock Hermite polynomial smoothing in beam-to-beam frictional contact.
\newblock {\em Computational Mechanics}, 40(5):815--826, 2007.

\bibitem{litewka2013}
P.~Litewka.
\newblock {Enhanced multiple-point beam-to-beam frictionless contact finite
  element}.
\newblock {\em Computational Mechanics}, 52(6):1365--1380, 2013.

\bibitem{litewka2015}
P.~Litewka.
\newblock {Frictional beam-to-beam multiple-point contact finite element}.
\newblock {\em Computational Mechanics}, 56(2):243--264, 2015.

\bibitem{litewka2002}
P.~Litewka and P.~Wriggers.
\newblock {Contact between 3D beams with rectangular cross-sections}.
\newblock {\em International Journal for Numerical Methods in Engineering},
  53:2019--2041, 2002.

\bibitem{litewka2002b}
P.~Litewka and P.~Wriggers.
\newblock {Frictional contact between 3D beams}.
\newblock {\em Computational Mechanics}, 28(1):26--39, 2002.

\bibitem{meier2014}
C.~Meier, A.~Popp, and W.~A. Wall.
\newblock {An objective 3D large deformation finite element formulation for
  geometrically exact curved Kirchhoff rods}.
\newblock {\em Computer Methods in Applied Mechanics and Engineering},
  278:445--478, 2014.

\bibitem{meier2015}
C.~Meier, A.~Popp, and W.~A. Wall.
\newblock {A locking-free finite element formulation and reduced models for
  geometrically exact Kirchhoff rods}.
\newblock {\em Computer Methods in Applied Mechanics and Engineering},
  290:314--341, 2015.

\bibitem{mueller2014}
K.W. M\"uller, R.~F. Bruinsma, O.~Lieleg, A.~R. Bausch, W.~A. Wall, and A.~J.
  Levine.
\newblock {Rheology of Semiflexible Bundle Networks with Transient Linkers}.
\newblock {\em Physical Review Letters}, 112:238102, 2014.

\bibitem{popp2009}
A.~Popp, M.~W. Gee, and W.~A. Wall.
\newblock {A finite deformation mortar contact formulation using a
  primal–dual active set strategy}.
\newblock {\em International Journal for Numerical Methods in Engineering},
  79(11):1354--1391, 2009.

\bibitem{popp2010}
A.~Popp, M.~Gitterle, M.~W. Gee, and W.~A. Wall.
\newblock {A dual mortar approach for 3D finite deformation contact with
  consistent linearization}.
\newblock {\em International Journal for Numerical Methods in Engineering},
  83(11):1428--1465, 2010.

\bibitem{popp2014}
A.~Popp and W.~A. Wall.
\newblock {Dual mortar methods for computational contact mechanics -- overview
  and recent developments}.
\newblock {\em GAMM-Mitteilungen}, 37(1):66--84, 2014.

\bibitem{romero2004}
I.~Romero.
\newblock {The interpolation of rotations and its application to finite element
  models of geometrically exact rods}.
\newblock {\em Computational Mechanics}, 34:121--133, 2004.

\bibitem{romero2008}
I.~Romero.
\newblock {A comparison of finite elements for nonlinear beams: the absolute
  nodal coordinate and geometrically exact formulations}.
\newblock {\em Multibody System Dynamics}, 20(1):51--68, 2008.

\bibitem{romero2002}
I.~Romero and F.~Armero.
\newblock {An objective finite element approximation of the kinematics of
  geometrically exact rods and its use in the formulation of an
  energy–momentum conserving scheme in dynamics}.
\newblock {\em International Journal for Numerical Methods in Engineering},
  54(12):1683--1716, 2002.

\bibitem{simo1985}
J.~C. Simo.
\newblock {A Finite Strain Beam Formulation. The Three-Dimensional Dynamic
  Problem. Part I}.
\newblock {\em Computer Methods in Applied Mechanics and Engineering},
  49:55--70, 1985.

\bibitem{simo1986}
J.~C. Simo and L.~Vu~Quoc.
\newblock {A Three Dimensional Finite Strain Rod Model Part II: Computational
  Aspects}.
\newblock {\em Computer Methods in Applied Mechanics and Engineering},
  58:79--116, 1986.

\bibitem{sonneville2014}
V.~Sonneville, A.~Cardona, and O.~Brüls.
\newblock {Geometrically exact beam finite element formulated on the special
  Euclidean group}.
\newblock {\em Computer Methods in Applied Mechanics and Engineering},
  268(0):451 -- 474, 2014.

\bibitem{wohlmuth2001}
B.~I. Wohlmuth.
\newblock {\em {Discretization methods and iterative solvers based on domain
  decomposition}}.
\newblock Springer-Verlag Berlin Heidelberg, 2001.

\bibitem{wohlmuth2011}
B.~I. Wohlmuth.
\newblock {Variationally consistent discretization schemes and numerical
  algorithms for contact problems}.
\newblock {\em Acta Numerica}, 20:569--734, 2011.

\bibitem{wriggers2006}
P.~Wriggers.
\newblock {\em {Computational Contact Mechanics}}.
\newblock Springer, 2006.

\bibitem{wriggers1997}
P.~Wriggers and G.~Zavarise.
\newblock {On contact between three-dimensional beams undergoing large
  deflections}.
\newblock {\em Communications in Numerical Methods in Engineering},
  13(6):429--438, 1997.

\bibitem{zavarise2000}
G.~Zavarise and P.~Wriggers.
\newblock {Contact with friction between beams in 3-{D} space}.
\newblock {\em International Journal for Numerical Methods in Engineering},
  49(8):977--1006, 2000.

\bibitem{zupan2003}
D.~Zupan and M.~Saje.
\newblock {Finite-element formulation of geometrically exact three-dimensional
  beam theories based on interpolation of strain measures}.
\newblock {\em Computer Methods in Applied Mechanics and Engineering},
  192(49--50):5209--5248, 2003.

\end{thebibliography}
%
%%---------------------------------------------------------------------------
%
\end{document}